%% file: main.tex
\documentclass[sn-mathphys-ay]{sn-jnl}
\makeatletter
\renewenvironment{table}[1][]%
  {\begin{tableorg}[#1]\centering\tablebodyfont%
   \renewcommand\footnotetext[2][]{%
     {\removelastskip\vskip3pt%
      \let\tablebodyfont\tablefootnotefont%
      \hskip0pt\if!##1!\else{\smash{$^{##1}$}}\fi##2\par}}%
  }%
  {\end{tableorg}}
\makeatother
\usepackage[english]{babel}
\usepackage[title]{appendix}
\usepackage{color}
\usepackage[table]{xcolor}%
\usepackage{colortbl}
\usepackage{textcomp}%
\usepackage{manyfoot}%
\usepackage{graphicx}
\usepackage[normal]{caption}
\usepackage{subcaption}
\usepackage{booktabs}
\usepackage{array}
\usepackage{tabularx}
\usepackage{float}
\usepackage{tikz}
\usetikzlibrary{arrows, decorations.pathmorphing, decorations.markings, backgrounds, positioning, fit, petri}
\usepackage{quantikz}
\usepackage{pdflscape}
\usepackage{multirow}%
\usepackage{comment}
\usepackage{algorithm2e}

\usepackage{lscape}

\usepackage{amsmath,amsthm,amssymb,amsfonts}%
\usepackage{bm}
\usepackage{eucal}
\usepackage{braket,mleftright}
\usepackage[version=3]{mhchem}
\usepackage{siunitx}
\theoremstyle{plain}

\theoremstyle{plain}

\theoremstyle{plain}

\begin{document}

\title{Advancing Machine Learning Applications in Quantum Few-Body Systems}

\author*[1]{\fnm{Jin} \sur{Ziqi}}\email{ziqi.jin@u.nus.edu}

\author*[1,2]{\fnm{Paolo} \sur{Recchia}}\email{paolo\_re@nus.edu.sg}

\author[3]{\fnm{Mario} \sur{Gattobigio}}\email{mario.gattobigio@inphyni.cnrs.fr}

\affil[1]{\orgdiv{School of Computing}, \orgname{National University of Singapore}, \orgaddress{\street{13 Computing Drive}, \city{Singapore}, \postcode{117417}, \state{Singapore}}}

\affil[2]{\orgdiv{Centre for Quantum Technologies}, \orgname{National University of Singapore}, \orgaddress{\street{3 Science Drive 2, Block S15}, \city{Singapore}, \postcode{117543}, \state{Singapore}}}

\affil[3]{\orgdiv{INPHYNI}, \orgname{ Universit\'e C\^ote d'Azur, CNRS}, \orgaddress{\street{17 rue Julien Lauprêtre}, \city{Nice}, \postcode{06200}, \state{France}}}

\abstract{
This paper presents a general neural network framework for solving quantum few-body systems, extending prior methods to handle diverse particle masses, interaction types, and system configurations. Our architecture, which combines an adaptive step size with the Metropolis-Adjusted Langevin Algorithm for Monte Carlo sampling, accurately approximates the ground-state wave functions of systems featuring harmonic confinement, Gaussian two-body interactions, and including three-body forces. In ten-particle systems, it achieves lower relative energy errors (with respect to the reference values) than previous machine-learning methods. Leveraging GPU-accelerated computation, the method scales favorably with system size while maintaining robust convergence, reduced hyperparameter sensitivity, and stable training. Beyond accurate energy estimation, the model captures spatial distributions and correlation structures, offering physical insights about inter-particle structure. By unifying applicability across identical and nonidentical particles, the proposed approach establishes a versatile computational tool for exploring complex few-body quantum systems, with significant implications for advancing computational models in few-body quantum systems.}


\maketitle

\section{Introduction}
\label{sec:introduction}
\input{sections/intro}

\section{Related works}
\label{sec:relate_works}
\input{sections/related_works}

\section{Background}
\label{sec:background}
\input{sections/background}

\section{Methodology}
\label{sec:methodology}
\input{sections/methodology}

\section{Results}
\label{sec:results}
\input{sections/results}

\section{Conclusions}
\label{sec:conclusions}
\input{sections/conclusions}

\bibliography{references}

\appendix
\input{sections/appendix}

\end{document}

%% file: sections/intro.tex
The Schrödinger equation governing the dynamics of quantum few-body systems generally lacks analytical solutions once more than two particles are involved, even when the interaction potentials are explicitly known. To address this challenge, a variety of numerical approximation methods have been developed. Among the most prominent are variational techniques~\cite{cohen2019quantum}, which rely on the Variational Principle\cite{griffiths2018, cohen2019quantum}. The principle states that the expectation value of the Hamiltonian with respect to any trial state provides an upper bound to the ground-state energy, which is exactly obtained when the trial state coincides with the true ground state.

The accuracy of variational methods crucially depends on the choice of the parametrized trial state, often referred to as the ansatz, and on the optimization of its parameters. In the Quantum Few-Body literature, various ansatz have been proposed, leading to diverse optimization schemes, for instance, the use of Hyperspherical Harmonics (HH)~\cite{de1983potential, marcucci2020hyperspherical} or Gaussian bases as in the Stochastic Variational Method (SVM)~\cite{suzuki1998stochastic} or Gradient Variational Method (GVM)~\cite{recchia2024}.

More recently, Machine Learning approaches have been explored in which the ansatz is represented by a Neural Network and the energy, i.e., the expectation value of the Hamiltonian, is estimated through Monte Carlo integration. This line of research was pioneered by Carleo and collaborators~\cite{carleo2017solving}, who applied Neural Network representations to quantum many-body spin systems. In those systems, the Hamiltonian typically acts on internal degrees of freedom such as spin, which are discrete and confined to finite-dimensional Hilbert spaces, making them well-suited for Restricted Boltzmann Machine (RBM) type ansatz.

By contrast, few-body systems are governed primarily by Hamiltonians involving external degrees of freedom, such as the spatial positions of the particles. These degrees of freedom are continuous, resulting in infinite-dimensional Hilbert spaces and posing distinct computational challenges that require different machine-learning architectures. Additionally, few-body systems feature a wide variety of interactions: from two-body and three-body couplings to higher-order terms, often coexisting with external potentials. Their behavior is also highly sensitive to particle-specific features, including variations in mass, spin, and other quantum numbers, all of which must be treated with precision. These factors explain why, even for systems composed of only a few particles, the computational demands remain considerable.

An early attempt to apply neural-network ansatz to few-body systems was made by Saito~\cite{saito2018}, who employed a simple Fully Connected Neural Network (FNN) to reproduce the ground-state energy of a few identical bosonic systems with reasonable accuracy. However, the approach was limited to systems with identical masses and exhibited strong sensitivity to hyperparameter choices, such as the initialization of particle distances and the sampling step in the Monte Carlo random walks, a drawback common to variational methods for few-body physics, including the Stochastic and Gradient-based approaches (see~\cite{recchia2024}).

The specific objectives of this paper are fourfold. The first is to generalize Neural Network models to handle particles of different masses, thereby broadening the scope of physical scenarios that can be addressed. The second is to reduce hyperparameter sensitivity through adaptive techniques, such as dynamically adjusting the sampling standard deviation, which improves stability and robustness. The third objective is to enhance training reliability by modifying activation functions, network depth, and sampling methods to reduce oscillatory behavior and convergence issues. Finally, the fourth is to demonstrate scalability by showing that the framework can efficiently simulate systems of up to at least twenty particles using double-precision floating-point arithmetic with GPU acceleration, highlighting its potential for high-performance applications.

The remainder of this work is organized into five main parts. The next Section~\ref{sec:relate_works} surveys the existing literature on Quantum Few-Body systems and machine learning approaches, with particular attention to methods developed by Saito and related work on neural quantum states. Section~\ref{sec:background} provides a brief introduction to the necessary background, including the variational method for solving the Schrödinger equation, Monte Carlo integration for calculating the expectation value, and Jacobi coordinate transformations, thereby establishing the theoretical foundation for our methods.  Section~\ref{sec:methodology} presents the computational framework in detail, describing the neural network architecture, Metropolis sampling processes, and adaptive mechanisms tailored to few-body systems. Section~\ref{sec:results} discusses the experimental results, analyzing convergence, training performance, hyperparameter sensitivity, and scalability. Section~\ref{sec:conclusions} concludes by summarizing the main findings and outlining possible avenues for future research.

%% file: sections/related_works.tex
\subsection{Pre-Machine Learning methods}

The quantum few-body problem has been the subject of extensive research, owing to its importance in nuclear, atomic, and molecular physics. Exact analytical solutions exist only for very restricted cases, such as certain two-body systems, which makes approximation methods indispensable for larger systems. Broadly, the methods can be divided into non-variational and variational approaches. 

Among the non-variational approaches, the Faddeev equations~\cite{faddeev1965mathematical} provide an exact framework for three-body problems, later generalized to the Faddeev–Yakubovsky~\cite{faddeev1993quantum} formalism for systems with more than three particles. While these methods offer rigorous solutions, their factorial scaling with the number of particles limits practical applications to systems of at most five particles~\cite{lazauskas2019faddeev, braaten2006universality}. Another example is the Numerov integration method~\cite{numerov1927note}, which enables accurate solutions for low-dimensional problems but does not scale well to higher dimensions~\cite{kuenzer2016pushing, kuenzer2019four}. These limitations motivate the development of more flexible variational. 

Variational methods, rooted in the Ritz principle, form a cornerstone of few-body physics. A wide variety of basis function families have been employed to construct trial wavefunctions. The HH basis has long been studied and provides an orthonormal and complete representation~\cite{kievsky:2008_J.Phys.G}. However, it suffers from slow convergence, requiring a large number of basis functions and heavy numerical integration of potential terms. In contrast, the Gaussian basis has become the preferred choice in the Stochastic Variational Method (SVM)~\cite{suzuki1998stochastic}, due to its analytic tractability and practical overcompleteness. The SVM employs stochastic search to optimize nonlinear basis parameters, a strategy that proved effective across many few-body problems. However, it is also prone to numerical instabilities, particularly due to singularities in the overlap matrix, and becomes inefficient as the number of particles increases.  

More recently, researchers have revisited long-standing criticisms of the SVM’s stochastic parameter search. In particular, Suzuki and Varga anticipated the potential of replacing the “gambling” nature of random search with deterministic optimization. This idea has been re-examined in the light of modern gradient-based algorithms widely used in machine learning, such as stochastic gradient descent and Adam. A new family of GVMs has been proposed, directly optimizing the nonlinear parameters of Gaussian basis functions with gradient descent~\cite{recchia2024}. Empirical benchmarks demonstrate that GVMs and hybrid SVM–GVM schemes outperform traditional SVM in systems with more than three particles, showing improved stability and scalability while reducing the impact of singularities and local minima.

Although mainly used in many–body systems, Quantum Monte Carlo approaches can also be applied to study quantum few–body systems. Over the years, several variants of Monte Carlo techniques have been proposed, including Variational Monte Carlo (VMC) \cite{hammond1994monte}, Diffusion Monte Carlo \cite{anderson1975random}, Path Integral Monte Carlo \cite{barker1979quantum}, and Auxiliary Field Monte Carlo \cite{blankenbecler1981monte, ceperley1977monte}.
Since VMC is directly linked with the proposed machine learning approach, we briefly outline the fundamentals of the VMC approach before focusing on its modern developments.

In VMC, the ground-state energy is expressed as the expectation value of the Hamiltonian over a trial wave function whose parameters are optimized variationally. The corresponding integral, defined over a high-dimensional configuration space, is evaluated through stochastic sampling techniques. Typical sampling strategies include Metropolis sampling \cite{metropolis1953equation,hastings1970monte}, importance sampling based on the Fokker–Planck equation \cite{hammond1994monte}, and correlated sampling \cite{bar1988mh}. The choice of the sampling method and of the variational ansatz is crucial to ensure accuracy and numerical stability.

Once the sampling procedure is established, the variational parameters of the trial wave function are optimized by minimizing either the expectation value of the energy, its variance, or a linear combination of both \cite{umrigar1988optimized}. Recent studies have demonstrated that neural-network-based wave-function ansätze can significantly enhance the expressivity and precision of VMC simulations, extending their applicability to more complex systems~\cite{carleo2017solving, zen2019transfer}.

\subsection{Machine Learning methods}

In recent years, the application of machine learning to quantum physics has attracted considerable attention. Neural networks, in particular, have proven to be an expressive ansatz for approximating complex quantum states in many-body systems. A landmark contribution by Carleo and Troyer \cite{carleo2017solving} showed that Neural Network Quantum States (NNQS), based on RBMs, could represent ground states of many-body spin systems with remarkable accuracy.

Since then, a wide range of Neural Network architectures have been explored for quantum state representation. RBMs, as initially proposed, remain effective for spin systems with short-range interactions. Deep feed-forward neural networks were later employed by Hermann and Noé \cite{Hermann2020} to approximate ground states of molecular systems with high efficiency. Convolutional Neural Networks were introduced by Choo et al. \cite{choo2019} to study the frustrated spin-1/2 J1-J2 Heisenberg model, achieving state-of-the-art accuracy for ground-state energies except in the maximally frustrated regime. More recently, Zen et al. \cite{zen2020transfer} demonstrated that transfer learning can significantly improve both optimization time and accuracy in scaling Neural Network Quantum States, especially when leveraging GPU-accelerated implementations.

Beyond many-body systems, Neural Networks have also been applied to quantum few-body problems. Saito \cite{saito2018} proposed a simple feed-forward architecture to approximate the ground-state wavefunctions of bosonic systems with identical particles. Using a single hidden layer with hyperbolic tangent activations, the method was successfully applied to systems of three to ten bosons subject to harmonic confinement and Gaussian two-body interactions, capturing key inter-particle correlations. Despite its novelty, the approach faced important limitations, including high sensitivity to hyperparameter choices, restricted applicability to identical-particle systems, limited scalability, and difficulties in generalizing to loosely bound systems such as helium~\cite{recchia2022subleading}.

These advances highlight both the promise and the challenges of Neural Networks in quantum few-body physics. 
This work seeks to develop a generalized neural network framework for quantum few-body systems. The proposed approach aims to reduce hyperparameter sensitivity, extend applicability to heterogeneous particle configurations, and enhance training stability through adaptive sampling techniques. Building on Saito’s foundational work and incorporating more recent advances in neural architectures and sampling strategies, our goal is to establish a more robust and broadly applicable framework for quantum few-body simulations.


%% file: sections/background.tex
\subsection{Variational Methods for the Schr\"odinger Equation}

In this section, we introduce the variational framework used throughout this work, 
with particular emphasis on Monte Carlo integration in the neural network setting. 
Unless otherwise specified, the notation follows the convention of Saito~\cite{saito2018}. 
We consider a system of $N$ interacting particles, where the three-dimensional position 
and momentum of the $i$-th particle are denoted by 
$\bm{x}_i = (x^{(x)}, x^{(y)}, x^{(z)})$ and 
$\bm{p}_i = (p^{(x)}, p^{(y)}, p^{(z)})$, respectively, and the corresponding mass is $m_i$. The collection of all particle coordinates is written as $X = (\bm{x}_1, \dots, \bm{x}_N)$.

For such a system, the dynamics are governed by the Hamiltonian operator,
\begin{equation}
\label{eq:hamiltonian}
    \hat{H} = \sum_{i=1}^N -\frac{\hbar^2}{2m_i} \frac{\partial^2}{\partial \bm{x}_i^2} + \sum_{i=1}^NU(\bm{x}_i)  
    + \sum_{j>i=1}^N V(\bm{x}_i,\bm{x}_j) 
    + \sum_{k>j>i=1}^N W(\bm{x}_i,\bm{x}_j,\bm{x}_k),
\end{equation}
where $\hbar$ denotes the reduced Planck constant, and $V$ and $W$ represent the two-body and three-body interaction potentials, respectively. We also include the possibility of an external potential $U$, which will primarily be considered of harmonic-like shape. The associated Hilbert space is infinite-dimensional and admits both continuous and discrete spectra. In practice, we are concerned with the bound states of the Hamiltonian, reducing the time-independent Schr\"odinger equation to the eigenvalue problem
\begin{equation}
\label{eq:sch_eigen}
    \hat{H} \psi_n = E_n \psi_n,
\end{equation}
with a discrete set of eigenvalues $E_0 \leq E_1 \leq E_2 \leq \dots$ and the corresponding eigenfunctions $\psi_l$. Of particular interest are the ground state energy $E_0$ and corresponding wavefunction $\psi_0$, which provide the primary target in quantum few-body calculations. For this reason and for the sake of a simplified notation, we will identify $\psi_0$  as $\psi$ hereafter.

Directly solving this problem is generally intractable. Instead, one employs the Ritz variational principle~\cite{cohen2019quantum, griffiths2018}, which states that for any trial state $\phi$ in the Hilbert space, the functional
\begin{equation}
\label{eq:functional}
    E[\phi] = \frac{\braket{\phi|H|\phi}}{\braket{\phi|\phi}}
\end{equation}
provides an upper bound to the ground state energy, i.e.,
\begin{equation}
E[\phi] \geq E_0,    
\end{equation}
with equality if and only if $\phi$ coincides with the exact ground state $\psi$. The variational method thus reduces the problem of solving the Schr\"odinger equation to minimizing $E[\phi]$ over a suitably parameterized family of trial wavefunctions. 
In this work, neural network models supply such trial functions, with the expectation value estimated via Monte Carlo integration, enabling tractable approximations in the high-dimensional configuration space.

\subsection{Jacobi Coordinates}
\label{sec:jac}

In most physical scenarios, two-body and three-body interactions depend only on the relative coordinates  $\bm{x}_{ij} = \bm{x}_i - \bm{x}_j$. As a result, the Hamiltonian in Eq.~\eqref{eq:hamiltonian} is invariant under global translations, and the contribution of the center-of-mass coordinate can be decoupled. This observation motivates the introduction of Jacobi coordinates, a convenient change of variables that separates 
the overall center-of-mass motion from the internal relative dynamics.

The transformation from particle coordinates 
$X = (\bm{x}_1, \ldots, \bm{x}_N)$ to Jacobi coordinates 
$R = (\bm{r}_1, \bm{r}_2, \ldots, \bm{r}_N)$ can be written as a matrix operation,
\begin{equation}
\label{eq:transf_jacobi}
R = AX,
\end{equation}
where $A$ is the transformation matrix. Different choices of $A$ correspond to different sets of Jacobi coordinates. We can, for instance, choose a Jacobi set such that the first coordinate $\bm{r}_1$ represents the center of mass of the system,
\begin{align}
\bm{r}_1 = \sum_{j=1}^N A_{1j}\bm{x}_j = \sum_{j=1}^N \frac{m_j}{M_N}\bm{x}_j,
\end{align}
with $M_N$ denoting the total mass of the system (see Appendix~\ref{app:jac_trans}). The remaining coordinates $(\bm{r}_2, \bm{r}_3, \ldots, \bm{r}_N)$ are constructed to describe the relative motion between particles, with suitable mass-dependent scaling factors that ensure orthogonality between the center-of-mass and relative subspaces. Finally, the transformation is defined such that its Jacobian equals unity, guaranteeing that expectation values in Eq.~\eqref{eq:functional} remains invariant under the change of variables. The construction of the transformation matrix $A$ is provided in Appendix~\ref{app:jac_trans} and follows the convention adopted in~\cite{saito2018}.

This formulation allows us to rewrite the Hamiltonian in terms of Jacobi coordinates, thereby simplifying the description of the system by isolating the physically relevant internal degrees of freedom while removing the redundant center-of-mass motion. The Hamiltonian can then be rewritten as
\begin{align}
    \hat{H} = \hat{H}_1 + \hat{H}',
\end{align}
where $\hat{H}_1$ operates only on the center-of-mass coordinate $\bm{r}_1$ and $\hat{H}'$ operates on the relative coordinates $R' = (\bm{r}_2,\dots,\bm{r}_N)$.

A key advantage of this decomposition is that for many physical systems, the center-of-mass solution is known analytically. This allows us to factorize the total ground state wavefunction as
\begin{align}
    \psi(R) = \psi(\bm{r}_1, R') = \psi_1(\bm{r}_1) \cdot \psi'(R'),
\end{align}
where $\psi_1(\bm{r}_1)$ is the known center-of-mass wavefunction and $\psi'(R')$ is the wavefunction for the internal dynamics that we need to compute.

With this factorization, our computational task reduces to finding $\psi'(R')$ that minimizes the expectation value of $\hat{H}'$. The wavefunction $\psi'(R')$ is precisely what our neural network aims to approximate.

\subsection{Monte Carlo Evaluation of the Expectation Value}
\label{sec:monte_carlo}

To evaluate the expectation value of the Hamiltonian $\hat{H}'$ in Eq.~\eqref{eq:functional}, one must compute a high-dimensional integral over the relative coordinates $R'$. Following Saito~\cite{saito2018}, we approximate this functional using the neural network representation of $\psi'(R')$ combined with Monte Carlo sampling~\cite{hastings1970monte}. Since our neural network outputs a real-valued wavefunction\footnote{The approach can be extended to complex-valued wavefunctions by adding an additional output unit to the neural network, as discussed in~\cite{saito2018}.}, the expectation value is expressed as
\begin{align}
\braket{\hat{H}'} 
&=  \frac{\braket{\psi'(R')|\hat{H}'|\psi'(R')}}{\braket{\psi'(R')|\psi'(R')}} \\
&= \frac{\int \psi'(R') \hat{H}' \psi'(R') \, dR'}{\int \psi'^2(R') \, dR'} \\
&= \int P(R') \, \psi'^{-1}(R') \hat{H}' \psi'(R') \, dR' \\
&\simeq \frac{1}{N_{\text{sample}}} \sum_{i=1}^{N_{\text{sample}}} 
       \psi'^{-1}(R'_i)\hat{H}'\psi'(R'_i),
\label{eq:hamiltonian:prob_distribution}
\end{align}
where 
\begin{equation}
P(R') = \frac{\psi'^2(R')}{\int \psi'^2(R')\, dR'}
\end{equation}
denotes the probability density from which the configurations are sampled. The Metropolis–Hastings algorithm~\cite{Metropolis1953,hastings1970monte} is employed to generate a sequence of configurations $(R'_1, R'_2, \ldots, R'_{N_{\text{sample}}})$, each drawn with probability proportional to $\psi'^2(R')$. In this way, the otherwise intractable integral is replaced by a statistical average over samples.

This Monte Carlo integration scheme lies at the core of our variational method, providing a computationally tractable means of estimating expectation values in high-dimensional Hilbert spaces. The precision of the energy estimate depends both on the number of samples collected and on the efficiency of the sampling procedure. As detailed in Section~\ref{sec:sampling}, the central contribution of the present work is to enhance the accuracy and stability of training by leveraging advanced sampling strategies rather the vanilla Metropolis-Hasting.

%% file: sections/methodology.tex
This section details the methods employed in our study to simulate Quantum Few-Body systems. We first present the Neural Network architecture used to approximate the system's wavefunction, including the input transformation from particles' Jacobi coordinates to inter-particle distances. We then describe our sampling strategy based on the Metropolis algorithm in the space of Jacobi coordinates.

\subsection{The Multilayer Perceptron (MLP) architecture}
Neural Networks offer a powerful framework for representing quantum wavefunctions due to their universal approximation capabilities and computational efficiency. In our approach, we employ the Multilayer perceptron (MLP) to approximate the wavefunction, $\psi'(R')$, which encapsulates the dynamics of the Quantum Few-Body system. This section details the structure of the Neural Network, including the input representation and architecture design. 



A multilayer perceptron can be viewed as a composition of affine transformations followed by nonlinear activation functions. For our quantum wavefunction approximation, the network maps the input vector of interparticle distances $\Delta \in \mathbb{R}^M,~ M = \binom{N}{2} = \frac{N(N-1)}{2}$ to a scalar output representing the wavefunction. The computation of the interparticle distances $\Delta$ is reported in the following Section~\ref{sec:input_delta}, and the details of the transformation are in Appendix~\ref{app:transf_D}.

For a network with $L$ hidden layers, the forward propagation is defined as:

\textbf{Input layer:}
\begin{align}
\mathbf{h}^{(0)} = \Delta
\end{align}

\textbf{Hidden layers} ($l = 1, 2, \ldots, L$):
\begin{align}
\mathbf{z}^{(l)} &= \mathbf{W}^{(l)} \mathbf{h}^{(l-1)} + \mathbf{b}^{(l)} \\
\mathbf{h}^{(l)} &= \sigma^{(l)}(\mathbf{z}^{(l)})
\end{align}

\textbf{Output layer:}
\begin{align}
z^{(L+1)} &= \mathbf{w}^{(L+1)} \mathbf{h}^{(L)} + b^{(L+1)} \\
\psi'(R') &= \exp(z^{(L+1)})
\end{align}

where:
\begin{itemize}
\item $\mathbf{W}^{(l)} \in \mathbb{R}^{n_l \times n_{l-1}}$ are the weight matrices for layer $l$
\item $\mathbf{b}^{(l)} \in \mathbb{R}^{n_l}$ are the bias vectors for layer $l$  
\item $n_l$ is the number of neurons in layer $l$ (with $n_0 = M$ and $n_{L+1} = 1$)
\item $\sigma^{(l)}(\cdot)$ is the activation function for layer $l$
\item $\mathbf{w}^{(L+1)} \in \mathbb{R}^{n_L}$ and $b^{(L+1)} \in \mathbb{R}$ are the output layer parameters
\end{itemize}

The exponential activation in the output layer ensures that the wavefunction amplitude $\psi'(R')$ is strictly positive, which is appropriate for ground state wavefunctions of bosonic systems where we can choose a real, positive function.

\autoref{fig:mlp_architecture} illustrates the Neural Network architecture used for wavefunction approximation:
\begin{figure}[htbp]
\centering
\begin{tikzpicture}[
 node distance=1.5cm and 2.5cm,
 neuron/.style={circle, draw, minimum size=8mm, fill=blue!20},
 input_neuron/.style={circle, draw, minimum size=8mm, fill=green!20},
 output_neuron/.style={circle, draw, minimum size=8mm, fill=red!20},
 layer/.style={rectangle, draw, dashed, minimum width=1.5cm, minimum height=4cm, fill=gray!10},
 >=stealth
]
\node[input_neuron] (i1) at (0,2.5) {$\Delta_1$};
\node[input_neuron] (i2) at (0,1.5) {$\Delta_2$};
\node[input_neuron] (i3) at (0,0.5) {$\Delta_3$};
\node at (0,-0.5) {$\vdots$};
\node[input_neuron] (i4) at (0,-1.5) {$\Delta_M$};
\node[neuron] (h1_1) at (3,2.5) {};
\node[neuron] (h1_2) at (3,1.5) {};
\node[neuron] (h1_3) at (3,0.5) {};
\node at (3,-0.5) {$\vdots$};
\node[neuron] (h1_4) at (3,-1.5) {};
\node[neuron] (h2_1) at (5,2.5) {};
\node[neuron] (h2_2) at (5,1.5) {};
\node[neuron] (h2_3) at (5,0.5) {};
\node at (5,-0.5) {$\vdots$};
\node[neuron] (h2_4) at (5,-1.5) {};
\node at (6.5,0.5) {\large $\cdots$};
\node[neuron] (h4_1) at (8,2.5) {};
\node[neuron] (h4_2) at (8,1.5) {};
\node[neuron] (h4_3) at (8,0.5) {};
\node at (8,-0.5) {$\vdots$};
\node[neuron] (h4_4) at (8,-1.5) {};
\node[output_neuron] (o1) at (10.5,0.5) {$\psi'$};
\node[above] at (0,3.3) {$\mathbf{h}^{(0)}$};
\node[above] at (3,3.3) {$\mathbf{h}^{(1)}$};
\node[above] at (5,3.3) {$\mathbf{h}^{(2)}$};
\node[above] at (6.5,3.3) {$\cdots$};
\node[above] at (8,3.3) {$\mathbf{h}^{(L)}$};
\node[above] at (10.5,1.8) {\textbf{Output}};
\foreach \i in {1,2,3,4} {
 \foreach \j in {1,2,3,4} {
     \draw[->] (i\i) -- (h1_\j);
 }
}
\foreach \i in {1,2,3,4} {
 \foreach \j in {1,2,3,4} {
     \draw[->] (h1_\i) -- (h2_\j);
 }
}
\foreach \i in {1,2,3,4} {
 \draw[->] (h4_\i) -- (o1);
}
\node[below] at (10.5,-0.5) {$\exp(\cdot)$};
\end{tikzpicture}
\caption{Architecture of the multilayer perceptron used for quantum wavefunction approximation. The network takes interparticle distances as input and outputs the wavefunction amplitude. The exponential output activation ensures positive wavefunction values suitable for bosonic ground states.}
\label{fig:mlp_architecture}
\end{figure}
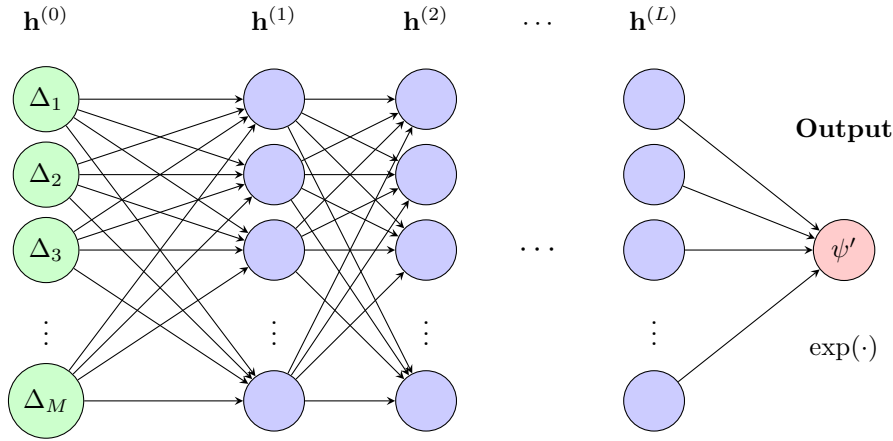

\subsubsection{Input representation}
\label{sec:input_delta}

A critical aspect of Neural Network design is determining an appropriate input representation that captures the essential physics of the system.

In standard coordinate space, the interparticle distances for a system of $N$ particles are defined as:
\begin{align}
   \Delta = \left\{ \|\bm{x}_i - \bm{x}_j\| \;\middle|\; 1 \leq i < j \leq N \right\},
   \label{eq:standard_distances}
\end{align}
where $\bm{x}_i$ represents the position of particle $i$ in 3D space, and $X = (\bm{x}_1, \bm{x}_2, \ldots, \bm{x}_N)$ denotes the full set of particle coordinates. This gives us a total of $M = \binom{N}{2} = \frac{N(N-1)}{2}$ interparticle distances.

However, in our implementation, we work with Jacobi coordinates $R'$ rather than standard coordinates $X$. As shown in Section~\ref{sec:background}, the Jacobi coordinate system eliminates the center-of-mass motion and provides a more efficient representation for few-body quantum systems. To obtain the interparticle distances from Jacobi coordinates, we use a linear transformation:
\begin{align}
   \Delta = \|D \cdot R'\|_2,
   \label{eq:jacobi_to_distances}
\end{align}
where $D \in \mathbb{R}^{M \times (N-1)}$ is the transformation matrix that converts Jacobi coordinates to interparticle distances (detailed in Appendix~\ref{app:transf_D}), $R' \in \mathbb{R}^{(N-1) \times 3}$ represents the Jacobi coordinates, and $\|\cdot\|_2$ denotes the Euclidean norm applied row-wise.

These interparticle distances $\Delta$ serve as the direct input to our Neural Network, providing a representation that respects the translational and rotational symmetries of the physical system while containing the essential geometric information needed to approximate the quantum wavefunction $\psi'(R')$.

\subsubsection{Network Architecture Variants}

We implement two architectural variants:

\begin{enumerate}
\item \textbf{Variant A:} Single hidden layer with $n_1 = 64$ nodes and $\sigma^{(1)} = \tanh$, which is similar to the Neural Network model utilized in Saito's paper~\cite{saito2018}
\item \textbf{Variant B:} Five hidden layers with $n_l = 64$ nodes each and $\sigma^{(l)} = \text{GELU}$ for $l = 1,\ldots,5$
\end{enumerate}

The GELU (Gaussian Error Linear Unit) activation function\cite{hendrycks2023gaussianerrorlinearunits}, as implemented in PyTorch, is defined as:
\begin{align}
\text{GELU}(x) = 0.5 \cdot x \cdot \left(1 + \tanh\left(\sqrt{\frac{2}{\pi}} \cdot (x + 0.044715 \cdot x^3)\right)\right)
\end{align}

This is a computationally efficient approximation to the exact GELU function $\text{GELU}(x) = x \cdot \Phi(x)$, where $\Phi(x)$ is the cumulative distribution function of the standard normal distribution~\cite{hendrycks2023gaussianerrorlinearunits}. The approximation uses a tanh-based formula that provides nearly identical results while being more computationally efficient for Neural Network training.

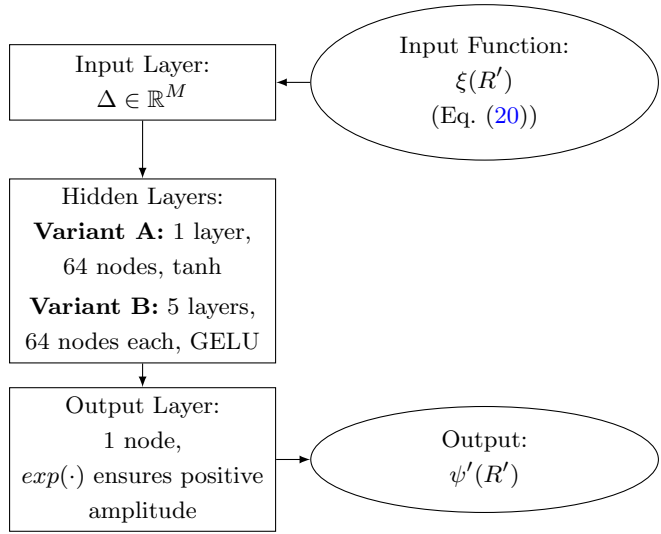
\begin{figure}[htbp]
\centering
\begin{tikzpicture}[
  node distance=2.5cm,
  auto,
  >=latex,
  every node/.style={font=\small}
]
\usetikzlibrary{shapes}
\node[draw, rectangle,
      minimum width=3.5cm, 
      minimum height=1cm] (input) {%
  \parbox{3.2cm}{\centering Input Layer:\\
  \(\Delta \in \mathbb{R}^{M}\)}
};
\node[draw, rectangle, 
      minimum width=3.5cm, 
      minimum height=1.5cm, 
      below of=input] (hidden) {%
  \parbox{3.2cm}{\centering Hidden Layers:\\
  \textbf{Variant A:} 1 layer, 64 nodes, tanh\\[0.5ex]
  \textbf{Variant B:} 5 layers, 64 nodes each, GELU}
};
\node[draw, rectangle, 
      minimum width=3.5cm, 
      minimum height=1cm, 
      below of=hidden] (output) {%
  \parbox{3.2cm}{\centering Output Layer:\\
  1 node, \\ \(exp(\cdot)\) ensures positive amplitude}
};
\draw[->] (input) -- (hidden);
\draw[->] (hidden) -- (output);
\node[draw, ellipse,
      right of=input,
      xshift=2cm] (infunc)
{%
  \parbox{3cm}{\centering Input Function:\\
  \(\xi(R')\)\\ (Eq.~\eqref{eq:jacobi_to_distances})}
};
\draw[->] (infunc.west) -- (input.east);
\node[draw, ellipse,
      right of=output,
      xshift=2cm] (outfunc)
{%
  \parbox{3cm}{\centering Output: \\
  \(\psi'(R')\)}
};
\draw[->] (output.east) -- (outfunc.west);
\end{tikzpicture}
\caption{%
  Schematic diagram of the Neural Network architecture.
  The network accepts inter-particle distances as input, processes them through hidden layers, 
  and outputs a single amplitude value using an exponential function.
}
\label{fig:nn_arch}
\end{figure}

\subsection{Sampling} 
\label{sec:sampling}

Efficient and accurate sampling plays a crucial role in the successful training of Neural Network wavefunctions for quantum few-body systems. The quality of samples directly impacts the convergence behavior of the Neural Network parameters and the reliability of quantum expectation values. High-quality sampling strategies help mitigate oscillatory behavior during training, leading to more stable optimization trajectories and ultimately more accurate physical predictions. Furthermore, proper sampling ensures that the configuration space is explored effectively, allowing the Neural Network to learn a faithful representation of the true wavefunction.

\subsubsection{General Metropolis Framework}

As shown in Section~\ref{sec:background}, our sampling framework targets the probability distribution:
\begin{align}
P(R') = \frac{\psi'^2(R')}{\int \psi'^2(R')dR'} \propto \psi'^2(R'),
\label{eq:probability_distribution}
\end{align}
where the Neural Network approximates the wavefunction $\psi'(R')$.

The general Metropolis-Hastings algorithm \cite{Metropolis1953} provides the foundation for both sampling methods implemented in our framework. The algorithm follows these steps:

\begin{enumerate}
    \item \textbf{Initialize:} Sample the initial state $R'_0$ from a uniform distribution $\mathcal{U}(-\rho, \rho)$ across all dimensions of $R'$, where $\rho$ is a boundary parameter.
    
    \item \textbf{Propose a Move:} Given the current state $R'_t$, propose a new candidate $R'_c$ according to a proposal distribution $q(R'_c | R'_t)$.
    
    \item \textbf{Apply Boundary Conditions:} Clamp the proposed coordinate to the predefined range:
    \begin{align}
        R'_c = \max(\min(R'_c, \rho), -\rho) \quad \text{(element-wise)}
    \end{align}
    
    \item \textbf{Compute Acceptance Probability:} Calculate the Metropolis-Hastings acceptance probability:
    \begin{align}
        \alpha = \min\left(1, \frac{P(R'_c) q(R'_t | R'_c)}{P(R'_t) q(R'_c | R'_t)}\right)
    \end{align}
    
    \item \textbf{Accept or Reject:} Draw a uniform random number $u \sim \mathcal{U}(0,1)$ and update:
    \begin{align}
        R'_{t+1} =
        \begin{cases}
            R'_c & \text{if } u < \alpha,\\
            R'_t & \text{otherwise}.
        \end{cases}
    \end{align}
\end{enumerate}

The specific form of the proposal distribution $q(R'_c | R'_t)$ distinguishes the different sampling methods implemented in our framework.

\subsubsection{Random Walk Metropolis Sampling}

Random Walk (RW) Metropolis sampling uses a symmetric proposal distribution, which significantly simplifies the acceptance probability calculation.

\textbf{Proposal Distribution:} The candidate state is proposed as:
\begin{align}
    R'_c = R'_t + \epsilon \mathcal{N}(0, I),
\end{align}
where $\epsilon$ is the step size parameter and $\mathcal{N}(0, I)$ is a multivariate normal distribution with mean vector $0$ and identity covariance matrix.

\textbf{Simplified Acceptance Probability:} Since the proposal distribution is symmetric ($q(R'_c | R'_t) = q(R'_t | R'_c)$), the acceptance probability simplifies to:
\begin{align}
    \alpha = \min\left(1, \frac{P(R'_c)}{P(R'_t)}\right) = \min\left(1, \frac{\psi'^2(R'_c)}{\psi'^2(R'_t)}\right).
\end{align}

This ratio quantifies whether the candidate configuration is more or less probable than the current sample configuration.

\subsubsection{Metropolis-Adjusted Langevin Algorithm (MALA) Sampling}

To enhance sampling quality, we implement the Metropolis-Adjusted Langevin Algorithm (MALA) \cite{mala1996}, which incorporates gradient information of the target distribution to guide the proposal distribution. This approach helps reduce oscillations and improves sampling efficiency compared to RW Metropolis.

\textbf{Gradient Computation:} At each step, we compute the gradient of the log probability:
\begin{align}
    \nabla \log P(R'_t) = \nabla \log \psi'^2(R'_t) = 2\nabla \log |\psi'(R'_t)|,
\end{align}
where the gradient is efficiently computed using automatic differentiation through the Neural Network via:
\begin{enumerate}
    \item Forward pass through the Neural Network to obtain $\psi'(R'_t)$
    \item Compute $\log(\psi'^2(R'_t)) = 2\log|\psi'(R'_t)|$
    \item Apply automatic differentiation to compute $\nabla \log \psi'^2(R'_t)$
\end{enumerate}

\textbf{Proposal Distribution:} The candidate state incorporates a drift term based on the gradient:
\begin{align}
    R'_c = R'_t + \frac{\epsilon^2}{2} \nabla \log P(R'_t) + \epsilon \mathcal{N}(0, I),
\end{align}
where $\epsilon$ is the step size parameter.

\textbf{Asymmetric Acceptance Probability:} Due to the gradient-based drift term, the proposal distribution is asymmetric, requiring computation of both forward and reverse proposal densities. We define:
\begin{align}
\mu_{\text{forward}} &= R'_t + \frac{\epsilon^2}{2}\nabla \log P(R'_t)\\
\mu_{\text{reverse}} &= R'_c + \frac{\epsilon^2}{2}\nabla \log P(R'_c).
\end{align}

The proposal densities are:
\begin{align}
\log q(R'_c|R'_t) &= -\frac{1}{2\epsilon^2}\|R'_c - \mu_{\text{forward}}\|^2 + C\\
\log q(R'_t|R'_c) &= -\frac{1}{2\epsilon^2}\|R'_t - \mu_{\text{reverse}}\|^2 + C,
\end{align}
where $C$ is a normalization constant that cancels in the acceptance ratio.

The full acceptance probability becomes:
\begin{align}
\alpha = \min\left(1, \frac{\psi^2(R'_c)}{\psi^2(R'_t)} \exp\left(-\frac{1}{2\epsilon^2}\|R'_t - \mu_{\text{reverse}}\|^2 + \frac{1}{2\epsilon^2}\|R'_c - \mu_{\text{forward}}\|^2\right)\right).
\end{align}

This gradient-guided approach leverages the probability landscape structure to propose moves that are more likely to be in regions of high probability, leading to improved sampling efficiency compared to RW Metropolis sampling \cite{XIFARA201414}.

\subsection{Implementation Considerations}

\subsubsection{Sampling Steps in Training}
An important practical consideration in the implementation of both RW Metropolis and MALA is determining the number of sampling steps $T$ performed during each training iteration. The sampling step parameter defines how many times the respective sampling algorithm is executed to generate a new sample configuration for each network parameter update. This represents a critical trade-off between computational efficiency and sample quality: more sampling steps allow thorough exploration of the configuration space and reduce autocorrelation between successive samples, but increase computational overhead per training iteration.

\subsubsection{Adaptive Parameter Adjustment}
Both sampling methods employ adaptive mechanisms to dynamically adjust the step size parameter $\epsilon$ for optimal sampling efficiency. The Adaptive Random Walk (ARW) method tunes $\epsilon$ based on acceptance statistics, while the Metropolis Adjusted Langevin Algorithm (MALA) relies on inherent step-size adaptation throughout training.

The running acceptance rate $\bar{a}$ is estimated using an exponential moving average:
\begin{align}
\bar{a} \leftarrow (1-\gamma)\bar{a} + \gamma a_t,
\end{align}
where $a_t$ is an indicator variable equal to 1 if the proposed move is accepted and 0 otherwise, $\gamma$ (set to 0.01) is a smoothing parameter, and $\bar{a}_0$ is initialized to 0 at the first training iteration.

The step size is then updated according to an exponential rule based on the deviation from the target acceptance rate:
\begin{align}
\epsilon \leftarrow \epsilon \exp\Bigl(\eta (\bar{a} - \alpha_{\text{target}})\Bigr),
\end{align}
where $\alpha_{\text{target}} = 0.234$ for Random Walk Metropolis and $\alpha_{\text{target}} = 0.574$ for MALA, and $\eta$ denotes the adaptation learning rate (typically set to 0.003 for both methods).

To ensure stability, adaptive adjustment is activated only after an initial warm-up phase $\tau$ (typically 1000 steps), since premature updates during the early stages—when the neural network parameters are rapidly evolving—could destabilize the sampling process.

\subsection{Pseudocode Implementation}
For readers interested in the algorithmic details, we provide complete pseudocode descriptions of both the Random Walk Metropolis and MALA algorithms with adaptive parameter adjustment in Appendix~\ref{appendix:sampling}. These pseudocode listings include initialization procedures, the main sampling loops, and the adaptive parameter update mechanisms, providing sufficient detail for implementation.

In our experiments, these adaptive sampling methods have shown significant improvements in both sampling efficiency and the overall stability of the Neural Network training process, particularly for quantum systems with complex interaction landscapes.

\subsection{Slow Introduction of Interactions}
\label{sec:slow_intro}

We implement a gradual introduction of interaction potentials to enhance the training stability and convergence of Neural Networks in our quantum few-body systems. This method, inspired by the work of Saito \cite{saito2018}, allows the Neural Network to gradually adapt to the full complexity of the interactions. While Saito's original implementation used linear ramping, our approach employs a power-law scaling that introduces interactions more rapidly at the beginning of training and then slows down as training progresses. This approach is motivated by our observation that energy optimization typically shows rapid improvement in early stages and requires more careful fine-tuning in later stages.

For a training process with a total of $n_0$ steps (set to 1000), we define a scaling factor $\lambda$ as:
\begin{align}
\lambda = \min\left(\left(\frac{t}{n_0}\right)^{\beta}, 1.0\right),
\end{align}
where $t$ represents the current training step and $\beta = 0.3$ is the power parameter controlling the introduction rate. Using a power value less than 1 ensures faster introduction at the beginning of training followed by a gradual slowdown as training progresses toward step $n_0$. In our general Hamiltonian formulation, we separate the terms that do not require gradual introduction ($\hat{H}_0$) from those that benefit from slow introduction ($\hat{H}_{\text{int}}$):
\begin{align}
\hat{H}' = \hat{H}_0 + \lambda\hat{H}_{\text{int}}.
\end{align}
The specific forms of $\hat{H}_0$ and $\hat{H}_{\text{int}}$ vary for each system configuration and are detailed in the following section.

%% file: sections/results.tex
To establish the effectiveness of our approach, we conducted comprehensive comparisons across four distinct Neural Network configurations and sampling methodologies as illustrated in the previous Section~\ref{sec:methodology}. The configurations tested, along with some of their hyperparameters, are summarized in Table~\ref{tab:nn_configs} 




\begin{table}[h]
   \centering
   \caption{Neural network configurations and sampling methods employed in comparative analysis. HL: Hidden Layers, NPL: Nodes per Layer, AF: Activation Function}
   \begin{tabular}{lcccccc}
       \hline
       \textbf{ID} & \textbf{Configuration} & \textbf{HL} & \textbf{NPL} & \textbf{Parameters} & \textbf{AF} &\textbf{Sampling} \\
       \hline
       I & \texttt{GELU-MALA} & 5 & 64 & 16,961 & GELU & MALA \\
       II & \texttt{GELU-ARW} & 5 & 64 & 16,961 & GELU & ARW \\
       III & \texttt{GELU-RW} & 5 & 64 & 16,961 & GELU & RW \\
       IV & \texttt{tanh-ARW} & 1 & 64 & 321 & tanh & ARW \\
       \hline
   \end{tabular}
   \label{tab:nn_configs}
\end{table}

These configurations were systematically evaluated across multiple quantum few-body systems (detailed in the following Section~\ref{sec:physical_sys}), to assess convergence stability, computational efficiency, and accuracy of ground-state energy predictions.

\subsection{Application to Different System Configurations}
\label{sec:physical_sys}

We implement the slow introduction method across all three system configurations by decomposing each Hamiltonian into a base component $\hat{H}_0$ and an interaction component $\hat{H}_{\text{int}}$, as specified in Section~\ref{sec:slow_intro}. The effective total Hamiltonian during training becomes:
\begin{align}
   \hat{H}' = \hat{H}_0 + \lambda\hat{H}_{\text{int}},
\end{align}
where $\lambda$ controls the interaction strength and is gradually increased during training according to our slow introduction methodology. Below, we present the specific forms of $\hat{H}_0$ and $\hat{H}_{\text{int}}$ for each system configuration.

\subsubsection{System A (Saito-like): Harmonic Confinement with Two-Body Gaussian Interaction and identical particles}
\label{sub:systemII}

The first system we investigated corresponds to the configuration studied by Saito~\cite{saito2018}. For this system, which features harmonic confinement, the base Hamiltonian includes both kinetic and potential energy terms:
\begin{align}
\label{eq:base_harm}
\hat{H}_0 = \sum_{i=1}^{N} \left(-\frac{\hbar^2}{2m} \frac{\partial^2}{\partial \bm{r}_i^2} + \frac{\gamma}{2} m \omega^2 \bm{r}_i^{2}\right),
\end{align}
where $m$ denotes the mass of the identical particles, $\omega$ is the frequency of the harmonic trap and $\gamma$ is its strength.

The interaction term is defined as
\begin{align}
\label{eq:2B_gauss}
\hat{H}_{\text{int}} = \sum_{i<j=1}^N V_{0}\exp\left[-\frac{\Delta_{ij}^2}{r_{0}^{2}}\right],
\end{align}
where the constant \(V_0\) and \(r_0\) are the strength and the range of the Gaussian interaction, respectively.

For consistency and to facilitate comparison with Ref.~\cite{saito2018, Yan_2014}, we rescale the Hamiltonian $\hat{H}$ so that all physical quantities become dimensionless. Energies, such as the base Hamiltonian in Eq.~\eqref{eq:base_harm} and the constant $V_{0}$ in Eq.~\eqref{eq:2B_gauss} are expressed in units of $E_\text{scale} = \hbar^{2}/(m r_{0}^{2})$, while lengths (e.g., $r_{0}$) are measured in units of the harmonic-oscillator characteristic length $l_{0} = \sqrt{\frac{\hbar}{m\omega}}\,$. The mass term  $\hbar^{2}/m$, $\omega$, and the coupling strength $\gamma$ in Eq.~\eqref{eq:base_harm} are chosen accordingly so that the resulting dimensionless Hamiltonian for the harmonic system takes a simple and convenient form 
\begin{align}
    \hat{H}' = \sum_{i=2}^{N} \left(-\frac{1}{2} \frac{\partial^2}{\partial \bm{r}_i^2} + \frac{1}{2}  \bm{r}_{i}^2\right) + \lambda\hat{H}_\text{int}
\end{align}




\subsubsection{System B: Two-Body and Three-Body Gaussian Interactions}
\label{sub:systemIII}

For the most complex system in our study, we examine quantum few-body systems with both two-body and three-body Gaussian-shaped interactions. These interaction potentials are particularly suitable for modeling clusters of identical helium atoms, as demonstrated in Ref.~\cite{recchia2024,recchia2022subleading, gattobigio:2011_Phys.Rev.A}.

For these systems, the base Hamiltonian contains only the kinetic energy term:
\begin{align}
\label{eq:ham_B}
   \hat{H}_0 = \sum_{i=2}^{N} \left(-\frac{\hbar^2}{2m_\text{$^4$He}} \frac{\partial^2}{\partial \bm{r}_i^2}\right).
\end{align}
where $m_\text{$^4$He}$ inidcates the mass of the $^4$He atom.

The interaction Hamiltonian incorporates both two-body and three-body interactions:
\begin{align}
\label{eq:hint_B}
   \hat{H}_{\text{int}} &= \sum_{i<j=1}^{N} V_{0}\exp\left[-\frac{\Delta_{ij}^2}{r_{0}^{2}}\right] \nonumber \\
   &+ \sum_{i<j<k=1}^{N} W_{0}\exp\left[-\frac{\Delta_{ij}^2+\Delta_{ik}^2+\Delta_{jk}^2}{\rho_{0}^{2}}\right]
\end{align}
where \(W_0\) and \(\rho_0\) denote the parameters of the three-body interaction. As in the previous system, we rescale the Hamiltonian parameters to ensure consistency and to enable direct comparison with Ref.~\cite{recchia2024,recchia2022subleading,gattobigio:2011_Phys.Rev.A}. In this case, we work in atomic units. The base Hamiltonians in Eqs.~\eqref{eq:ham_B} and the strength parameters $V_{0}$ and $W_{0}$ in Eq.~\eqref{eq:hint_B} are rescaled using the mass term $\frac{\hbar^{2}}{m_\text{$^4$He}} = \SI{43.2813}{K\, a_{0}^{2}} \,$. With this choice, energies are expressed in Kelvin (K), while lengths are given in atomic units a$_{0}$. The corresponding parameter values are reported in Table~\ref{tab:slow_intro_params}.

\begin{table}[htbp]
\centering
\caption{Parameters for slow introduction method.  
}
\begin{tabular}{llccccccc}
\toprule
\textbf{System} & \textbf{Particle Case} & \textbf{N} & \boldmath$\rho$ & \boldmath$\epsilon$ & \boldmath$V_0$ & \boldmath$r_0$ & \boldmath$W_0$ & \boldmath$\rho_0$ \\
\midrule
    A & Identical & 3--10 & 1 $l_0$ & 0.1 $l_0$ & 2.684 $E_\text{scale}$ & 0.1 $l_0$ & -- & -- \\
\midrule
    B & Identical & 3--10 & 60 a$_0$ & 6 a$_0$ & -1.227 K & 10.03 a$_0$ & 2.79 K & 7.5 a$_0$ \\
\midrule
\multirow{2}{*}{C} 
    & Identical & 2 & 500 a$_0$ & 15 a$_0$ & -1.227 K & 10.03 a$_0$ & -- & -- \\
    & Different Eq.~\eqref{eq:m_different} & 3 & 150 a$_0$ & 15 a$_0$ & Eq.~\eqref{eq:V_0} & Eq.~\eqref{eq:r_0} & -- & -- \\
\bottomrule
\end{tabular}
\label{tab:slow_intro_params}
\end{table}


\subsubsection{System C: two particles and three heterogeneous particles with a Two-Body Gaussian Interaction}
\label{sub:systemI}

As in the previous system B, the base Hamiltonian contains only the kinetic energy 
\begin{align}
\label{eq:h0_sysc}
  \hat{H}_0 = \sum_{i=2}^{N} \left(-\frac{\hbar^2}{2m_i} \frac{\partial^2}{\partial \bm r_i^2}\right),
\end{align}
where $m_i$ are the masses of the particles, which in this context can differ. The interaction Hamiltonian incorporates Gaussian-type potentials between all particle pairs:
\begin{align}
\label{eq:hint_sysc}
  \hat{H}_{\text{int}} = \sum_{i<j=1}^N V_{0}^{ij}\exp\left[-\frac{\Delta_{ij}^2}{(r_{0}^{ij})^{2}}\right]
\end{align}
The parameters $V_0^{ij}$ and $r_0^{ij}$ denote, respectively, the strength and range of the Gaussian interaction associated with the $ij$th pair. In the case of identical particles, these parameters are identical for all pairs. For our three-particle system with different masses, we employed the parameters found in~\cite{Nielsen_1998}, which describe the $^4$He$_2$--$^3$He system:
\begin{align}
    & m_1   = m_2 = 1 ~  m_\text{$^4$He} \quad &\mbox{ and } & \quad m_3 = 0.75 ~ m_\text{$^4$He} \label{eq:m_different} \\
    & V_0^{13} = V_0^{23} = \SI{-0.8925}{K} \quad &\mbox{ and } & \quad V_0^{12} = \SI{-1.227}{K} \label{eq:V_0}\\
    & r_0^{13} = r_0^{23} = \SI{10.55}{a_0} \quad &\mbox{ and } & \quad  r_0^{12} = \SI{10.03}{a_0} \label{eq:r_0}.
\end{align}
Consistently with previous section, for Equations~\eqref{eq:h0_sysc} and~\eqref{eq:hint_sysc}, and for the constants in~\eqref{eq:V_0} and~\eqref{eq:r_0}, we use \(\frac{\hbar^2}{m_{\text{$^4$He}}}\) as the unit of energy and the atomic unit a$_0$ as the unit of length.

\subsection{Hyperparameter Configuration}
\label{sec:hyperparameter}

To ensure fair comparison across different architectures and sampling methods, we maintained consistent hyperparameter settings for all experiments in System A and System B described in \ref{sub:systemII} and \ref{sub:systemIII}. For the two more extended systems in System C described in \ref{sub:systemI}, we implemented specific hyperparameter adjustments, which we detail in their respective sections.

Our optimization employed the Adam optimizer \cite{kingma2014adam} with a learning rate of $1 \times 10^{-4}$ and no decay schedule, training for 40,000 iterations to assess convergence. The sampling configuration utilized 30,000 Monte Carlo samples with 5 sampling steps for MALA and 10 steps for random walk Metropolis methods. Target acceptance ratios were set to 0.234 for adaptive random walk Metropolis and 0.574 for the Metropolis-adjusted Langevin algorithm. Step size $\epsilon$ was adjusted for each training iteration after $\tau = 1000$ steps to maintain these target ratios.


The four configurations (\texttt{GELU-ARW}, \texttt{tanh-ARW}, \texttt{GELU-MALA}, and \texttt{GELU-RW}) were trained using identical hyperparameters for System A and System B, varying only in their sampling methods and neural network architectures as specified. This standardized approach enables us to directly attribute observed differences in results to the specific architectural and sampling method choices rather than to variations in hyperparameter settings.

\subsection{Implementation Details}
\label{ch:implement}

Our implementation is developed in Python 3.10.15 using PyTorch 2.3.0, which provides a robust framework for neural network-based quantum few-body simulations. This framework enables automatic differentiation and efficient gradient-based optimization required for our sampling methods.

The simulations were executed on a single NVIDIA A100 GPU with 40 GB memory, leveraging CUDA acceleration for efficient parallel computation during the training process. For numerical precision, we employed single precision (32-bit floating point, \texttt{float32}) for most computations to optimize computational efficiency, switching to double precision (64-bit floating point, \texttt{float64}) when necessary. 


Our implementation utilizes efficient matrix transformations for coordinate conversions and distance calculations, which are well-suited for GPU parallelization. Training progress is monitored at each update step, ensuring reproducibility and enabling systematic scalability across different system sizes.

\subsection{System A (Saito-like) Results}
\label{sub:harmonic_two_body}
\subsubsection{Benchmarking with three-particles system}
\label{sub:three_particle_benchmark}
To establish a reliable performance baseline for our methods, we first conducted comprehensive benchmarking using a three-particle system interacting with a harmonic oscillator and a two-body Gaussian interaction. We selected this system because reference energy values for 3-10 particles are available in the literature~\cite{Yan_2014}.

\begin{figure}[htbp]
    \centering
    \begin{subfigure}[b]{0.43\textwidth}
        \centering
        \includegraphics[width=\textwidth, height=6cm, keepaspectratio]{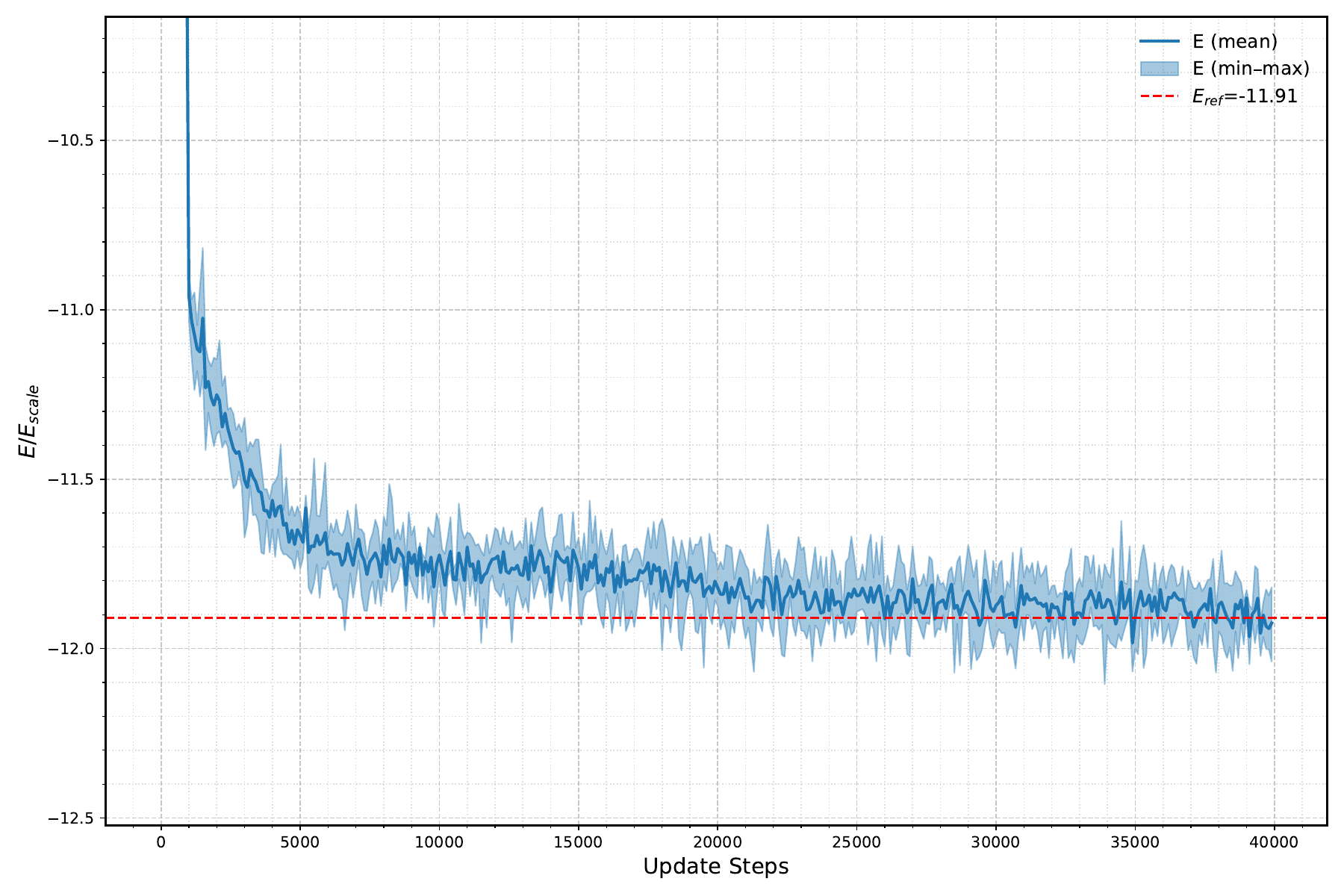}
        \caption{\texttt{GELU-MALA}}
        \label{fig:3p-gelu-mala}
    \end{subfigure}
    \hfill
    \begin{subfigure}[b]{0.43\textwidth}
        \centering
        \includegraphics[width=\textwidth, height=6cm, keepaspectratio]{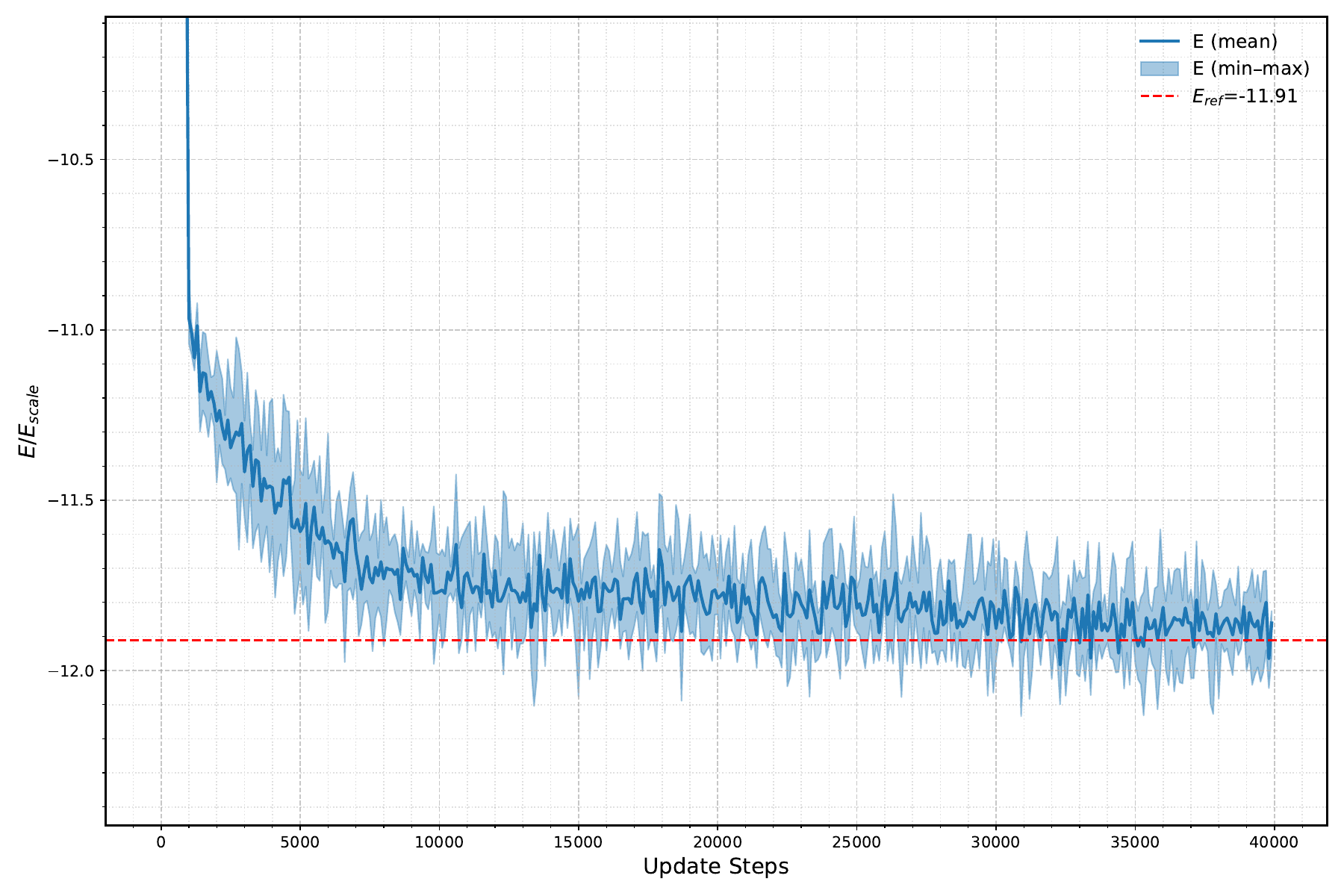}
        \caption{\texttt{GELU-ARW}}
        \label{fig:3p-gelu-adap}
    \end{subfigure}
    
    \begin{subfigure}[b]{0.43\textwidth}
        \centering
        \includegraphics[width=\textwidth, height=6cm, keepaspectratio]{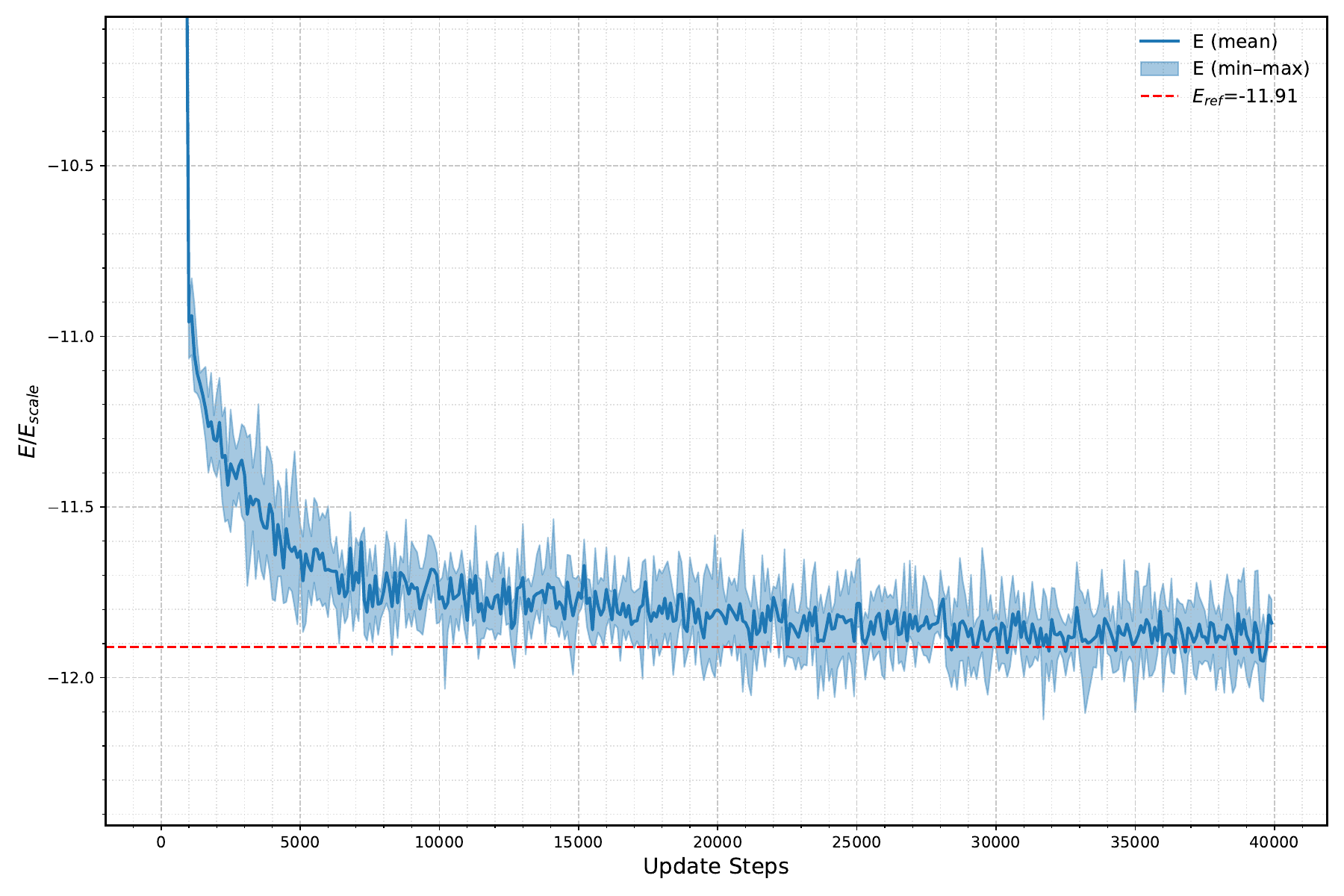}
        \caption{\texttt{GELU-RW}}
        \label{fig:3p-gelu-random}
    \end{subfigure}
    \hfill
    \begin{subfigure}[b]{0.43\textwidth}
        \centering
        \includegraphics[width=\textwidth, height=6cm, keepaspectratio]{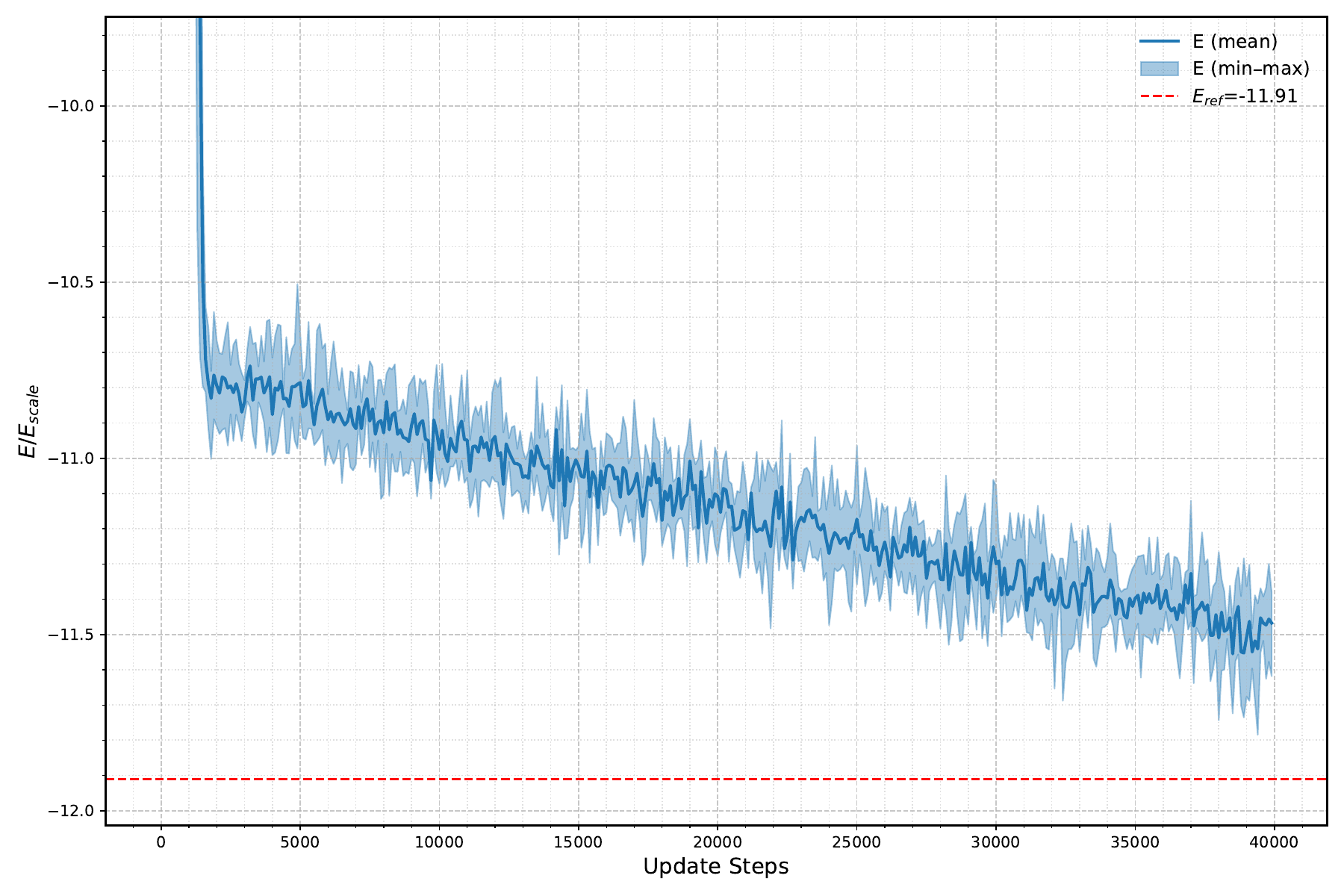}
        \caption{\texttt{tanh-ARW}}
        \label{fig:3p-tanh-adap}
    \end{subfigure}

    \caption{Energy training convergence for 3-particle harmonic potential systems with two-body interactions. The blue line represents the mean energy ($E_{\mathrm{mean}}$) across five independent trials, while the shaded blue region indicates the range between the minimum and maximum values among these trials. The red dashed line denotes the reference energy.}
    \label{fig:3particles-all-methods}
\end{figure}

Our benchmarking results reveal that for the three-particle system, GELU-based neural network architectures demonstrate comparable accuracy across different sampling methodologies. The final energy, denoted as $E_\text{final}/E_\text{scale}$, refers to the mean energy value obtained at the last training step ($40{,}000$ updates) across five independent runs. The \texttt{GELU-MALA} configuration (\autoref{fig:3p-gelu-mala}) achieved $E_\text{final}/E_\text{scale} = -12.0 \pm 0.4$, while the \texttt{GELU-ARW} approach (\autoref{fig:3p-gelu-adap}) reached $E_\text{final}/E_\text{scale} = -12.4 \pm 0.9$. These results provide statistical robustness to our findings. The \texttt{GELU-MALA} method exhibits slightly better agreement with the reference energy $E_\text{ref}/E_\text{scale} = -11.91$~\cite{Yan_2014}. The mean of relative errors\footnote{The relative errors were calculated by comparing $E_\text{final}/E_\text{scale}$ to the reference energy, $\frac{|E_\text{final} - E_\text{ref}|}{|E_\text{ref}|} \times 100\%$, following the same convention as Saito~\cite{saito2018}.} are approximately $0.76\%$ for \texttt{GELU-MALA} and $4.29\%$ for \texttt{GELU-ARW}, respectively.

These similar metrics from our five-run tests suggest that for the three-particle system, the choice of sampling algorithm may improve the final accuracy.

In contrast, the tanh activation network with adaptive random walk (\autoref{fig:3p-tanh-adap}) converged to a noticeably higher mean energy of $E_\text{final}/E_\text{scale} = -11.7 \pm 0.1$ across the five independent runs. This deviation is primarily attributable to our experimental design choice of maintaining consistent hyperparameters across all methods, particularly the learning rate, which may differ from the approach in \cite{saito2018}. While this higher energy value highlights potential sensitivity to optimization parameters, it does not detract from our focus on comparing different methods.

These benchmark results, based on multiple independent trials for statistical validity, serve as a foundational validation of our implementation and provide valuable insights into the comparative performance of different network architectures and sampling strategies in the three-particle system. While these smaller systems allow for effective method validation, the primary focus and scientific contribution of our work lies in addressing the more challenging higher-dimensional configurations with larger particle numbers and systems with different configurations, which we explore comprehensively in the following section.

\subsubsection{Scaling to Larger Systems: benchamrking with eight-particles}

\begin{figure}[htbp]
    \centering
    \begin{subfigure}[b]{0.48\textwidth}
        \centering
        \includegraphics[width=\textwidth, height=6cm, keepaspectratio]{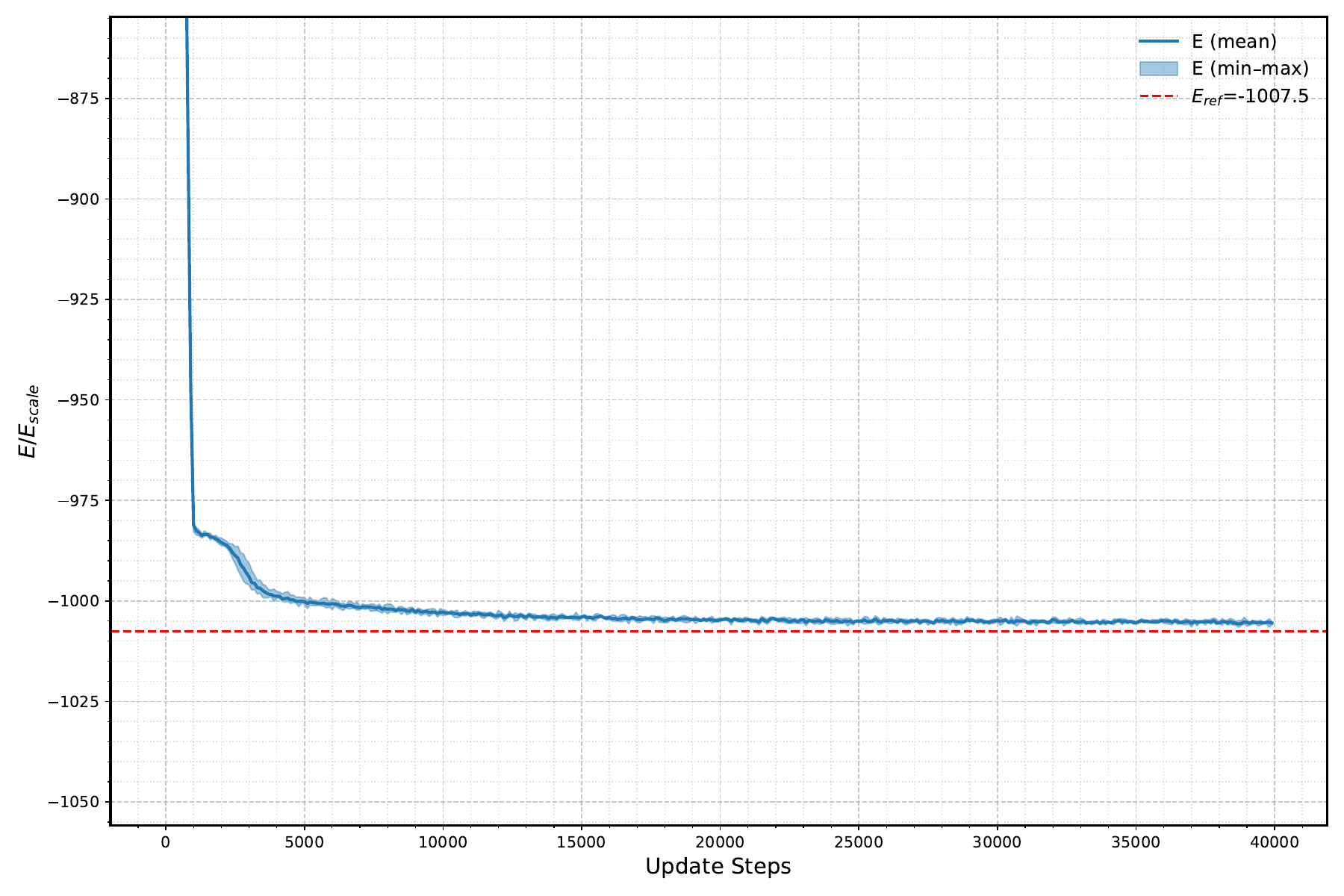}
        \caption{\texttt{GELU-MALA}}
        \label{fig:8p-gelu-mala}
    \end{subfigure}
    \hfill
    \begin{subfigure}[b]{0.48\textwidth}
        \centering
        \includegraphics[width=\textwidth, height=6cm, keepaspectratio]{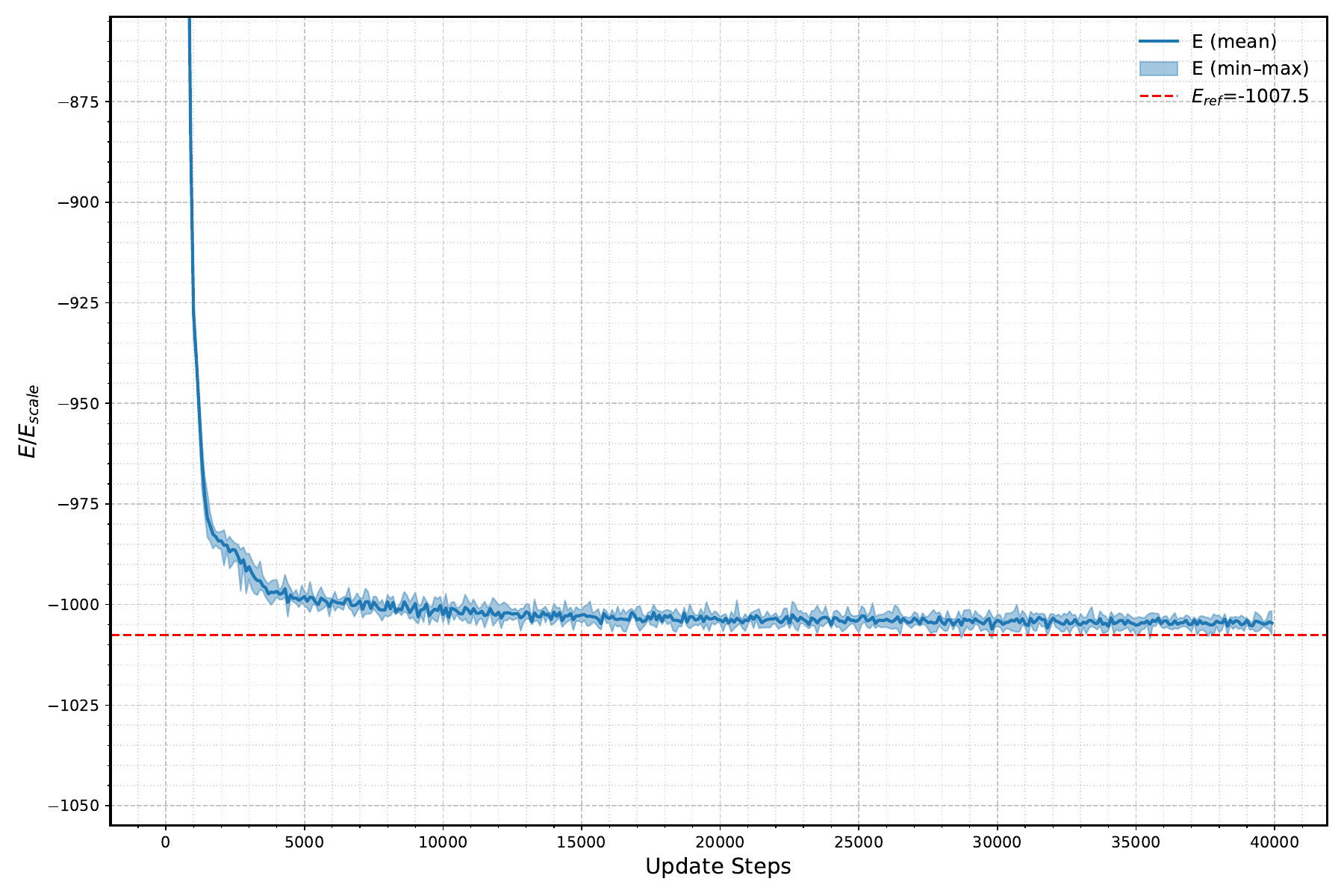}
        \caption{\texttt{GELU-ARW}}
        \label{fig:8p-gelu-adap}
    \end{subfigure}
    
    \begin{subfigure}[b]{0.48\textwidth}
        \centering
        \includegraphics[width=\textwidth, height=6cm, keepaspectratio]{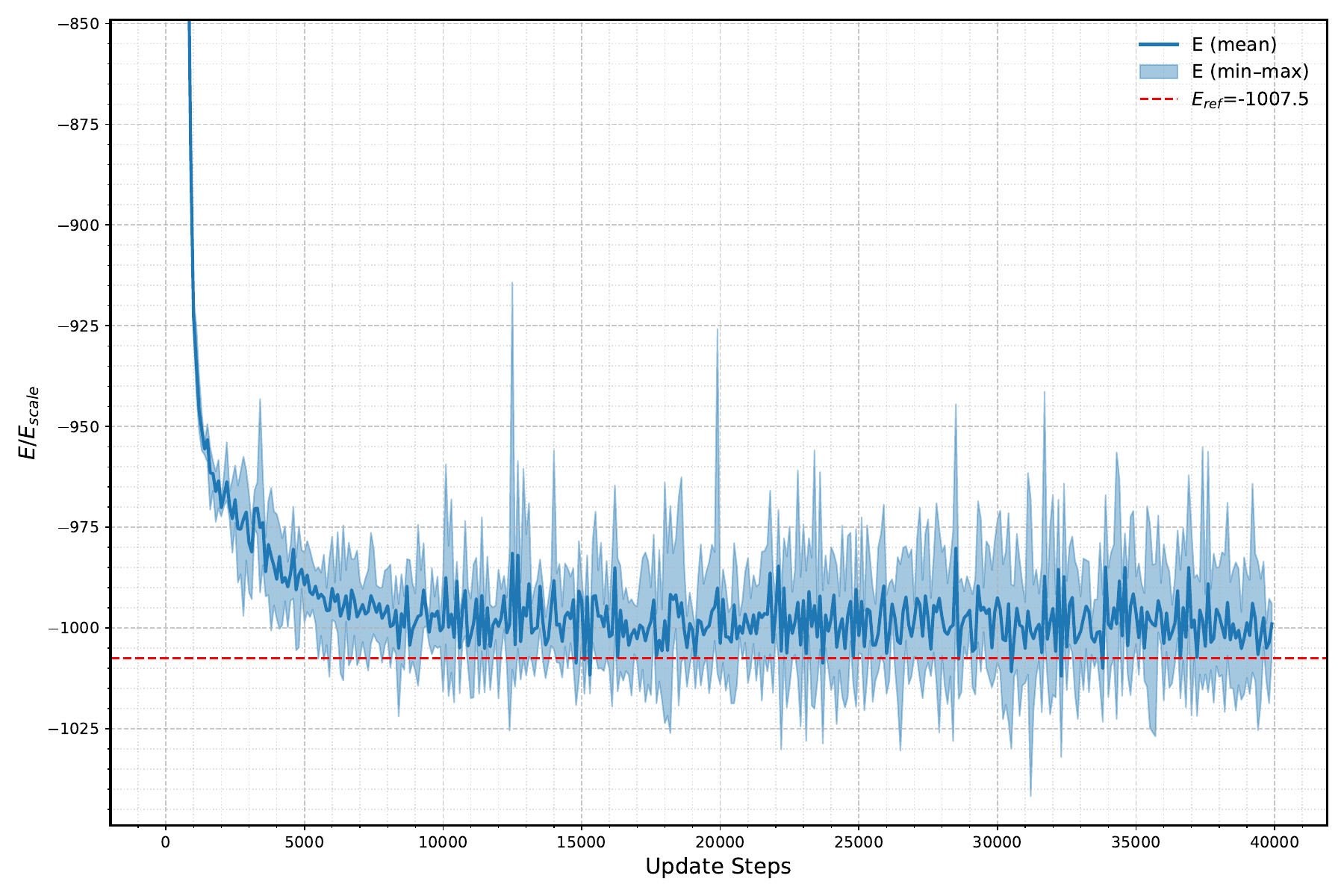}
        \caption{\texttt{GELU-RW}}
        \label{fig:8p-gelu-random}
    \end{subfigure}
    \hfill
    \begin{subfigure}[b]{0.48\textwidth}
        \centering
        \includegraphics[width=\textwidth, height=6cm, keepaspectratio]{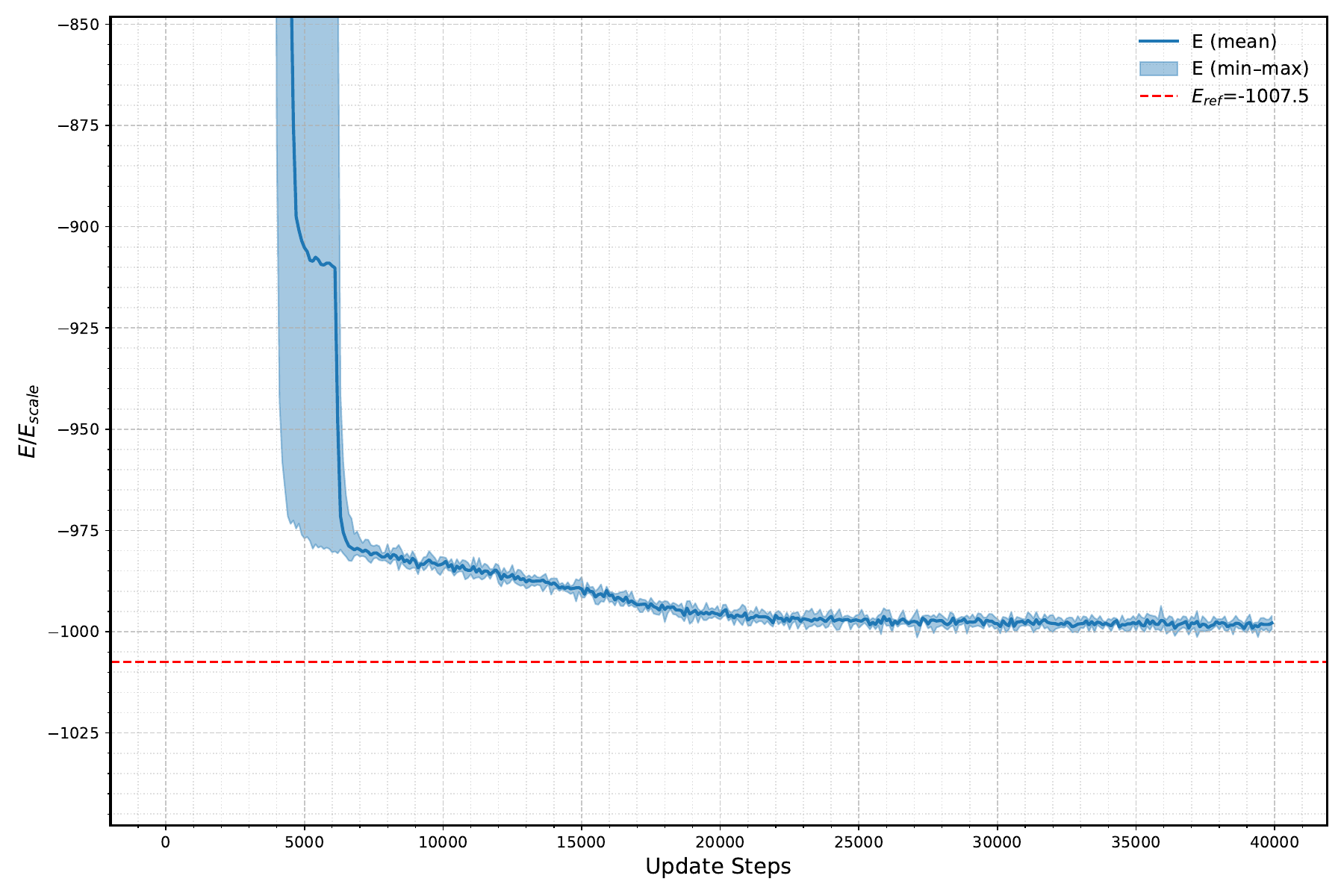}
        \caption{\texttt{tanh-ARW}}
        \label{fig:8p-tanh-adap}
    \end{subfigure}

    \caption{Energy training convergence for 8-particle harmonic potential systems with two-body interactions.}
    \label{fig:8particles-all-methods}
\end{figure}
Having established the reliability of our methods through 3-particle benchmarking, we extended our analysis to systems ranging from 4 to 10 particles, maintaining consistent hyperparameters across all simulations as detailed in \autoref{tab:slow_intro_params}. This scaling study allows us to evaluate the comparative performance and robustness of different configurations as system complexity increases.

Since configuration~III (\texttt{GELU-RW}) exhibits convergence difficulties starting from nine particles and fails to reach stable energy values within the allocated training iterations, in this section, we benchmark the proposed methods using eight particles. For a detailed examination of our method's performance, \autoref{fig:8particles-all-methods} presents the convergence patterns for 8-particle harmonic systems across multiple independent training runs. The figure compares different network architectures and sampling methods, revealing distinct convergence behaviors as the number of particles increases beyond the validated three-particle benchmark.

Our analysis reveals several important insights. First, all methods successfully converge to the reference energy, validating the overall effectiveness of the approach. However, the quality of convergence differs significantly between sampling methods and neural network model. The \texttt{GELU-MALA} (\autoref{fig:8p-gelu-mala}) demonstrates superior stability with minimal oscillations during convergence and achieves a close approximation to the reference energy. In stark contrast, the \texttt{GELU-RW} approach (\autoref{fig:8p-gelu-random}) exhibits pronounced oscillations throughout the training process, indicating substantial variance in the energy estimation.

\subsubsection{Statistical Analysis  and Computational Performance}

\begin{table}[htbp]
\centering
\renewcommand{\arraystretch}{1.25}
\setlength{\tabcolsep}{5pt}
\label{tab:energy_comparison_sigma}
\caption{
Energy comparison across different particle numbers ($N$) and configurations for the harmonic potential with two-body interactions. Each entry reports the sample mean $\mu$ and standard deviation $\sigma$ over five independent runs at the final training step (out of 40,000 updates). All numbers are in units of 
$E_\text{scale}=\hbar^2/(mr_0^2)$. Configurations: 
I = \texttt{GELU-MALA}, II = \texttt{GELU-ARW}, 
III = \texttt{GELU-RW}, IV = \texttt{tanh-ARW}.
}
\resizebox{\linewidth}{!}{\begin{tabular}{lcccccccc}
\toprule
$N$ & \textbf{3} & \textbf{4} & \textbf{5} & \textbf{6} & \textbf{7} & \textbf{8} & \textbf{9} & \textbf{10} \\
\midrule
$E_\text{ref}$~\cite{Yan_2014}
& $-11.91$ & $-70.00$ & $-19.20\times10$ & $-38.40\times10$ & $-65.44\times10$ & $-10.08\times10^2$ & $-14.48\times10^2$ & $-19.76\times10^2$ \\
\midrule
I ($\mu$)
& $\mathbf{-12.0}$ & $\mathbf{-69.5}$ & $\mathbf{-19.1\times10}$ & $\mathbf{-38.2\times10}$ & $\mathbf{-65.3\times10}$ & $\mathbf{-10.06\times10^2}$ & $-14.43\times10^2$ & $-19.72\times10^2$ \\
I ($\sigma$)
& $0.4$ & $0.7$ & $0.2\times10$ & $\mathbf{0.14\times10}$ & $0.4\times10$ & $\mathbf{0.03\times10^2}$ & $\mathbf{0.02\times10^2}$ & $\mathbf{0.04\times10^2}$ \\
\midrule
II ($\mu$)
& $-12.4$ & $-68$ & $-19.0\times10$ & $-38.7\times10$ & $-64.5\times10$ & $-10.0\times10^2$ & $\mathbf{-14.5\times10^2}$ & $\mathbf{-19.8\times10^2}$ \\
II ($\sigma$)
& $0.9$ & $2$ & $0.9\times10$ & $0.8\times10$ & $0.6\times10$ & $0.1\times10^2$ & $0.2\times10^2$ & $0.1\times10^2$ \\
\midrule
III ($\mu$)
& $-11.4$ & $-69$ & $-19.4\times10$ & $-37\times10$ & $-65\times10$ & $-10.2\times10^2$ & -- & -- \\
III ($\sigma$)
& $0.14$ & $2$ & $0.4\times10$ & $3\times10$ & $3\times10$ & $0.3\times10^2$ & -- & -- \\
\midrule
IV ($\mu$)
& $-11.7$ & $-69.0$ & $-18.9\times10$ & $-38.1\times10$ & $-64.8\times10$ & $-99.5\times10$ & $-14.39\times10^2$ & $-19.6\times10^2$ \\
IV ($\sigma$)
& $\mathbf{0.13}$ & $\mathbf{0.6}$ & $\mathbf{0.1\times10}$ & $0.2\times10$ & $\mathbf{0.2\times10}$ & $0.4\times10$ & $0.04\times10^2$ & $0.15\times10^2$ \\
\bottomrule
\end{tabular}}
\label{tab:energy_comparison_sigma}
\end{table}

\autoref{tab:energy_comparison_sigma} summarizes the final energies obtained for systems with $N=3$–$10$ particles under the harmonic potential with two-body interactions. Across all configurations, the \texttt{GELU-MALA} (Config.~I) consistently yields energies close to the reference values~\cite{Yan_2014}, indicating strong stability and convergence. The \texttt{GELU-ARW} (Config.~II) shows slightly higher deviations, while \texttt{GELU-RW} (Config.~III) and \texttt{tanh-ARW} (Config.~IV) exhibit larger fluctuations across particle numbers. Overall, the results confirm the robustness of the MALA-based sampling in achieving accurate ground-state energies.

\begin{figure}[htbp]
    \centering
    \begin{subfigure}[b]{0.48\textwidth}
        \centering
        \includegraphics[width=\textwidth]{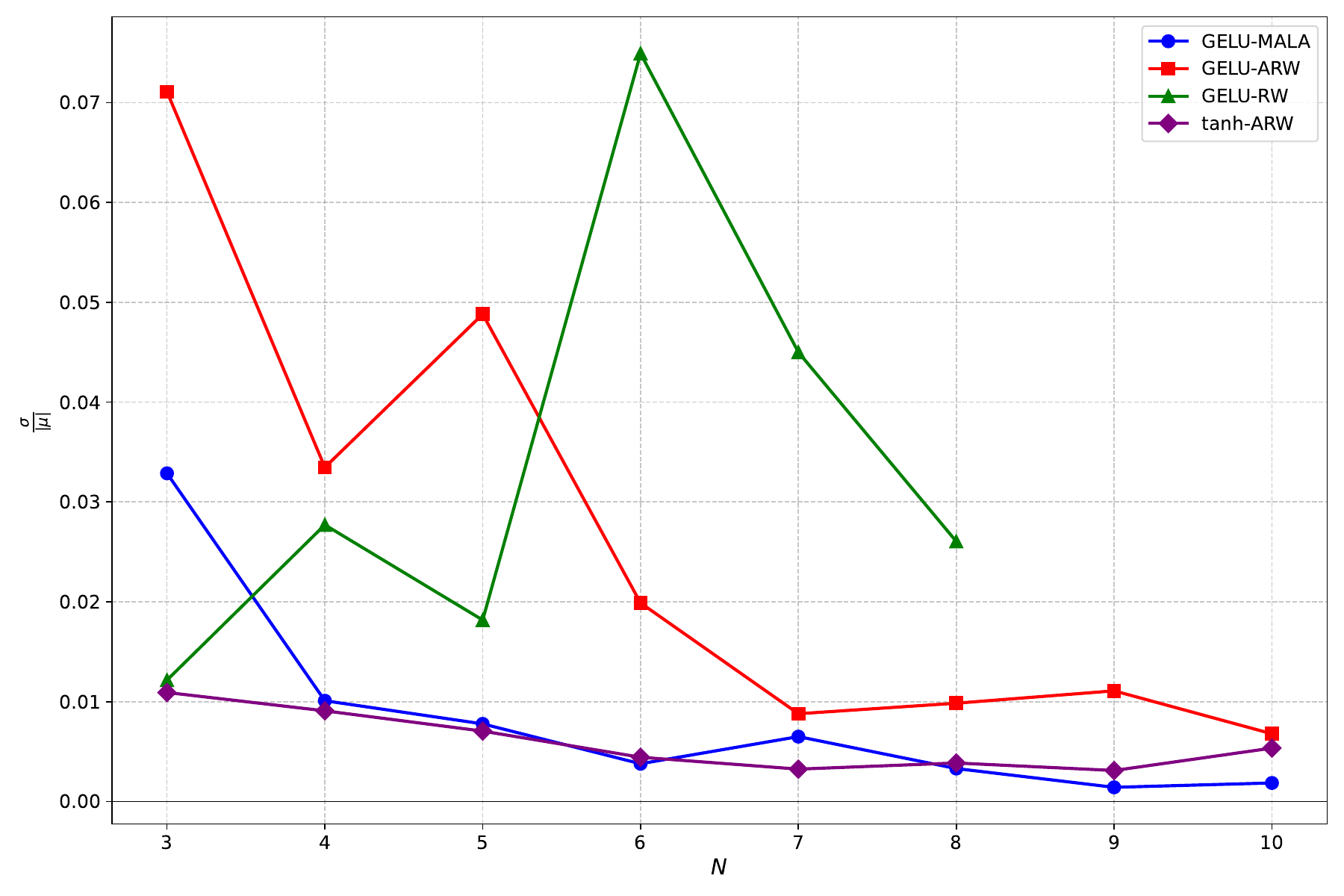}
        \caption{Coefficient of variation.}
        \label{fig:coefficient-variation-harm}
    \end{subfigure}
    \hfill
    \begin{subfigure}[b]{0.48\textwidth}
        \centering
        \includegraphics[width=\textwidth]{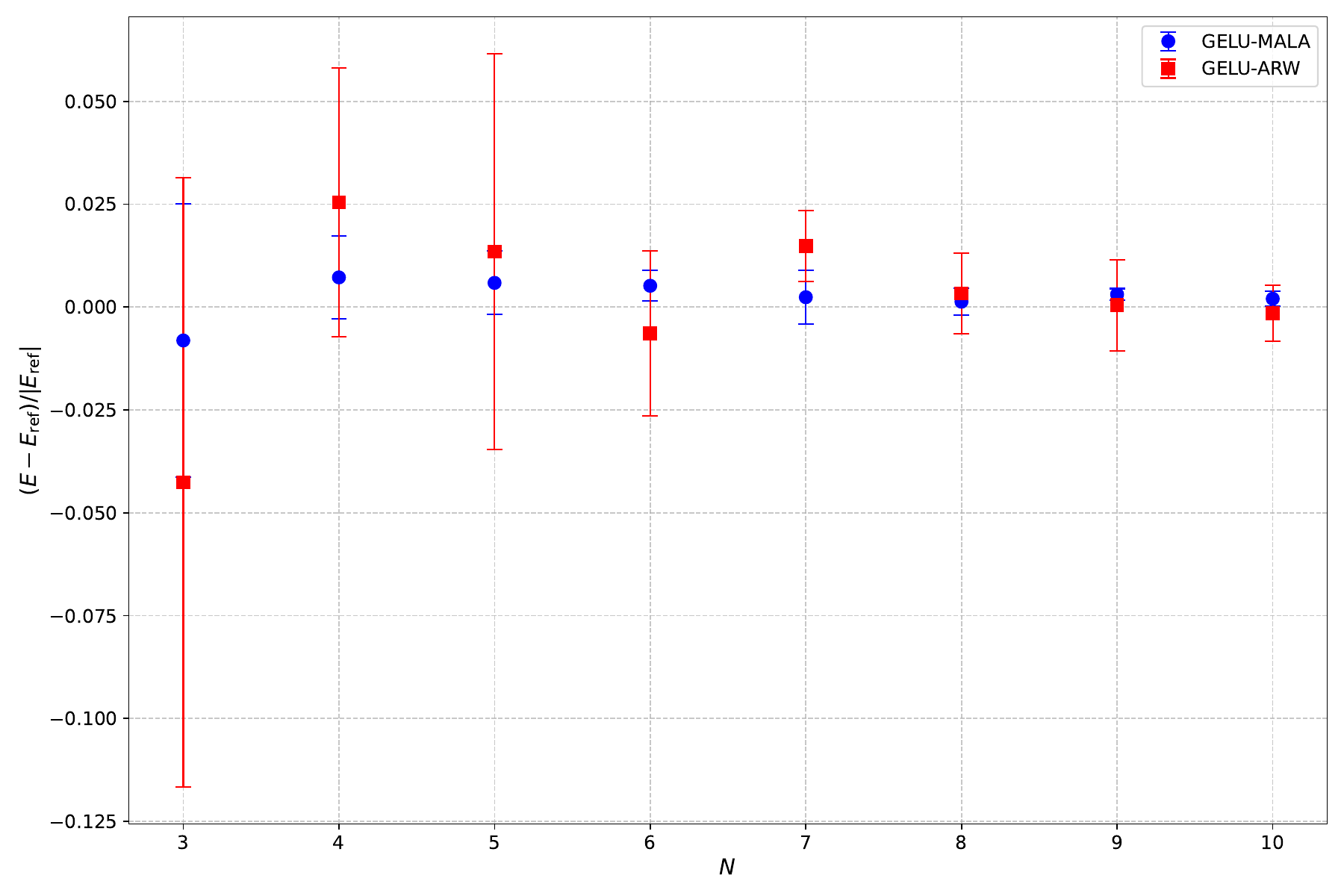}
        \caption{Relative error in energy estimation.}
        \label{fig:relative-error-harm}
    \end{subfigure}
    \caption{Performance metrics for harmonic potential systems. (a) Coefficient of variation of energy across particle numbers for 5 runs, showing improved consistency as system size increases for adaptive sampling methods. (b) Relative error in energy estimation across different particle numbers, comparing \texttt{GELU-MALA} (blue) and \texttt{GELU-ARW} (red) from five independent test runs.}
    \label{fig:combined-harm-metrics}
\end{figure}

\autoref{fig:coefficient-variation-harm} evaluated method consistency using the coefficient of variation (CV) $\frac{\sigma}{|\mu |}$ of energy values across particle counts. The \texttt{GELU-MALA} approach demonstrates exceptional stability with consistently decreasing relative standard deviation as particle count increases, reaching values below $\frac{\sigma}{|\mu |} = 0.01$ for systems with $N \geq 5$. The \texttt{tanh-ARW} method maintains similar stability with values consistently below $\frac{\sigma}{|\mu |}=0.01$ after $N=4$. The \texttt{GELU-ARW} approach exhibits moderate stability across all particle counts. In contrast, the \texttt{GELU-RW} method shows highly erratic behavior with significant fluctuations, including a pronounced spike at $N=6$ where the relative standard deviation exceeds $\frac{\sigma}{|\mu |}=0.07$, and it ultimately failed to converge for systems with $N > 8$.

The relative error analysis (\autoref{fig:relative-error-harm}) shows an inverse relationship between relative error and particle count, providing evidence for the robustness of the \texttt{GELU-MALA} framework. Referenced energy values $E_\text{ref}$ were obtained from \cite{Yan_2014}. Comparative analysis between \texttt{GELU-MALA} and \texttt{GELU-ARW} implementations reveals that the \texttt{MALA} approach exhibits notably reduced oscillatory behavior in the error profile. These results confirm that adaptive sampling methods generally produce more reliable energy estimates, particularly as system complexity increases, while the non-adaptive random walk approach becomes notably unstable for systems with more than $7$ particles.

\begin{figure}[htbp]
    \centering
    \includegraphics[width=0.5\textwidth]{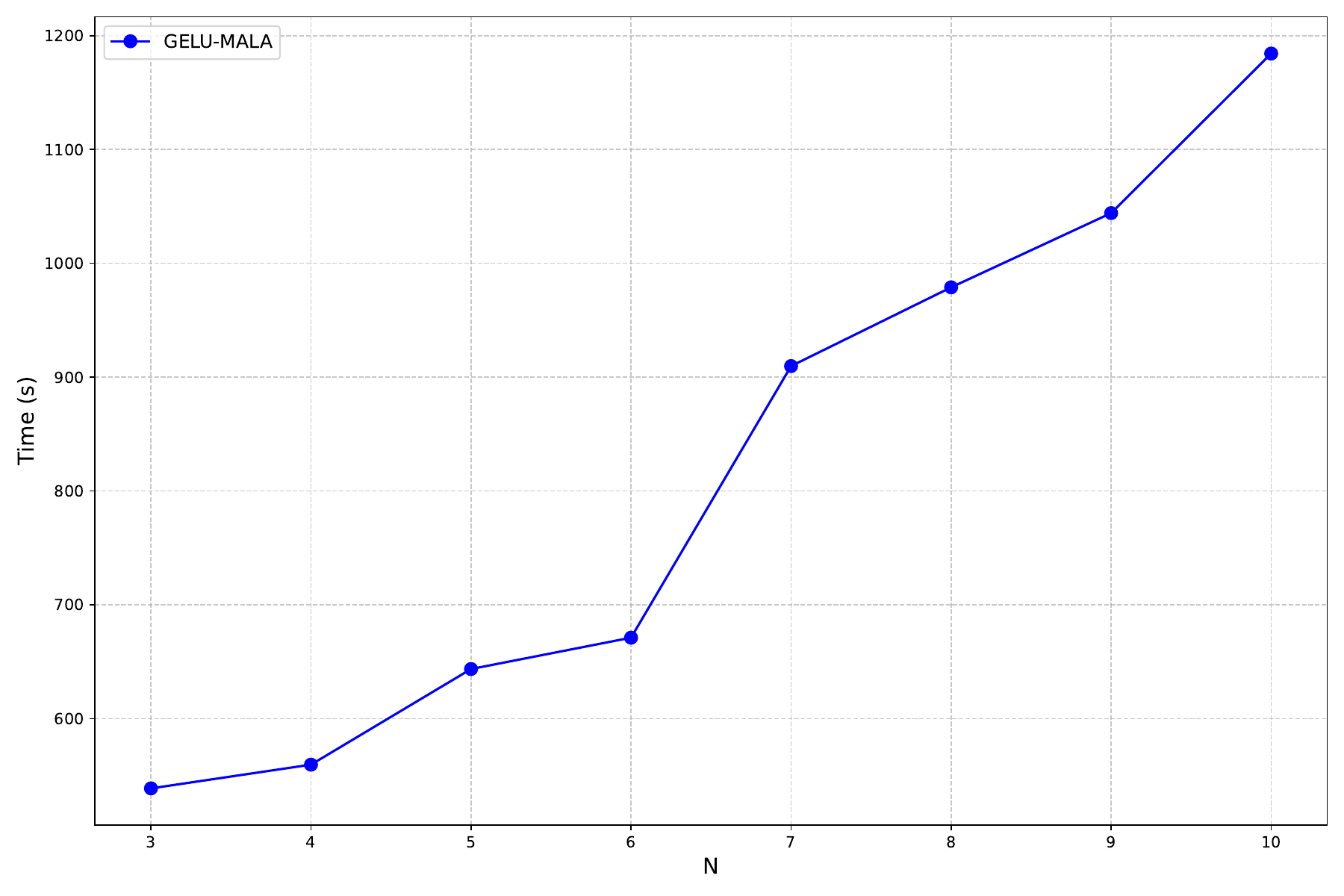}
    \caption{Performance metrics for harmonic potential systems A. Computation time of \texttt{GELU-MALA} with increasing particle counts. Total computation time remain manageable even for larger systems.}
    \label{fig:computation-time-harm}
\end{figure}

While computational requirements naturally increase with system size, our method maintains reasonable efficiency for harmonic potential systems, as shown in \autoref{fig:computation-time-harm}. The computational scaling remains practical even for 10-particle systems, allowing for effective simulation of more complex harmonic systems.

Unlike previous methods without adaptive adjustment that often required extensive parameter tuning for different system sizes, these adaptive approaches maintain consistent effectiveness using the same hyperparameters ($\rho = 1.0$, $\epsilon = 0.1$) across all particle counts. This uniform parameterization significantly reduces the need for system-specific optimization while delivering more reliable and accurate results, particularly for larger harmonic systems where original sampling methods without adaptive adjustment struggle to achieve stable convergence.

Detailed convergence plots for individual harmonic potential systems with two-body interactions across different particle counts (4 - 7 and 9 - 10) can be found in Appendix~\ref{appendix:systemA}. These plots provide a comprehensive view of the training trajectory for each system size and different methods, further demonstrating the consistency and reliability of \texttt{GELU-MALA} approach across the full range of particle counts tested.

\subsection{System B (helium cluster) Results}
\label{sub:systemiii}

Our method demonstrates good outcomes for systems of harmonic Confinement with Two-Body Interaction. In two-Body and Three-Body Interactions systems, we do not include the \texttt{tanh-ARW} because this original neural network model does not perform well. We conducted five individual tests on systems ranging from 3 to 10 particles, maintaining consistent hyperparameters across all simulations.

\subsubsection{Methods' Behavior Comparison}
\begin{figure}[htbp]
    \centering
    
    \begin{subfigure}[b]{0.31\textwidth}
        \centering
        \includegraphics[width=\textwidth, height=6cm, keepaspectratio]{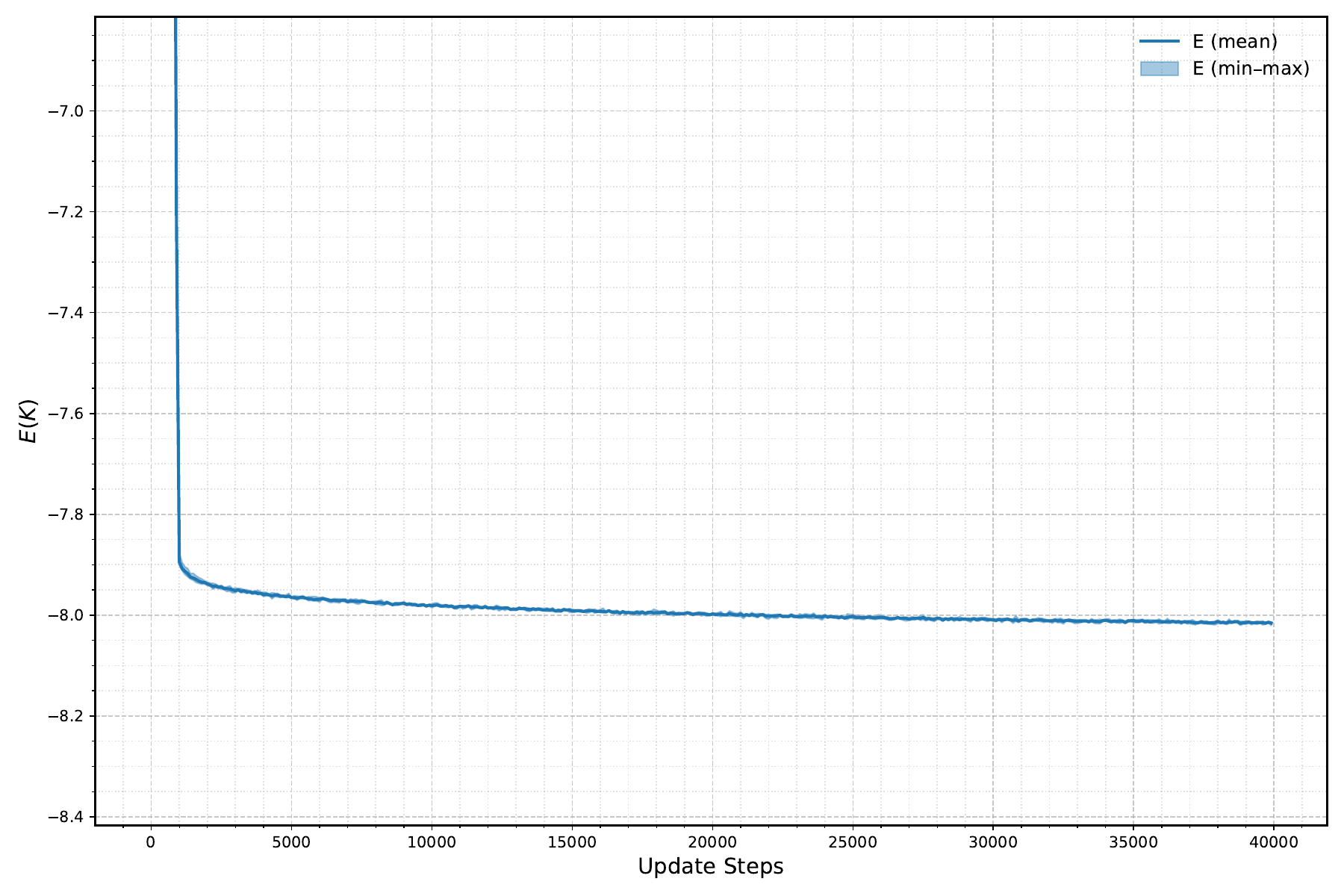}
        \caption{\texttt{GELU-MALA}}
        \label{fig:10p-tb-gelu-mala}
    \end{subfigure}
    \hfill
    \begin{subfigure}[b]{0.31\textwidth}
        \centering
        \includegraphics[width=\textwidth, height=6cm, keepaspectratio]{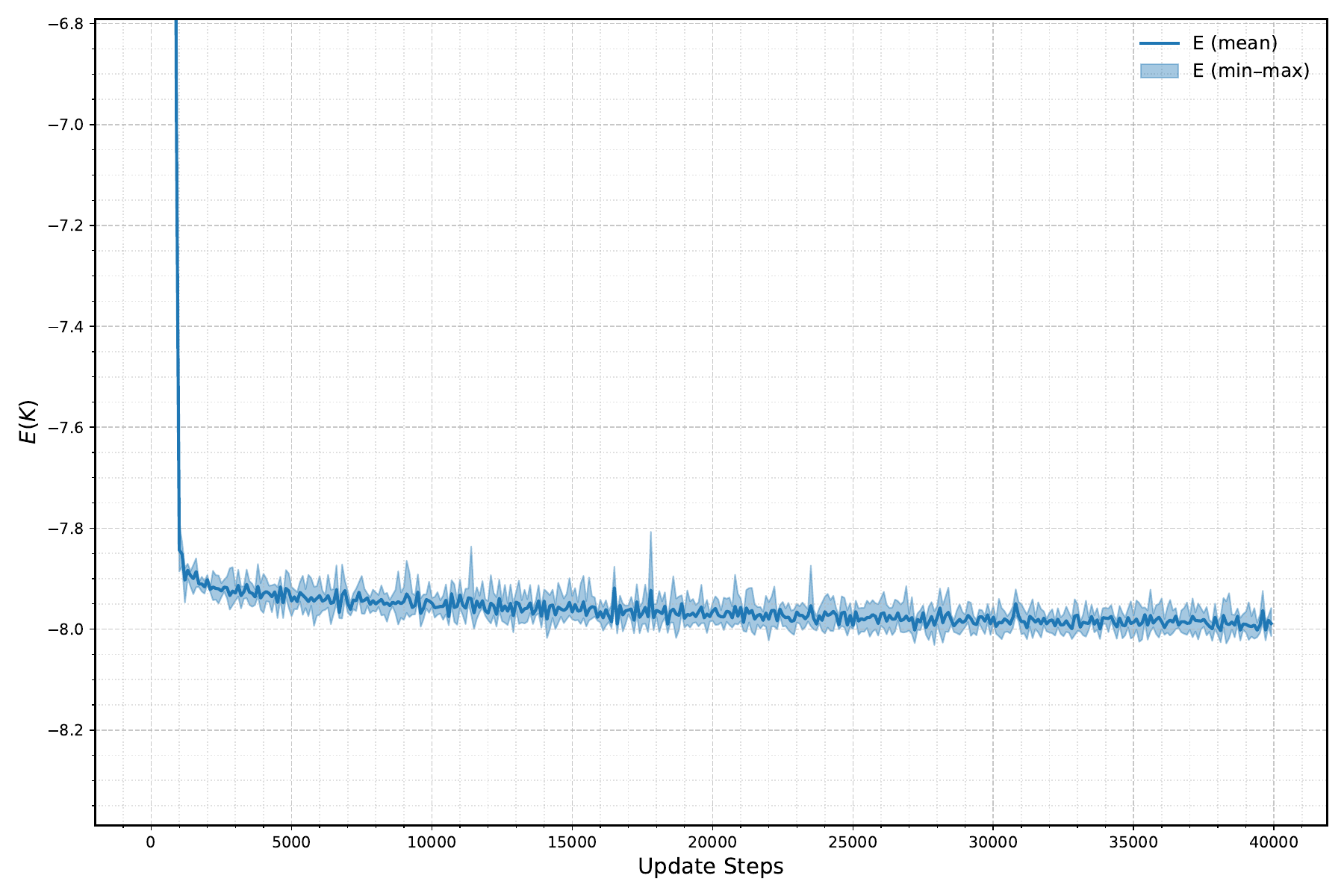}
        \caption{\texttt{GELU-ARW}}
        \label{fig:10p-tb-gelu-adap}
    \end{subfigure}
    \hfill
    \begin{subfigure}[b]{0.31\textwidth}
        \centering
        \includegraphics[width=\textwidth, height=6cm, keepaspectratio]{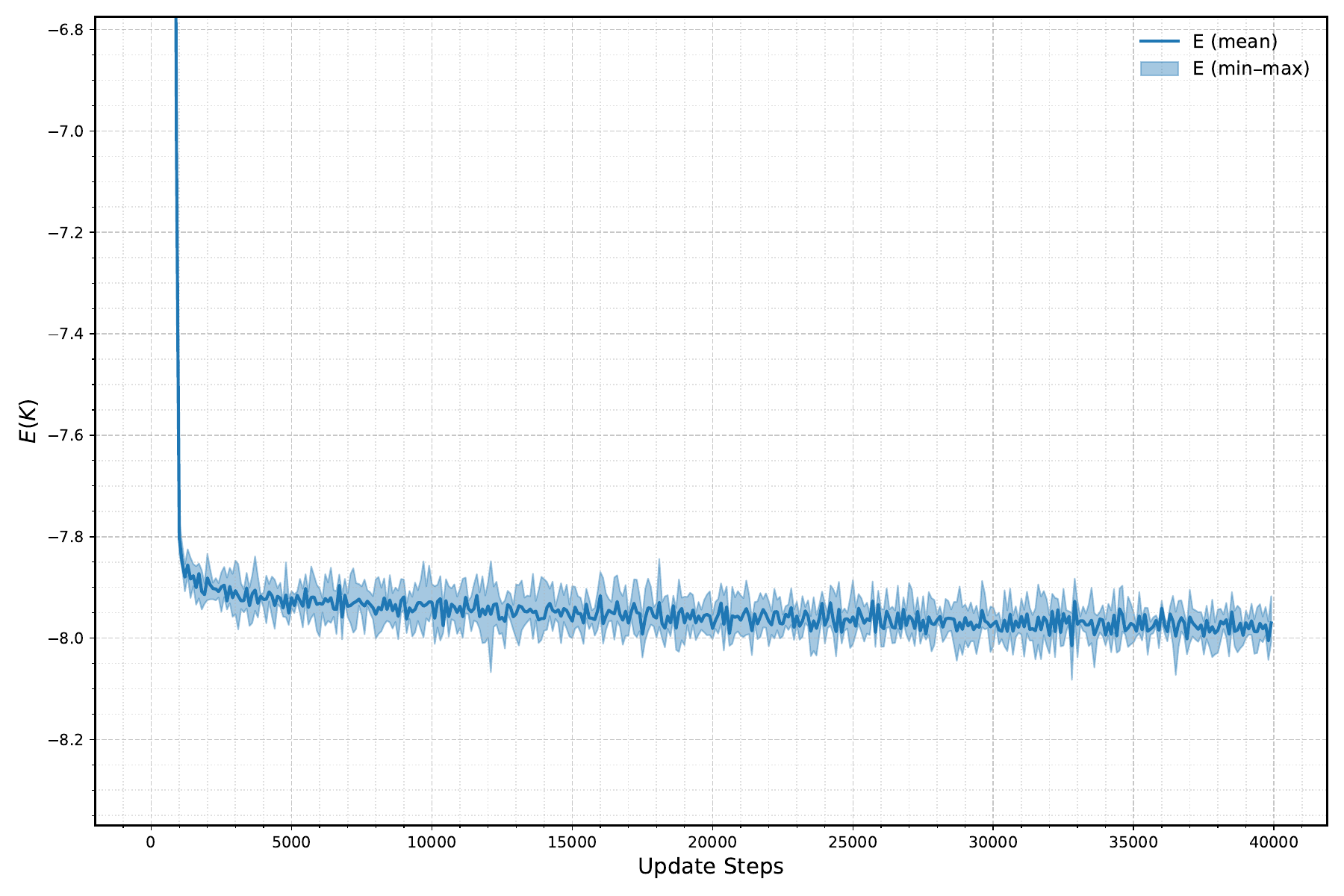}
        \caption{\texttt{GELU-RW}}
        \label{fig:10p-tb-gelu-random}
    \end{subfigure}
    
    \caption{Training convergence for ten-particles systems with two-body and three-body interactions.}
    \label{fig:10particles-tb-inte-methods}
\end{figure}

\autoref{fig:10particles-tb-inte-methods} provides a detailed analysis of convergence patterns for ten-particles systems across five independent training runs, offering insight into our method's performance with both two-body and three-body interactions. The comparative evaluation of various GELU-based approaches reveals distinctive convergence behaviors among the sampling methodologies. The data demonstrates that \texttt{GELU-MALA} exhibits superior performance compared to the alternative methods, as evidenced by its significantly reduced oscillation patterns during convergence.

\subsubsection{Statistical Analysis  and Computational Performance}

\autoref{tab:tb_inte_gelu_configs_by_N} summarizes the final trained energies for three-body interacting bosonic systems across particle numbers \(N=3\) to \(20\). Configuration I (\texttt{GELU-MALA}) consistently achieves the lowest variance and most stable convergence, while Configurations II and III exhibit larger fluctuations and degraded performance as \(N\) increases.

\begin{table}[htbp]
\centering
\renewcommand{\arraystretch}{1.25}
\setlength{\tabcolsep}{4pt}
\caption{Energy comparison across different particle numbers ($N$) and configurations for the three-body interaction system. Each entry reports the sample mean $\mu$ and standard deviation $\sigma$ over five independent runs at the final training step (out of 40{,}000 total updates). All the numbers are expressed in Kelvin (K) . Configurations: I = \texttt{GELU-MALA}, II = \texttt{GELU-ARW}, III = \texttt{GELU-RW}. For $N>10$, only configuration I is reported.}
\begin{tabular}{cc}

\begin{tabular}{lcccccc}
\toprule
$N$ 
& I ($\mu$) & I ($\sigma$)
& II ($\mu$) & II ($\sigma$)
& III ($\mu$) & III ($\sigma$) \\
\midrule
3  & $-0.126$ & $\mathbf{0.003}$ & $-0.124$ & $0.006$ & $-0.13$ & $0.01$ \\
4  & $-0.561$ & $\mathbf{0.003}$ & $-0.56$  & $0.01$  & $-0.55$ & $0.01$ \\
5  & $-1.29$  & $\mathbf{0.02}$  & $-1.3$   & $0.04$  & $-1.28$ & $0.03$ \\
6  & $-2.28$  & $\mathbf{0.003}$ & $-2.23$  & $0.02$  & $-2.25$ & $0.05$ \\
7  & $-3.47$  & $\mathbf{0.01}$  & $-3.5$   & $0.08$  & $-3.55$ & $0.06$ \\
8  & $-4.82$  & $\mathbf{0.01}$  & $-4.9$   & $0.2$   & $-4.83$ & $0.07$ \\
9  & $-6.35$  & $\mathbf{0.02}$  & $-6.2$   & $0.1$   & $-6.3$  & $0.1$  \\
10 & $-8.02$  & $\mathbf{0.02}$  & $-8.1$   & $0.2$   & $-8.0$  & $0.2$  \\
11 & $-9.80$  & $\mathbf{0.02}$  & --       & --      & --      & --     \\
\bottomrule
\end{tabular}
&
\begin{tabular}{lcccccc}
\toprule
$N$ 
& I ($\mu$) & I ($\sigma$)
& II ($\mu$) & II ($\sigma$)
& III ($\mu$) & III ($\sigma$) \\
\midrule
12 & $-11.65$ & $\mathbf{0.01}$ & -- & -- & -- & -- \\
13 & $-13.63$ & $\mathbf{0.03}$ & -- & -- & -- & -- \\
14 & $-15.67$ & $\mathbf{0.01}$ & -- & -- & -- & -- \\
15 & $-17.83$ & $\mathbf{0.02}$ & -- & -- & -- & -- \\
16 & $-20.04$ & $\mathbf{0.03}$ & -- & -- & -- & -- \\
17 & $-22.3$  & $\mathbf{0.03}$ & -- & -- & -- & -- \\
18 & $-24.62$ & $\mathbf{0.03}$ & -- & -- & -- & -- \\
19 & $-26.98$ & $\mathbf{0.06}$ & -- & -- & -- & -- \\
20 & $-29.41$ & $\mathbf{0.03}$ & -- & -- & -- & -- \\
\bottomrule
\end{tabular}

\end{tabular}
\label{tab:tb_inte_gelu_configs_by_N}
\end{table}

\begin{figure}[htbp]
    \centering
    \begin{subfigure}[b]{0.48\textwidth}
        \centering
        \includegraphics[width=\textwidth]{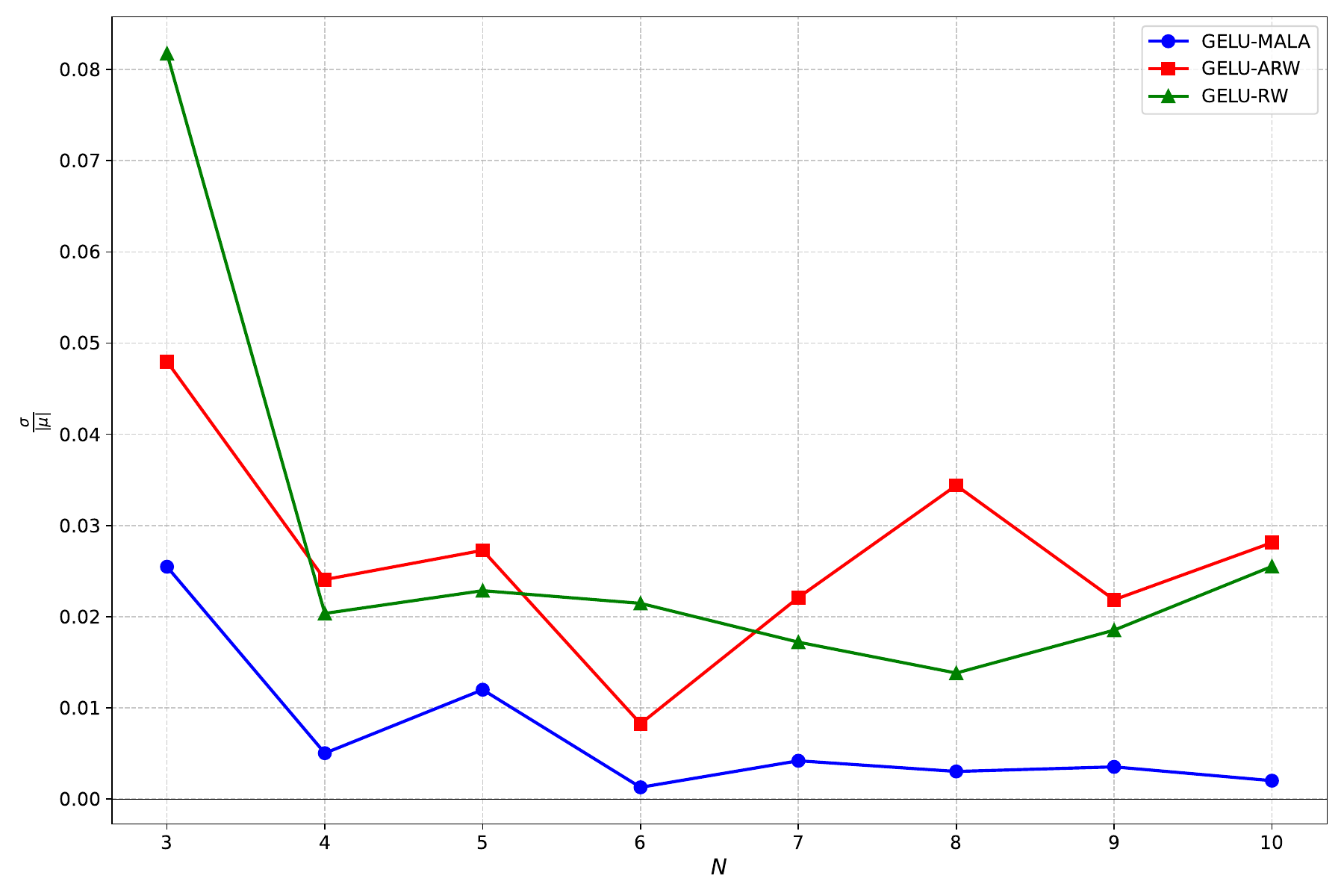}
        \caption{Coefficient of variation.}
        \label{fig:coefficient-variation-slow}
    \end{subfigure}
    \hfill
    \begin{subfigure}[b]{0.48\textwidth}
        \centering
        \includegraphics[width=\textwidth]{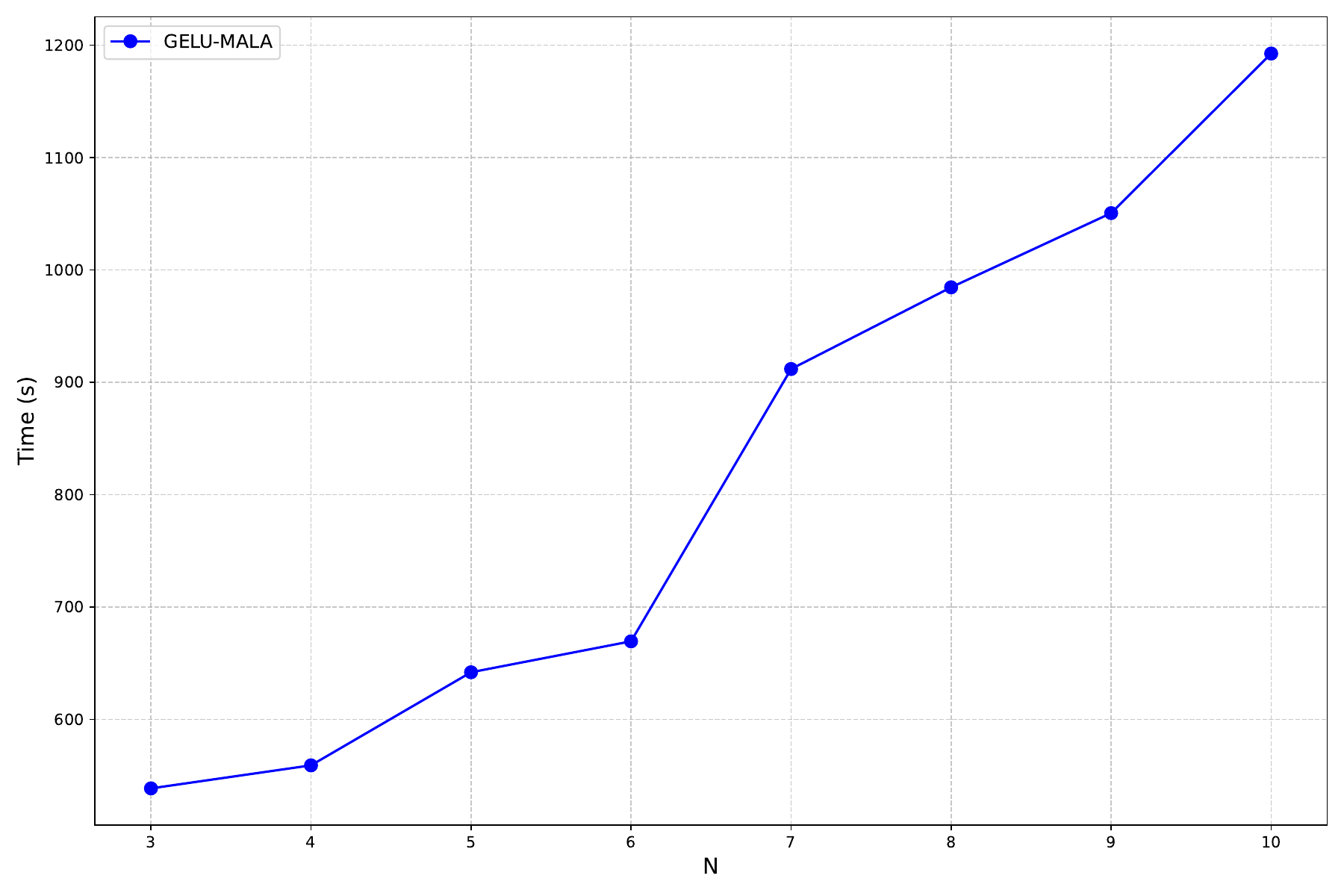}
        \caption{Mean total runtime.}
        \label{fig:computation-time-slow}
    \end{subfigure}
    \caption{Performance metrics for multi-particle systems. 
    (a)shows the coefficient of variation, and (b) shows the mean total runtime from five tests as the particle number increases from 3 to 10.}
    \label{fig:multi-performance}
\end{figure}

\autoref{fig:coefficient-variation-slow} validates the \texttt{GELU-MALA} approach by showing the coefficient of variation of energy values across different particle counts. After an initial higher variation at N=3, the coefficient stabilizes as system size increases, demonstrating that our method becomes more reliable for larger systems.

While computational requirements increase with system size, our method maintains reasonable efficiency even for larger systems. As shown in Figure \ref{fig:computation-time-slow}, \texttt{GELU-MALA} requires only about 1200 seconds for ten-particles systems—a manageable duration for machine learning applications.

Detailed convergence plots for individual harmonic potential systems with two-body interactions across different particle counts (3 - 9) can be found in the Appendix~\ref{appendix:systemB}. These plots provide a comprehensive view of the training trajectory for each system size and different methods, further demonstrating the consistency and reliability of \texttt{GELU-MALA} approach across the full range of particle counts tested.

Additional plots for systems with 11-20 particles using \texttt{GELU-MALA} are provided in Appendix~\ref{appendix:systemB}. These simulations were conducted using A100 GPUs with double precision floating-point arithmetic (\texttt{float64}).

\subsection{System C (Two Extended System) Results }
\label{sub:two_extended_system}
For the following two extended systems, we modify $\tau = 5000$ iterations, rather than the typical setting of $\tau = 1000$. This adjustment enhances the stability of training for these more challenging systems. Given their challenging nature and slower learning dynamics, we allow the network additional time to learn the fundamental features of the wave function before applying the adaptive methods. The detailed convergence plots for both two-particle and three-particle systems with different masses can be found in Appendix~\ref{appendix:systemC}
\subsubsection{Two particles with a Two-body interaction}

The two-particles case presents unique training challenges that require specialized parameter adjustments. The two-body system is weakly bound, which makes it more difficult to compute the ground state energy. Due to this weakly bound interaction, we needed to adjust the sampling range to a larger value. For this small system, we increased the sampling range from the standard 60.0 to 500.0, ensuring particles had sufficient freedom to explore the configuration space. We also enlarged the step size parameter to 15.0 and reduced $\eta$ to 0.001 instead of 0.003. Additionally, we adjusted the acceptance rate to 0.75 instead of 0.234 for RW and to 0.85 for MALA, since the original acceptance rates were optimized for high-dimensional problems. For the optimizer, we increased the learning rate from 0.0001 to 0.0005. 

\begin{table}[htbp]
\centering
\renewcommand{\arraystretch}{1.25}
\setlength{\tabcolsep}{6pt}
\caption{Comparison of mean energy $\mu$ and standard deviation $\sigma$ over five runs for two systems under two-body interactions. Each configuration corresponds to: I = \texttt{GELU-MALA}, II = \texttt{GELU-ARW}, III = \texttt{GELU-RW}.}
\label{tab:two_lowN_systems}
\begin{subtable}[t]{0.48\linewidth}
\centering
\caption{Two particles}
\label{tab:two_body_lowN}
\resizebox{\linewidth}{!}{\begin{tabular}{lcc}
\toprule
Config & $\mu$ (K) & $\sigma$ (K) \\
\midrule
$E_\text{ref}$~\cite{PhysRevA.84.052503} & $-1.304\times10^{-3}$ & -- \\
I   & $\mathbf{-1.1\times10^{-3}}$ & $\mathbf{0.5\times10^{-3}}$ \\
II  & $-2.1\times10^{-3}$ & $1.5\times10^{-3}$ \\
III & $-0.9\times10^{-3}$ & $0.6\times10^{-3}$ \\
\bottomrule
\end{tabular}}
\end{subtable}
\hfill
\begin{subtable}[t]{0.48\linewidth}
\centering
\caption{Three particles with different masses}
\label{tab:three_body_mass}
\resizebox{\linewidth}{!}{\begin{tabular}{lcc}
\toprule
Config & $\mu$ (K)& $\sigma$ (K) \\
\midrule
$E_\text{ref}$~\cite{Nielsen_1998} & $-18.8\times10^{-3}$ & -- \\
I   & $\mathbf{-18.3\times10^{-3}}$ & $\mathbf{0.9\times10^{-3}}$ \\
II  & $-17.9\times10^{-3}$ & $3.5\times10^{-3}$ \\
III & $-18.0\times10^{-3}$ & $4.2\times10^{-3}$ \\
\bottomrule
\end{tabular}}
\end{subtable}

\end{table}

\autoref{tab:two_body_lowN} presents the statistical error analysis across five independent training runs. The reported values represent the mean deviation of the computed energies relative to the reference value $E_\text{ref}$, while the accompanying $\sigma$ values quantify the standard deviation across these runs. The results indicate that the proposed approach can closely approximate $E_\text{ref}$, although the error fluctuations remain larger compared to systems with higher particle numbers.

\subsubsection{Three Particles with Different Masses with a Two-body Interaction}
Our investigation of the 3-particle system with heterogeneous masses reveals important insights into handling increased quantum complexity. For this configuration, we employed $\rho = 150$ and $\epsilon = 15.0$, parameters specifically calibrated for systems with different masses. All three sampling algorithms—MALA, random walk, and adaptive random walk—successfully converge to energy values consistent with the reference benchmarks established by Nielsen et al.~\cite{Nielsen_1998}. 

\autoref{tab:three_body_mass} shows that the three sampling methods \texttt{GELU-MALA}, \texttt{GELU-ARW}, and \texttt{GELU-RW} demonstrate comparable performance when applied to a three-particle system with different masses. The proximity of mean relative errors to zero supports our approach of treating mass as an explicit variable in the Jacobi transformation. 


\subsection{Parameter Sensitivity Analysis}

To assess the robustness of our proposed sampling methods, we conducted comprehensive parameter sensitivity analysis on a three-particle harmonic system. We systematically varied the range factor $\rho \in \{1, 5, 10\}$ and step size $\epsilon \in \{0.1, 0.5, 1.0\}$ across all sampling methods. The analysis reveals that adaptive methods (\texttt{GELU-MALA} and \texttt{GELU-ARW}) demonstrate superior stability and lower variance compared to the non-adaptive \texttt{GELU-RW} approach. Notably, adaptive methods benefit from initialization with smaller step sizes, as they can subsequently self-adjust to discover optimal sampling patterns. Detailed results and comprehensive analysis are provided in \autoref{appendix:parameter-sensitivity}.

%% file: sections/conclusions.tex
This work has demonstrated the successful application of Neural Network based machine learning techniques to the study of quantum few-body systems, achieving flexibility and accuracy beyond previous Machine Learning approaches such as the one proposed in~\cite{saito2018}. By enhancing previous methodologies to incorporate particles of unequal mass and both two-body and three-body interactions, we have developed a more generalizable and computationally efficient framework for modeling complex quantum few-body systems.

The proposed multilayer perceptron architecture, employing GELU activation functions and the Metropolis-Adjusted Langevin Algorithm (MALA) for sampling, consistently achieved stable convergence across systems ranging from three to ten particles. The framework demonstrated strong numerical robustness, with the coefficient of variation in the energy decreasing below 0.5\% for larger systems, highlighting its scalability and reliability.

GPU acceleration enables efficient parallelization, allowing the simulation of systems that were previously computationally prohibitive. The demonstrated stability, scalability, and minimal hyperparameter dependence make this approach a promising foundation for future applications of Neural Network quantum few-body. 

Looking ahead, several research avenues naturally emerge from this foundation. One promising direction is extending the model to approximate not only ground states but also excited states. Another is to adapt the proposed framework to fermionic or mixed few-body systems. The explicit integration of quantum symmetries into the network architecture (such as permutation invariance for bosonic systems and antisymmetry for fermionic systems) could enhance both accuracy and computational efficiency, potentially allowing the inclusion of spin degrees of freedom. Finally, the exploration of alternative neural architectures—such as graph-based or attention-driven models—may further improve the capacity to capture the intricate correlation structures inherent in few-body quantum dynamics.

\section*{Aknowledgments}

The authors gratefully acknowledge Prof. Stéphane Bressan, who initiated this research project and provided visionary ideas and early guidance during its initial stages, on which this work is partly based. His insight and mentorship laid the foundation for this research. We honor his memory and remember his contributions with deep gratitude.

The authors also sincerely thank Prof. Patrick Rebentrost for kindly taking over the supervision after Prof. Bressan’s passing, and for the valuable discussions that helped guide the project to completion.

\section*{Declarations}

The code supporting the findings of this study is available from the corresponding author upon reasonable request.

%% file: sections/appendix.tex
\newpage

\section{Jacobi transformation matrix and the interparticle distances}

In this Appendix, we provide the details of the transformation matrix $A$ in Equation~\eqref{eq:transf_jacobi} necessary for the calculation of Jacobi coordinates, and the transformation matrix $D$ in Equation~\eqref{eq:jacobi_to_distances} essential to calculate the interparticle distances $\Delta$, input of the Neural Network.

\subsection{Jacobi transformation matrix}
\label{app:jac_trans}

For a system of $N$ particles with masses $m_1, m_2, \ldots, m_N$, we define the cumulative mass:
\begin{align}
M_k = \sum_{j=1}^k m_j,\quad \text{for } k = 1,2,\dots,N.
\label{eq:M_k}
\end{align}
We also define reduced masses
\begin{align}
\mu_k = \frac{M_k\, m_{k+1}}{M_{k+1}}, \quad \text{for } k = 1,2,\dots,N-1.
\end{align}

The transformation from particle coordinates $X = (\bm{x}_1, \ldots, \bm{x}_N)$ to Jacobi coordinates $R = (\bm{r}_{\text{CM}}, \bm{r}_2, \ldots, \bm{r}_N)$, where $\bm{r}_{\text{CM}} =\bm r_1$ is the center-of-mass coordinate and $(\bm{r}_2, \ldots, \bm{r}_N)$ are the relative coordinates, can be expressed as a matrix operation in Equation~\eqref{eq:transf_jacobi}
\begin{align}
R = AX
\end{align}

The transformation matrix $A$ is constructed as follows:

1. For the first row (center-of-mass coordinate):
\begin{align}
A_{1j} = \frac{m_j}{M_N}, \quad \text{for } j = 1,2,\dots,N
\end{align}

2. For rows $i=2,3,\dots,N$ (relative coordinates), we set $k = N+1-i$ and define:
\begin{align}
A_{ij} =
\begin{cases}
-\sqrt{2\,\mu_k}\,\dfrac{m_j}{M_k}, & \text{for } j = 1,2,\dots,k, \\[1mm]
\sqrt{2\,\mu_k}, & \text{for } j = k+1, \\[1mm]
0, & \text{for } j > k+1.
\end{cases}
\end{align}

For the sake of an example we can visually illustrate that the matrix $A$ for the case of four particles is
\begin{align}
A = \begin{pmatrix} 
\frac{m_1}{M_4} & \frac{m_2}{M_4} & \frac{m_3}{M_4} & \frac{m_4}{M_4} \\
-\sqrt{2\mu_3}\frac{m_1}{M_3} & -\sqrt{2\mu_3}\frac{m_2}{M_3} & -\sqrt{2\mu_3}\frac{m_3}{M_3} & \sqrt{2\mu_3} \\
-\sqrt{2\mu_2}\frac{m_1}{M_2} & -\sqrt{2\mu_2}\frac{m_2}{M_2} & \sqrt{2\mu_2} & 0 \\
-\sqrt{2\mu_1}\frac{m_1}{M_1} & \sqrt{2\mu_1} & 0 & 0
\end{pmatrix}.
\end{align}

\subsection{Calculation of interparticle distances}
\label{app:transf_D}

To simplify the implementation of our Neural Network approach, we must express inter-particle distances in terms of Jacobi coordinates. Beginning with the inverse transformation $A$
\begin{equation}
X = A^{-1}R = A^{-1}\begin{pmatrix} \bm{r}_1 \\ R' \end{pmatrix}
\end{equation}

The displacement vector between particles $i$ and $j$ can be calculated as:
\begin{equation}
\vec{\bm{\delta}}_{ij} = \bm{x}_i - \bm{x}_j = \sum_{k=1}^{N} \left[(A^{-1})_{ik} - (A^{-1})_{jk}\right] \bm{r}_k
\end{equation}

A key insight is that the center-of-mass coordinate $\bm{r}_1$ contributes identically to all particle positions, thus canceling out in the difference:
\begin{equation}
\vec{\bm{\delta}}_{ij} = \sum_{k=2}^{N} \left[(A^{-1})_{ik} - (A^{-1})_{jk}\right] \bm{r}_k = D_{ij} \cdot R'
\end{equation}

In matrix form, the displacement coefficients for all particle pairs can be expressed as:
\begin{equation}
D_{ij} = (A^{-1})_{i,2:N} - (A^{-1})_{j,2:N}
\end{equation}

For a system with $N$ particles, we construct a coefficient matrix $D$ where each row corresponds to a unique particle pair $(i,j)$ with $1 \leq i < j \leq N$. Taking all such pairs, the complete matrix can be written as:
\begin{equation}
D = \begin{pmatrix} 
D_{12} \\
D_{13} \\
\vdots \\
D_{ij} \\
\vdots \\
D_{(N-1)N}
\end{pmatrix} = 
\begin{pmatrix} 
(A^{-1})_{1,2:N} - (A^{-1})_{2,2:N} \\
(A^{-1})_{1,2:N} - (A^{-1})_{3,2:N} \\
\vdots \\
(A^{-1})_{i,2:N} - (A^{-1})_{j,2:N} \\
\vdots \\
(A^{-1})_{N-1,2:N} - (A^{-1})_{N,2:N}
\end{pmatrix}
\end{equation}

The actual inter-particle distances form a vector $\Delta$ of dimension $\binom{N}{2} = \frac{N(N-1)}{2}$, where each element is computed as:
\begin{equation}
\Delta_p = |\vec{\bm{\delta}}_{ij}| = \sqrt{\vec{\bm{\delta}}_{ij} \cdot \vec{\bm{\delta}}_{ij}}
\end{equation}
with $p$ indexing the unique particle pairs.

This vector $\Delta$ provides the essential input features for our Neural Network model, completely characterizing the system's configuration independent of overall translation and rotation.

\section{Sampling Algorithm Implementations}
\label{appendix:sampling}

This appendix provides detailed pseudocode implementations for the two sampling methods described in Section~\ref{sec:sampling}. Both algorithms are designed to sample from the target probability distribution $P(R') \propto \psi'^2(R')$ in Jacobi coordinate space, with adaptive parameter adjustment mechanisms to maintain optimal sampling efficiency.

\subsection{Random Walk Metropolis Sampling}
\label{sec:appendix_rwm}

Algorithm~\ref{algo:rwm} presents the complete implementation of adaptive Random Walk Metropolis sampling. The algorithm maintains a running acceptance rate and dynamically adjusts the step size parameter $\epsilon$ to achieve the theoretically optimal acceptance rate of 0.234 for high-dimensional problems.

\begin{algorithm}[htbp]
\caption{Adaptive Random Walk Metropolis Sampling in Jacobi Coordinates}
\label{algo:rwm}
\KwData{
    $\{R'^{(i)}\}_{i=1}^{N_{\text{sample}}}$: Initial Jacobi coordinates \\
    $T$: Number of sampling steps \\
    $\epsilon$: Initial step size \\
    $\rho$: Boundary parameter for coordinate clamping \\
    $\eta$: Adaptation learning rate (default: 0.003) \\
    $\gamma$: Smoothing parameter for running acceptance rate (default: 0.01) \\
    $f_\theta$: Neural network wavefunction model \\
    $\tau$: Burn-in period for adaptive adjustment (default: 1000) \\
    $n$: Current training iteration \\
    $\bar{a}_{\text{prev}}$: Running acceptance rate from previous iteration
}

\tcp{Initialize running acceptance rate and coordinates}
\eIf{$n = 1$}{
    $\bar{a} \gets 0$ \tcp{Initialize acceptance rate}
    \For{$i \gets 1$ \KwTo $N_{\text{sample}}$}{
        $R'^{(i)}_0 \gets \mathcal{U}(-\rho, \rho)$ \tcp{Uniform initialization}
    }
}
{
    $\bar{a} \gets \bar{a}_{\text{prev}}$ \tcp{Continue from previous iteration}
}

\tcp{Main sampling loop}
\For{$t \gets 1$ \KwTo $T$}{
    $n_{\text{accepted}} \gets 0$ \tcp{Count accepted proposals in this step}
    
    \For{$i \gets 1$ \KwTo $N_{\text{sample}}$}{
        \tcp{Propose new configuration}
        $R'^{(i)}_c \gets R'^{(i)}_{t-1} + \epsilon \cdot \mathcal{N}(0, I)$ \;
        
        \tcp{Apply boundary conditions}
        $R'^{(i)}_c \gets \text{clamp}(R'^{(i)}_c, -\rho, \rho)$ \;
        
        \tcp{Evaluate wavefunctions}
        $\psi'^{(i)}_{\text{current}} \gets f_\theta(R'^{(i)}_{t-1})$ \;
        $\psi'^{(i)}_{\text{candidate}} \gets f_\theta(R'^{(i)}_c)$ \;
        
        \tcp{Compute acceptance probability}
        $\alpha^{(i)} \gets \min\left(1, \left(\frac{\psi'^{(i)}_{\text{candidate}}}{\psi'^{(i)}_{\text{current}}}\right)^2\right)$ \;
        
        \tcp{Accept or reject}
        $u^{(i)} \sim \mathcal{U}(0,1)$ \;
        \eIf{$u^{(i)} < \alpha^{(i)}$}{
            $R'^{(i)}_t \gets R'^{(i)}_c$ \tcp{Accept proposal}
            $n_{\text{accepted}} \gets n_{\text{accepted}} + 1$ \;
        }{
            $R'^{(i)}_t \gets R'^{(i)}_{t-1}$ \tcp{Reject proposal}
        }
    }
    
    \tcp{Update running acceptance rate}
    $a_t \gets n_{\text{accepted}} / N_{\text{sample}}$ \;
    $\bar{a} \gets (1-\gamma) \bar{a} + \gamma \, a_t$ \;
}

\tcp{Adaptive step size adjustment after burn-in}
\If{$n > \tau$}{
    $\epsilon \gets \epsilon \cdot \exp\bigl(\eta \, (\bar{a} - 0.234)\bigr)$ \;
}

\KwResult{
    $\{R'^{(i)}_T\}_{i=1}^{N_{\text{sample}}}$: Final sampled configurations \\
    $\epsilon$: Updated step size \\
    $\bar{a}$: Updated running acceptance rate
}
\end{algorithm}

\subsection{Metropolis-Adjusted Langevin Algorithm}
\label{sec:appendix_mala}

Algorithm~\ref{algo:mala} implements the Metropolis-Adjusted Langevin Algorithm, which incorporates gradient information to improve sampling efficiency. The algorithm targets an optimal acceptance rate of 0.574 and requires computation of both forward and reverse proposal probabilities due to the asymmetric nature of the gradient-guided proposal distribution.

\begin{algorithm}[htbp]
\caption{Metropolis-Adjusted Langevin Algorithm in Jacobi Coordinates}
\label{algo:mala}
\KwData{
    $\{R'^{(i)}\}_{i=1}^{N_{\text{sample}}}$: Initial Jacobi coordinates \\
    $T$: Number of sampling steps \\
    $\epsilon$: Initial step size \\
    $\rho$: Boundary parameter for coordinate clamping \\
    $\eta$: Adaptation learning rate (default: 0.003) \\
    $\gamma$: Smoothing parameter for running acceptance rate (default: 0.01) \\
    $f_\theta$: Neural network wavefunction model \\
    $\tau$: Burn-in period for adaptive adjustment (default: 1000) \\
    $n$: Current training iteration \\
    $\bar{a}_{\text{prev}}$: Running acceptance rate from previous iteration
}

\eIf{$n = 1$}{
    $\bar{a} \gets 0$ \;
    \For{$i \gets 1$ \KwTo $N_{\text{sample}}$}{
        $R'^{(i)}_0 \gets \mathcal{U}(-\rho, \rho)$ \;
    }
}
{
    $\bar{a} \gets \bar{a}_{\text{prev}}$ \;
}

\tcp{Main sampling loop}
\For{$t \gets 1$ \KwTo $T$}{
    $n_{\text{accepted}} \gets 0$ \;
    
    \For{$i \gets 1$ \KwTo $N_{\text{sample}}$}{
        $\psi'^{(i)}_{\text{current}} \gets f_\theta(R'^{(i)}_{t-1})$ \;
        $\nabla \log P(R'^{(i)}_{t-1}) \gets 2\nabla \log |\psi'^{(i)}_{\text{current}}|$ \;
        
        $R'^{(i)}_c \gets R'^{(i)}_{t-1} + \frac{\epsilon^2}{2} \nabla \log P(R'^{(i)}_{t-1}) + \epsilon \cdot \mathcal{N}(0, I)$ \;
        
        $R'^{(i)}_c \gets \text{clamp}(R'^{(i)}_c, -\rho, \rho)$ \;
        
        $\psi'^{(i)}_{\text{candidate}} \gets f_\theta(R'^{(i)}_c)$ \;
        $\nabla \log P(R'^{(i)}_c) \gets 2\nabla \log |\psi'^{(i)}_{\text{candidate}}|$ \;
        
        $\mu_{\text{forward}} \gets R'^{(i)}_{t-1} + \frac{\epsilon^2}{2}\nabla \log P(R'^{(i)}_{t-1})$ \;
        $\mu_{\text{reverse}} \gets R'^{(i)}_c + \frac{\epsilon^2}{2}\nabla \log P(R'^{(i)}_c)$ \;
        
        $\log q_{\text{forward}} \gets -\frac{1}{2\epsilon^2}\|R'^{(i)}_c - \mu_{\text{forward}}\|^2$ \;
        $\log q_{\text{reverse}} \gets -\frac{1}{2\epsilon^2}\|R'^{(i)}_{t-1} - \mu_{\text{reverse}}\|^2$ \;
        
        $\log \alpha^{(i)} \gets \min\left(0, 2\log|\psi'^{(i)}_{\text{candidate}}| - 2\log|\psi'^{(i)}_{\text{current}}| + \log q_{\text{reverse}} - \log q_{\text{forward}}\right)$ \;
        $\alpha^{(i)} \gets \exp(\log \alpha^{(i)})$ \;
        
        $u^{(i)} \sim \mathcal{U}(0,1)$ \;
        \eIf{$u^{(i)} < \alpha^{(i)}$}{
            $R'^{(i)}_t \gets R'^{(i)}_c$ \;
            $n_{\text{accepted}} \gets n_{\text{accepted}} + 1$ \;
        }{
            $R'^{(i)}_t \gets R'^{(i)}_{t-1}$ \;
        }
    }
    
    $a_t \gets n_{\text{accepted}} / N_{\text{sample}}$ \;
    $\bar{a} \gets (1-\gamma) \bar{a} + \gamma \, a_t$ \;
}

\If{$n > \tau$}{
    $\epsilon \gets \epsilon \cdot \exp\bigl(\eta \, (\bar{a} - 0.574)\bigr)$ \;
}

\end{algorithm}

\subsection{Implementation Notes}
\label{sec:appendix_notes}

\paragraph{Numerical Stability} In the MALA implementation, we compute acceptance probabilities in log-space to prevent numerical overflow when dealing with very large or small wavefunction ratios. The $\text{clamp}$ operation ensures coordinates remain within physically meaningful bounds.

\paragraph{Gradient Computation} The gradients $\nabla \log P(R')$ are computed using automatic differentiation through the neural network. This requires that the network architecture supports gradient computation with respect to input coordinates.

\paragraph{Batch Processing} Both algorithms process multiple sample configurations simultaneously ($N_{\text{sample}}$ samples per batch), which improves computational efficiency and provides better statistics for acceptance rate estimation.

\paragraph{Parameter Adaptation} The exponential adaptation rule ensures that step sizes adjust smoothly toward optimal acceptance rates. The delayed activation prevents instability during early training phases when the neural network parameters are changing rapidly.

\newpage
\section{System Detailed Results}
\label{appendix:systemresults}
\subsection{System A: 4-10 particles}
\label{appendix:systemA}

\begin{figure}[htbp]
    \centering
    \begin{subfigure}[b]{0.4\textwidth}
        \centering
        \includegraphics[width=\textwidth, height=6cm, keepaspectratio]{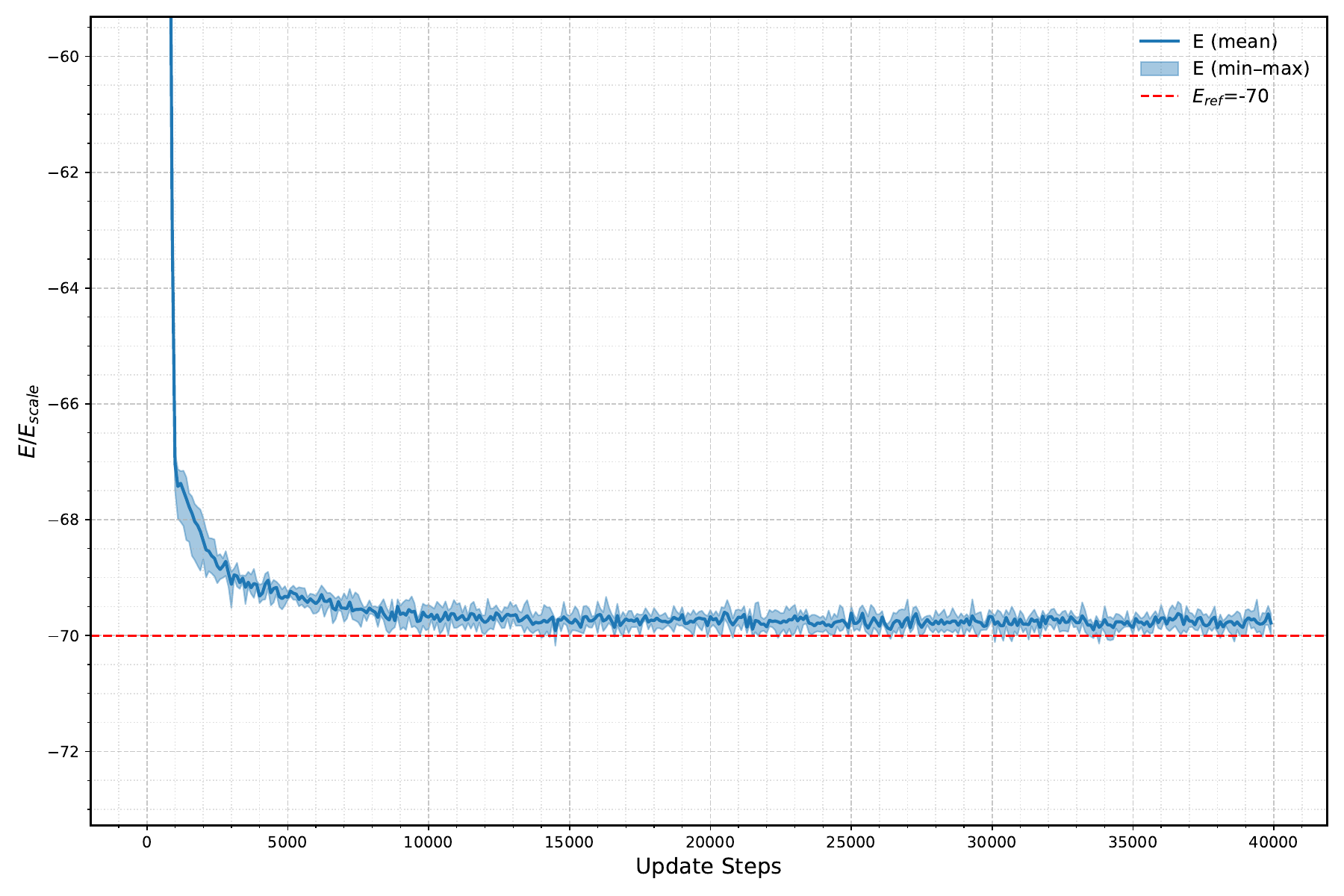}
        \caption{\centering
            \texttt{GELU-MALA}.\\
            $E/E_{scale} = 69.50 \pm 0.70$.\\
            r-range (99.9\% CI): $[-0.44,\,0.44]$.
        }
        \label{fig:4p-gelu-mala}
    \end{subfigure}
    \hfill
    \begin{subfigure}[b]{0.4\textwidth}
        \centering
        \includegraphics[width=\textwidth, height=6cm, keepaspectratio]{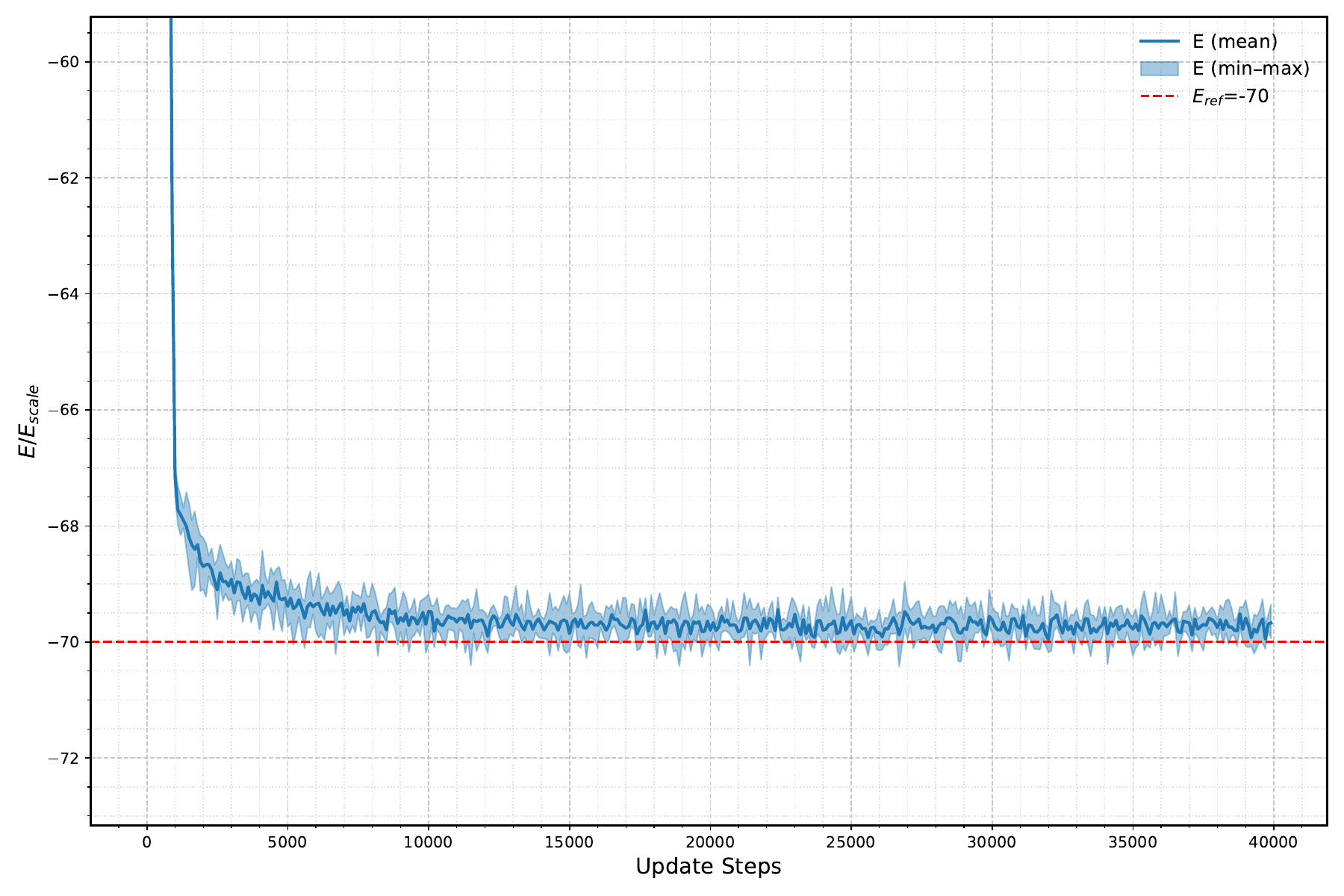}
        \caption{\centering
            \texttt{GELU-ARW}.\\
            $E/E_{scale} = 68.22 \pm 2.28$.\\
            r-range (99.9\% CI): $[-0.45,\,0.45]$.
        }
        \label{fig:4p-gelu-arw}
    \end{subfigure}

    \begin{subfigure}[b]{0.4\textwidth}
        \centering
        \includegraphics[width=\textwidth, height=6cm, keepaspectratio]{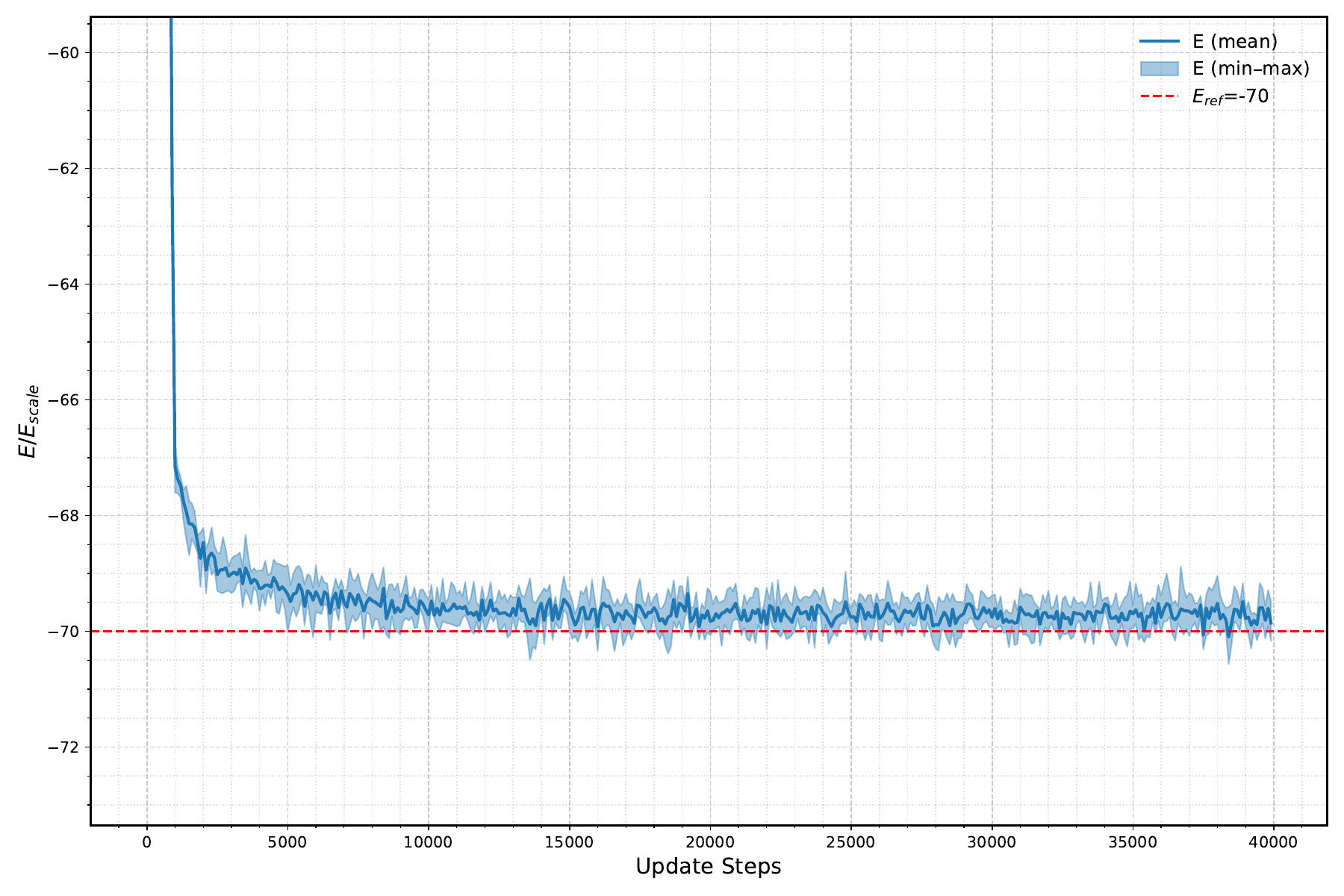}
        \caption{\centering
            \texttt{GELU-RW}.\\
            $E/E_{scale} = 69.16 \pm 1.92$.\\
            r-range (99.9\% CI): $[-0.44,\,0.44]$.
        }
        \label{fig:4p-gelu-rw}
    \end{subfigure}
    \hfill
    \begin{subfigure}[b]{0.4\textwidth}
        \centering
        \includegraphics[width=\textwidth, height=6cm, keepaspectratio]{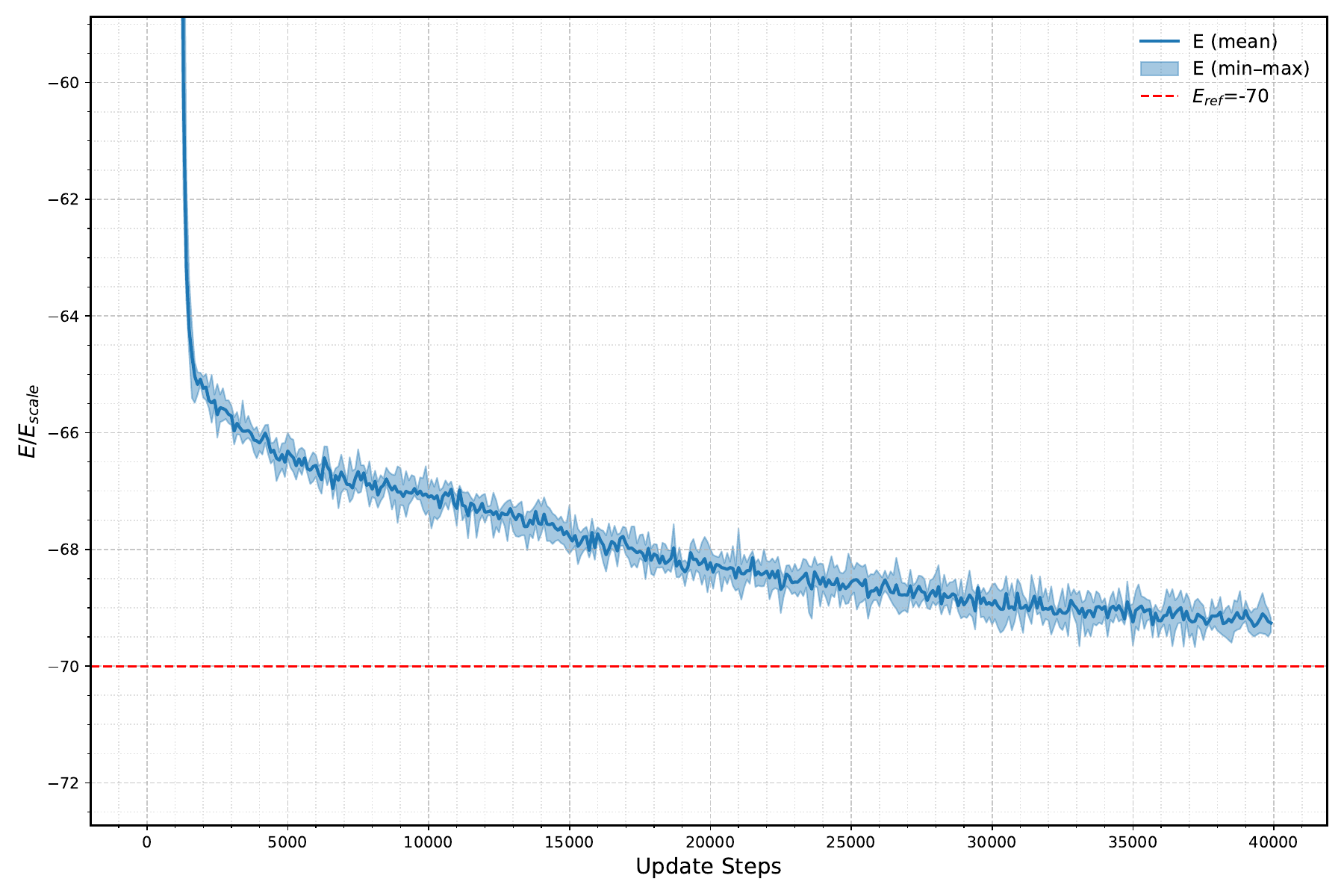}
        \caption{\centering
            \texttt{tanh-ARW}.\\
            $E/E_{scale} = 69.08 \pm 0.63$.\\
            r-range (99.9\% CI): $[-0.44,\,0.44]$.
        }
        \label{fig:4p-tanh-arw}
    \end{subfigure}

    \caption{4-particle harmonic-potential system with two-body interactions.
    The variable $r$ denotes the Jacobi coordinate; the reported r-range is the
    interval containing 99.9\% of sampled $r$ values.}
    \label{fig:4particles-all-methods}
\end{figure}

\begin{figure}[htbp]
    \centering
    \begin{subfigure}[b]{0.4\textwidth}
        \centering
        \includegraphics[width=\textwidth, height=6cm, keepaspectratio]{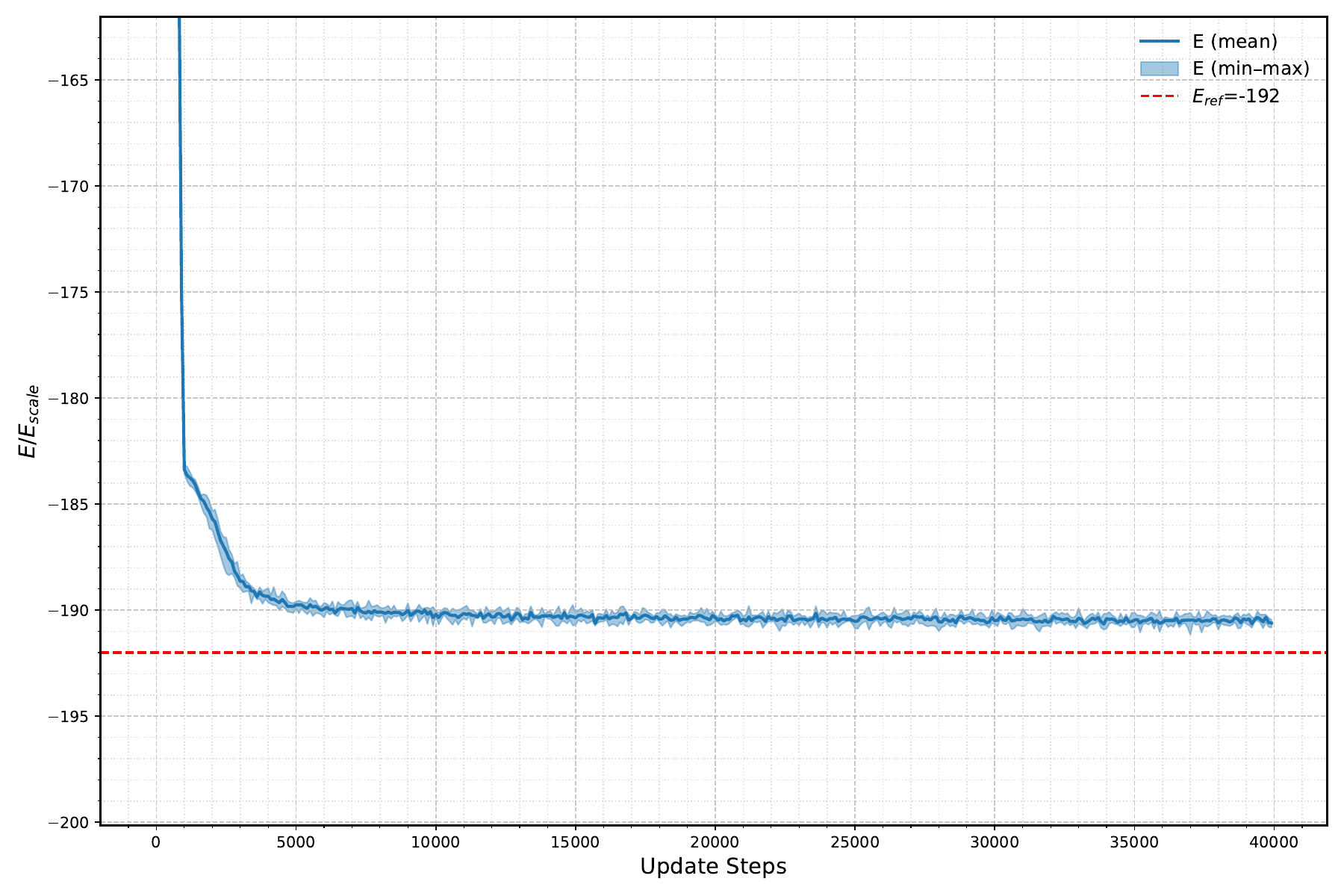}
        \caption{\centering
            \texttt{GELU-MALA}.\\
            $E/E_{scale} = -194.36 \pm 3.53$.\\
            r-range (99.9\% CI): $[-0.34,\,0.34]$.
        }
        \label{fig:5p-gelu-mala}
    \end{subfigure}
    \hfill
    \begin{subfigure}[b]{0.4\textwidth}
        \centering
        \includegraphics[width=\textwidth, height=6cm, keepaspectratio]{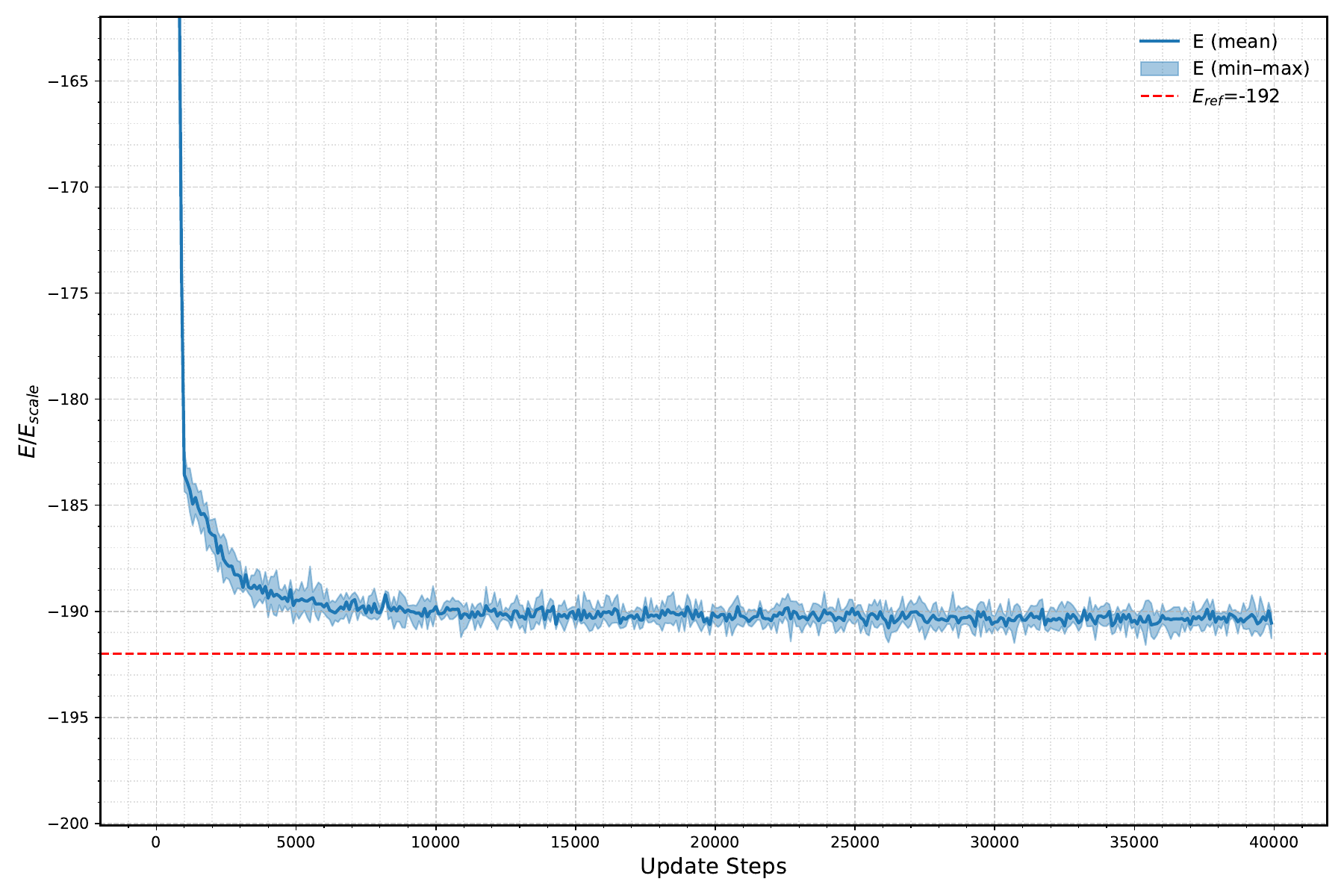}
        \caption{\centering
            \texttt{GELU-ARW}.\\
            $E/E_{scale} = -189.41 \pm 9.25$.\\
            r-range (99.9\% CI): $[-0.34,\,0.34]$.
        }
        \label{fig:5p-gelu-arw}
    \end{subfigure}

    \begin{subfigure}[b]{0.4\textwidth}
        \centering
        \includegraphics[width=\textwidth, height=6cm, keepaspectratio]{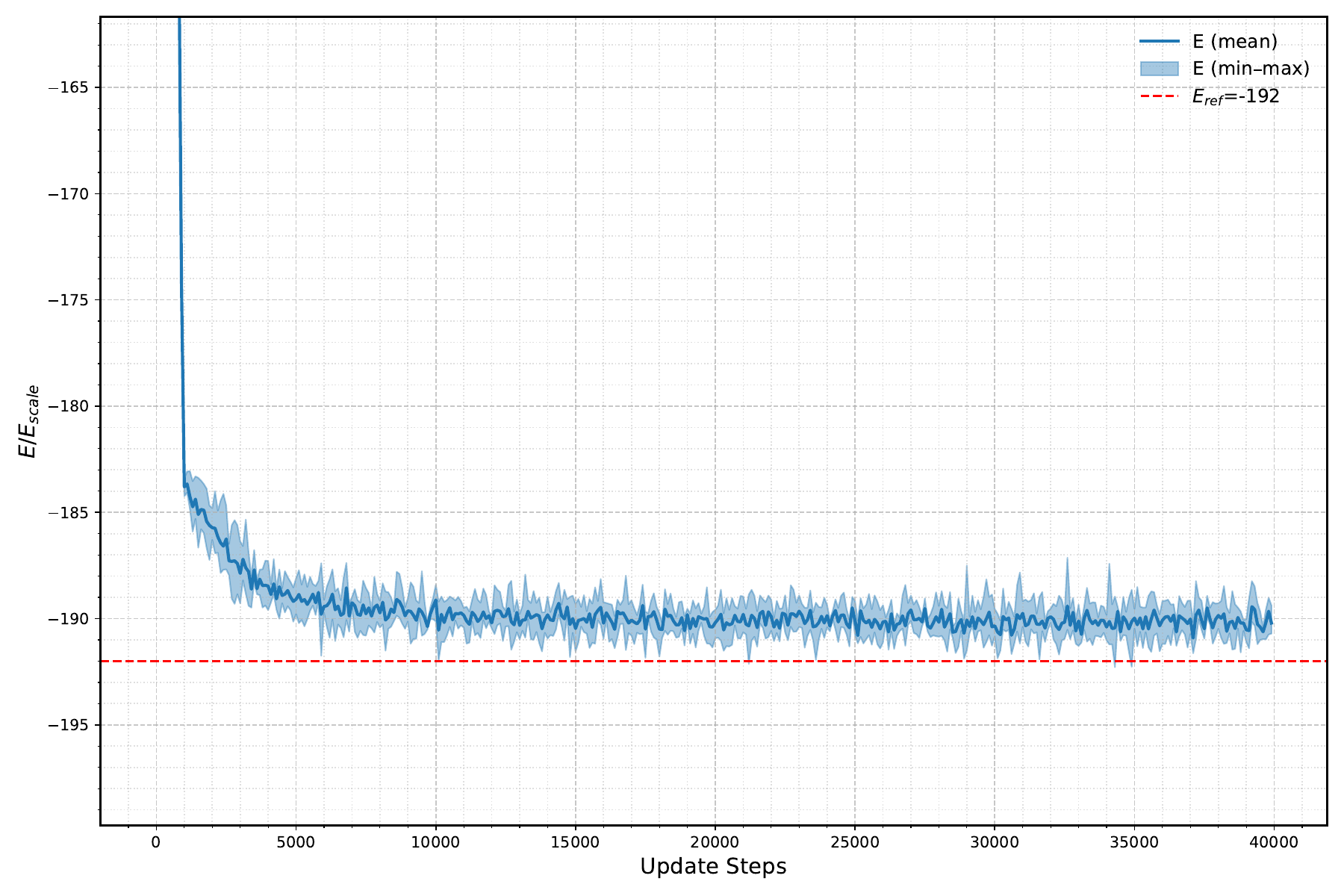}
        \caption{\centering
            \texttt{GELU-RW}.\\
            $E/E_{scale} = -190.87 \pm 1.48$.\\
            r-range (99.9\% CI): $[-0.34,\,0.34]$.
        }
        \label{fig:5p-gelu-rw}
    \end{subfigure}
    \hfill
    \begin{subfigure}[b]{0.4\textwidth}
        \centering
        \includegraphics[width=\textwidth, height=6cm, keepaspectratio]{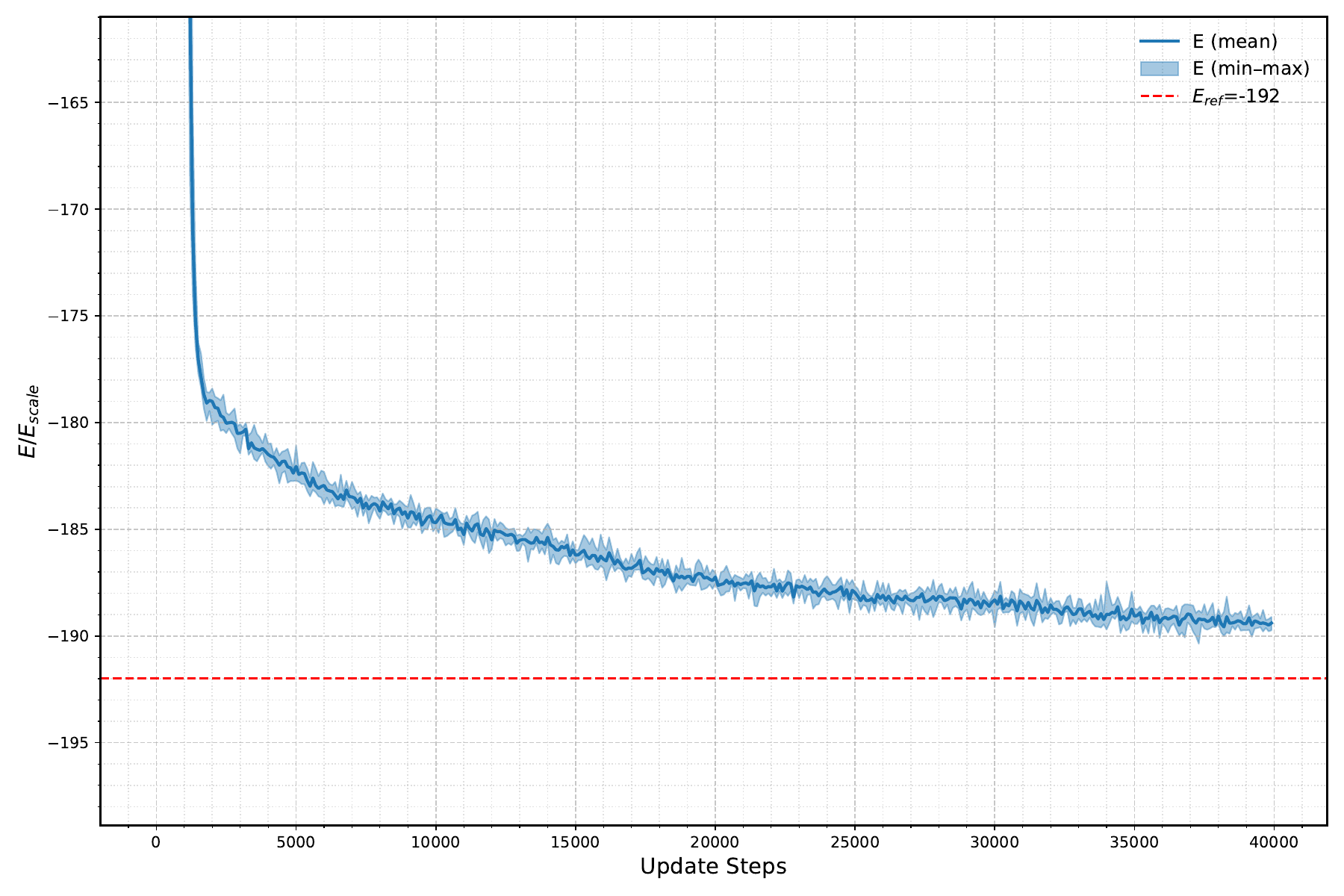}
        \caption{\centering
            \texttt{tanh-ARW}.\\
            $E/E_{scale} = -189.45 \pm 1.33$.\\
            r-range (99.9\% CI): $[-0.34,\,0.34]$.
        }
        \label{fig:5p-tanh-arw}
    \end{subfigure}

    \caption{5-particle harmonic-potential system with two-body interactions.}
    \label{fig:5particles-all-methods}
\end{figure}

\begin{figure}[htbp]
    \centering
    \begin{subfigure}[b]{0.4\textwidth}
        \centering
        \includegraphics[width=\textwidth, height=6cm, keepaspectratio]{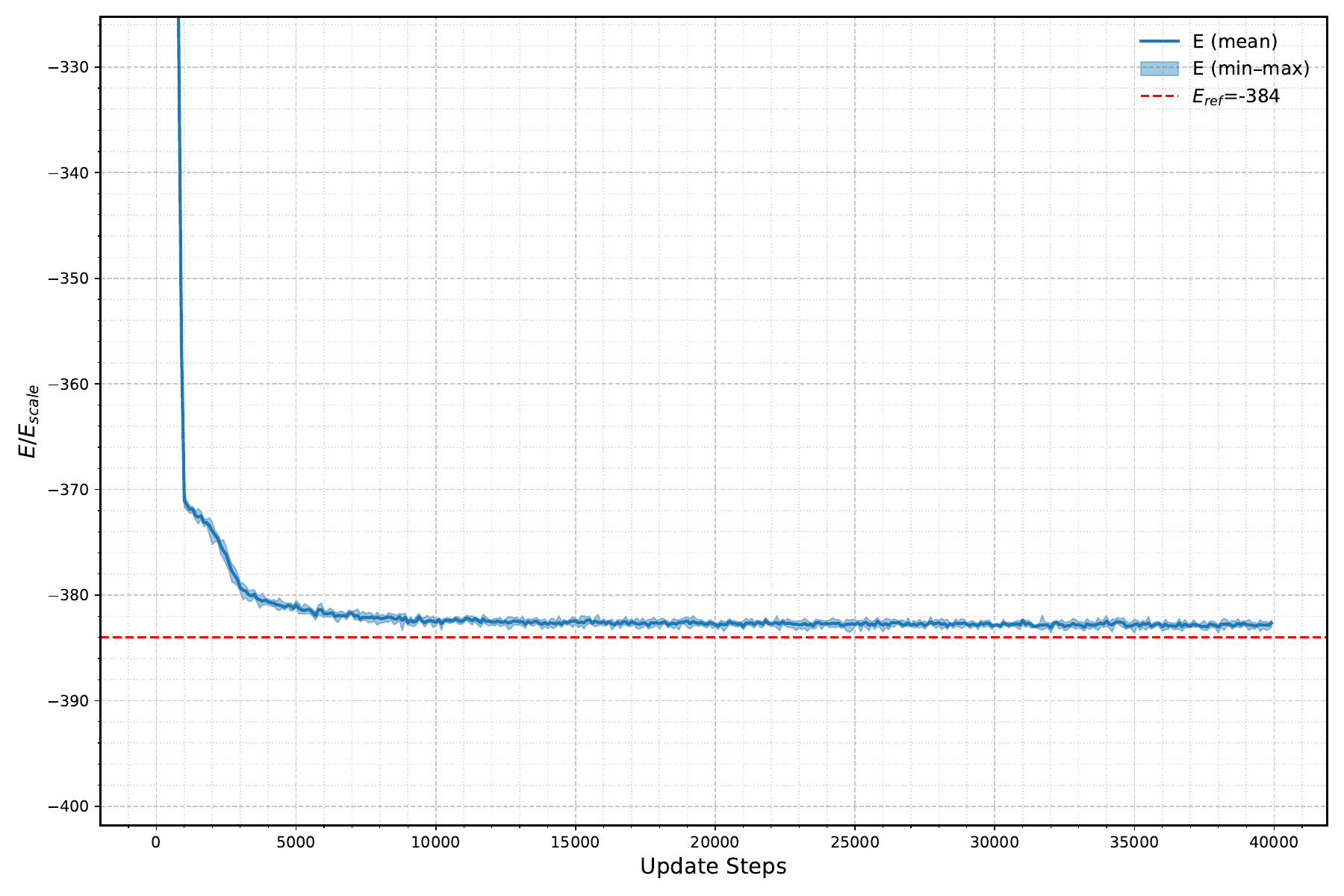}
        \caption{\centering
            \texttt{GELU-MALA}.\\
            $E/E_{scale} = -364.74 \pm 27.33$.\\
            r-range (99.9\% CI): $[-0.30,\,0.30]$.
        }
        \label{fig:6p-gelu-mala}
    \end{subfigure}
    \hfill
    \begin{subfigure}[b]{0.4\textwidth}
        \centering
        \includegraphics[width=\textwidth, height=6cm, keepaspectratio]{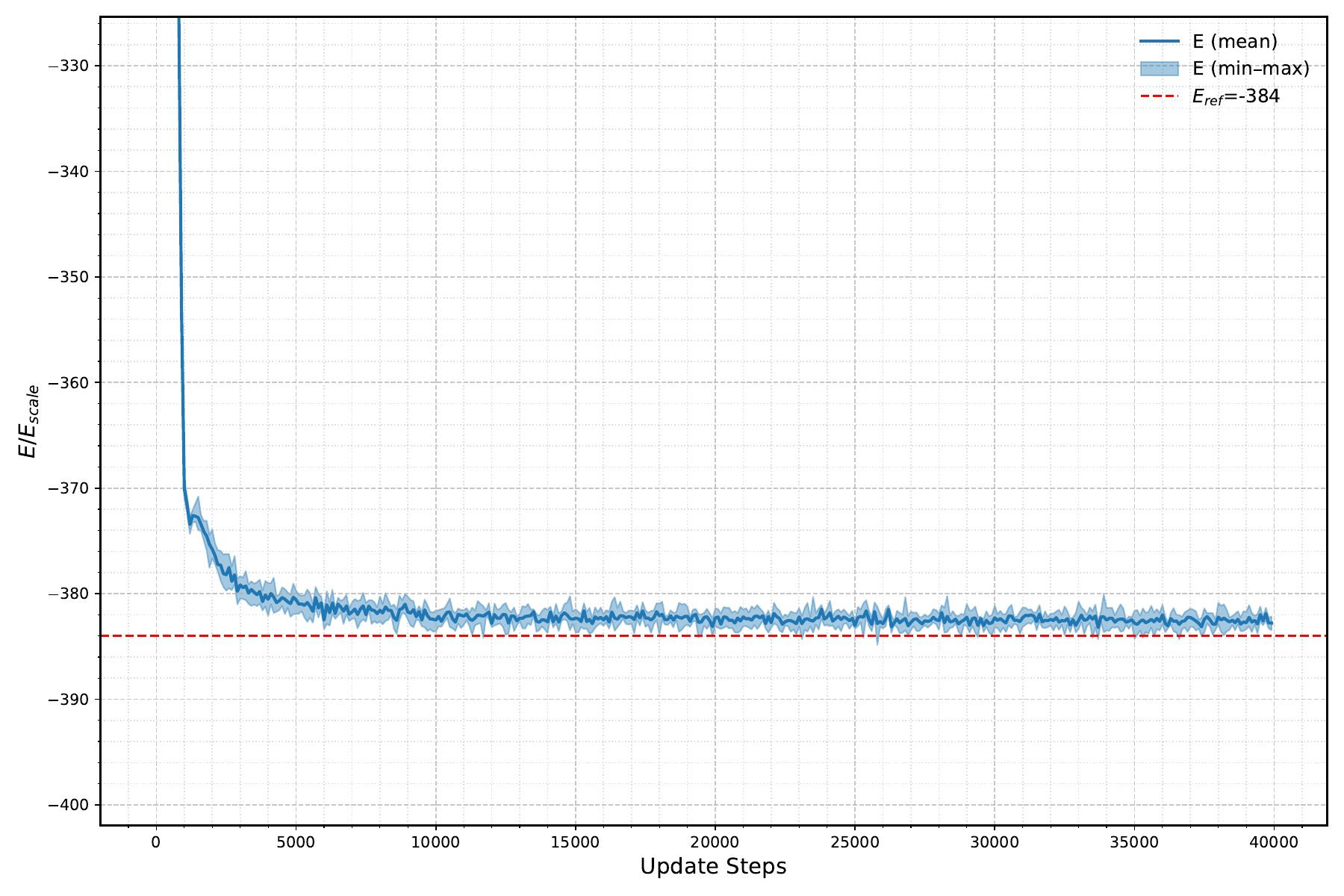}
        \caption{\centering
            \texttt{GELU-ARW}.\\
            $E/E_{scale} = -386.46 \pm 7.68$.\\
            r-range (99.9\% CI): $[-0.30,\,0.29]$.
        }
        \label{fig:6p-gelu-arw}
    \end{subfigure}

    \begin{subfigure}[b]{0.4\textwidth}
        \centering
        \includegraphics[width=\textwidth, height=6cm, keepaspectratio]{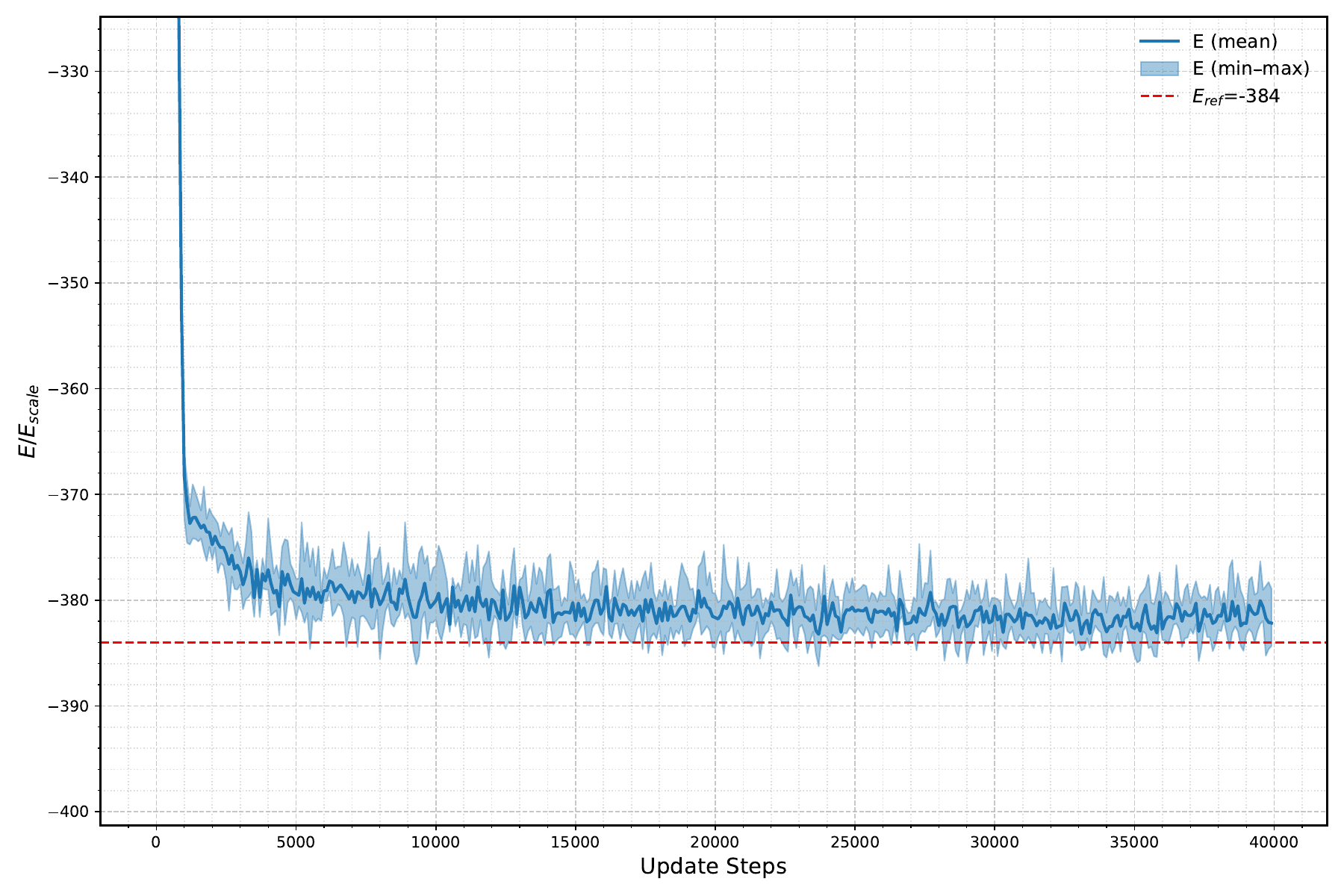}
        \caption{\centering
            \texttt{GELU-RW}.\\
            $E/E_{scale} = -382.02 \pm 1.44$.\\
            r-range (99.9\% CI): $[-0.30,\,0.30]$.
        }
        \label{fig:6p-gelu-rw}
    \end{subfigure}
    \hfill
    \begin{subfigure}[b]{0.4\textwidth}
        \centering
        \includegraphics[width=\textwidth, height=6cm, keepaspectratio]{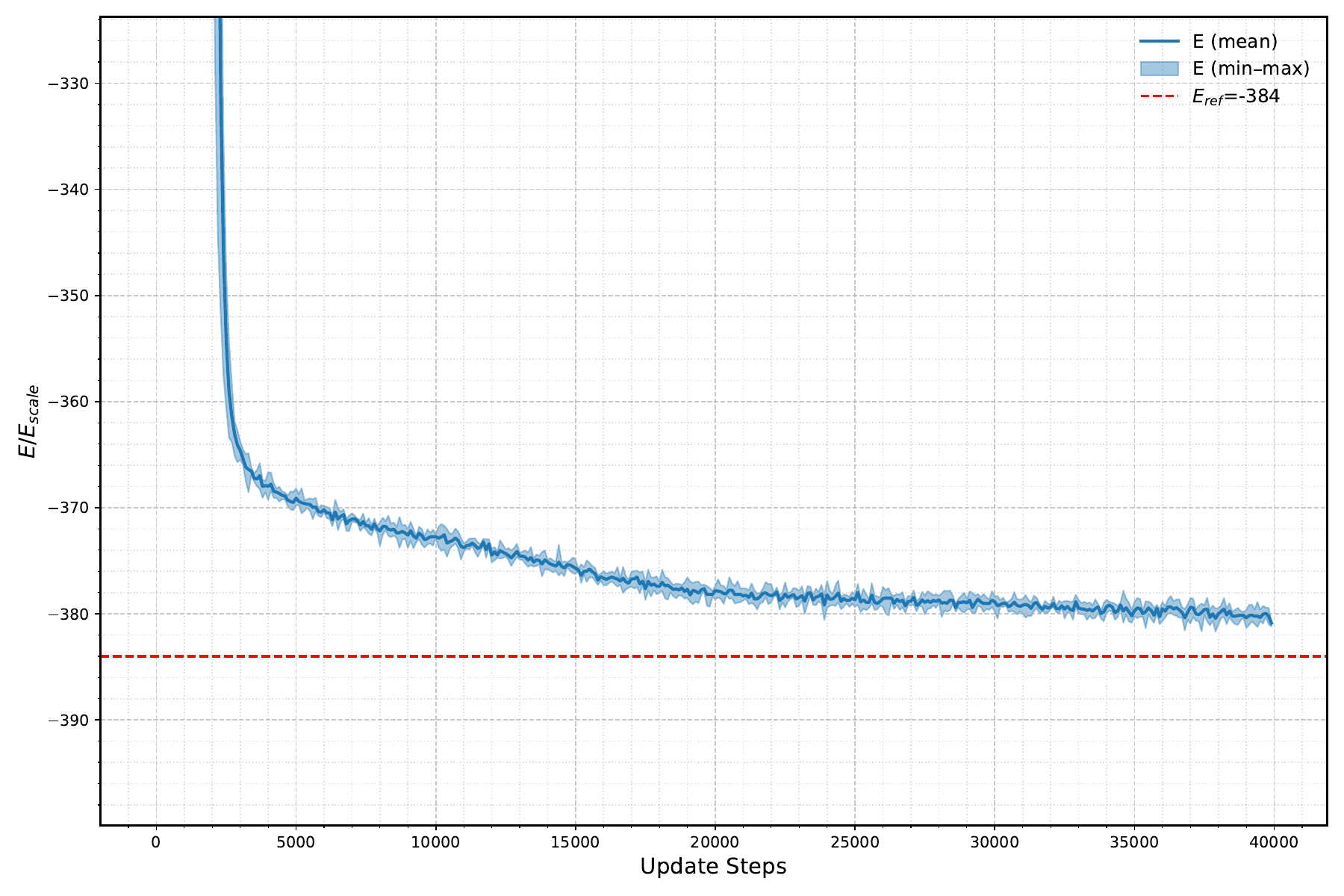}
        \caption{\centering
            \texttt{tanh-ARW}.\\
            $E/E_{scale} = -380.72 \pm 1.69$.\\
            r-range (99.9\% CI): $[-0.30,\,0.30]$.
        }
        \label{fig:6p-tanh-arw}
    \end{subfigure}

    \caption{6-particle harmonic-potential system with two-body interactions.}
    \label{fig:6particles-all-methods}
\end{figure}

\begin{figure}[htbp]
    \centering
    \begin{subfigure}[b]{0.4\textwidth}
        \centering
        \includegraphics[width=\textwidth, height=6cm, keepaspectratio]{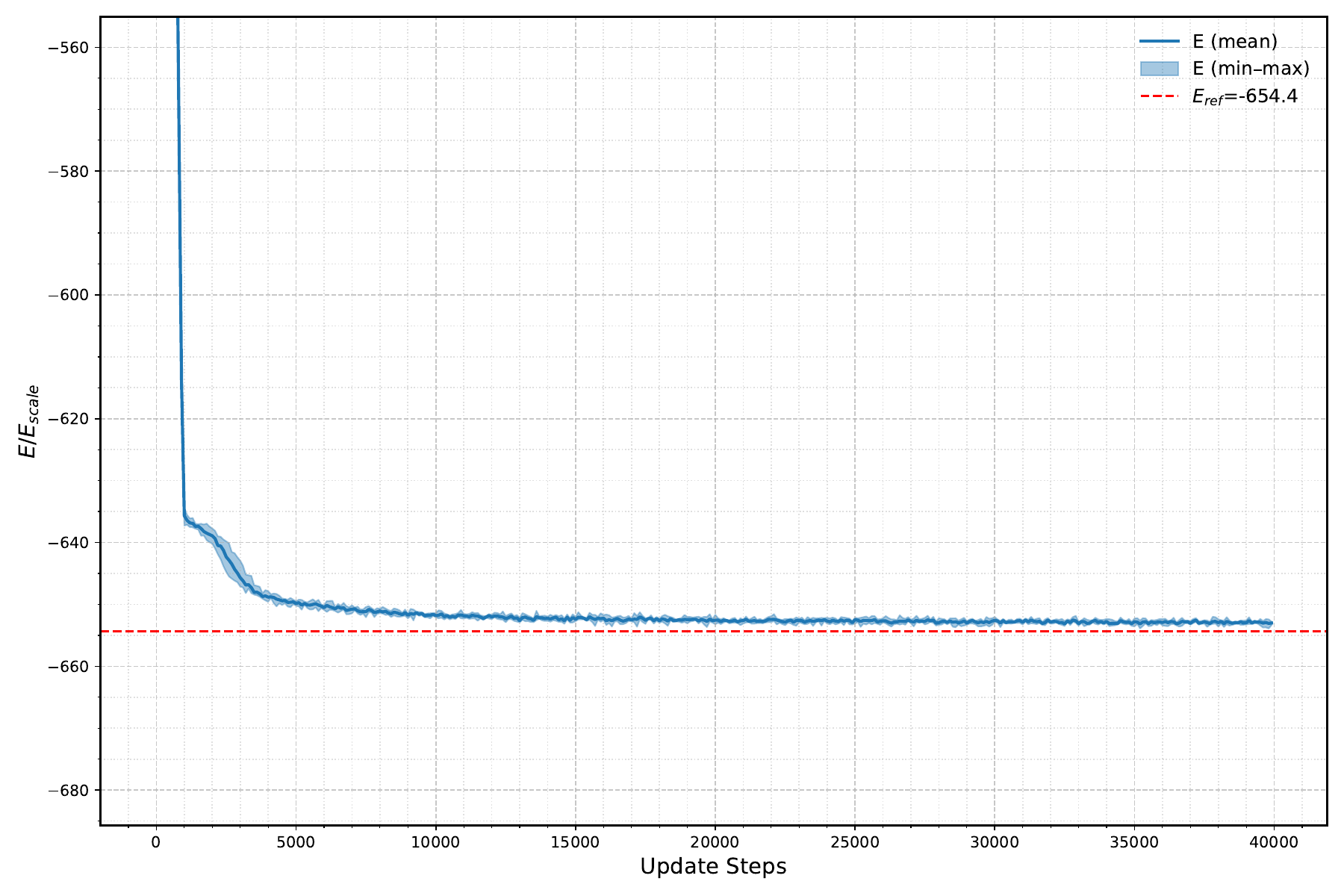}
        \caption{\centering
            \texttt{GELU-MALA}.\\
            $E/E_{scale} = -653.39 \pm 29.40$.\\
            r-range (99.9\% CI): $[-0.27,\,0.27]$.
        }
        \label{fig:7p-gelu-mala}
    \end{subfigure}
    \hfill
    \begin{subfigure}[b]{0.4\textwidth}
        \centering
        \includegraphics[width=\textwidth, height=6cm, keepaspectratio]{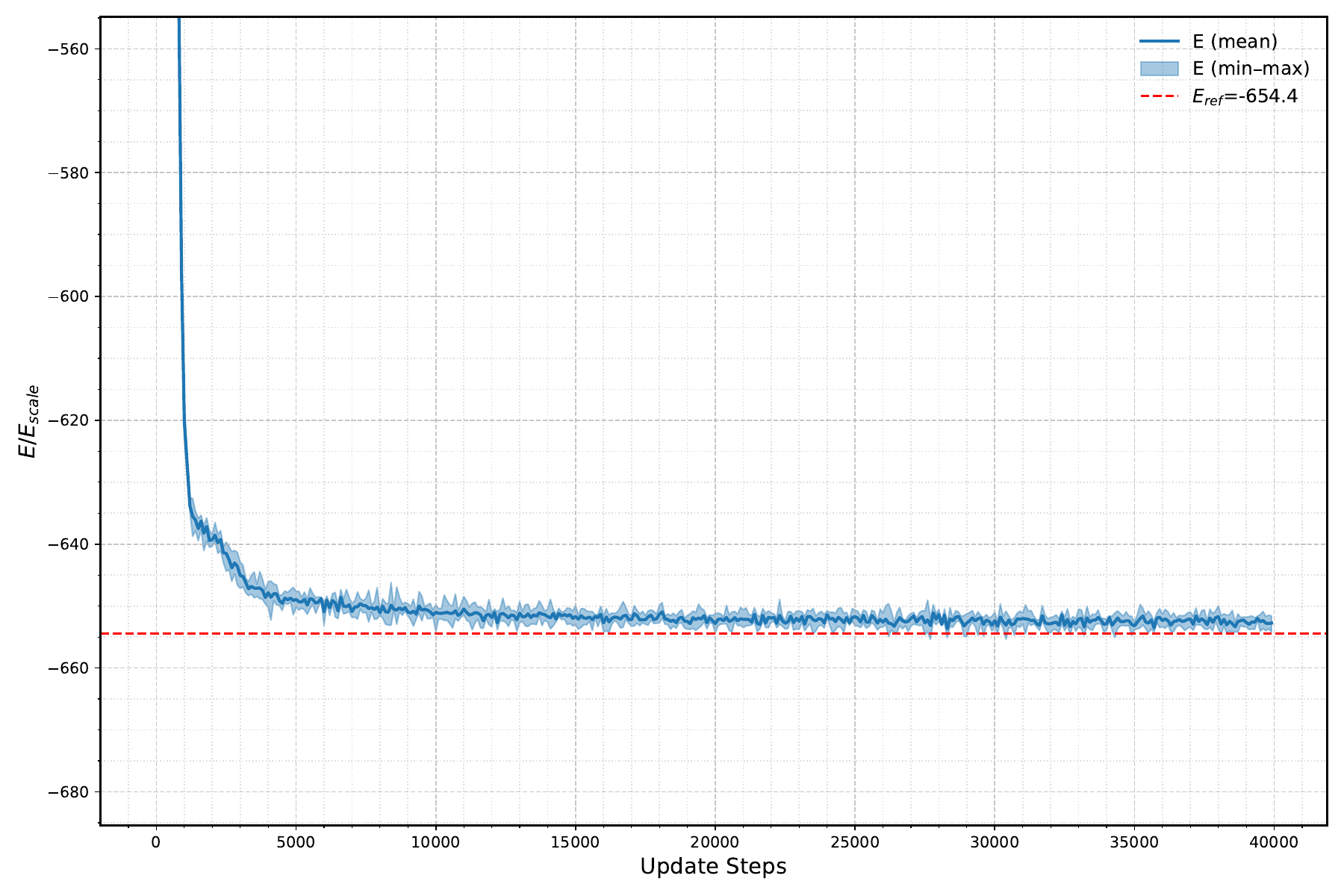}
        \caption{\centering
            \texttt{GELU-ARW}.\\
            $E/E_{scale} = -644.70 \pm 5.66$.\\
            r-range (99.9\% CI): $[-0.27,\,0.27]$.
        }
        \label{fig:7p-gelu-arw}
    \end{subfigure}

    \begin{subfigure}[b]{0.4\textwidth}
        \centering
        \includegraphics[width=\textwidth, height=6cm, keepaspectratio]{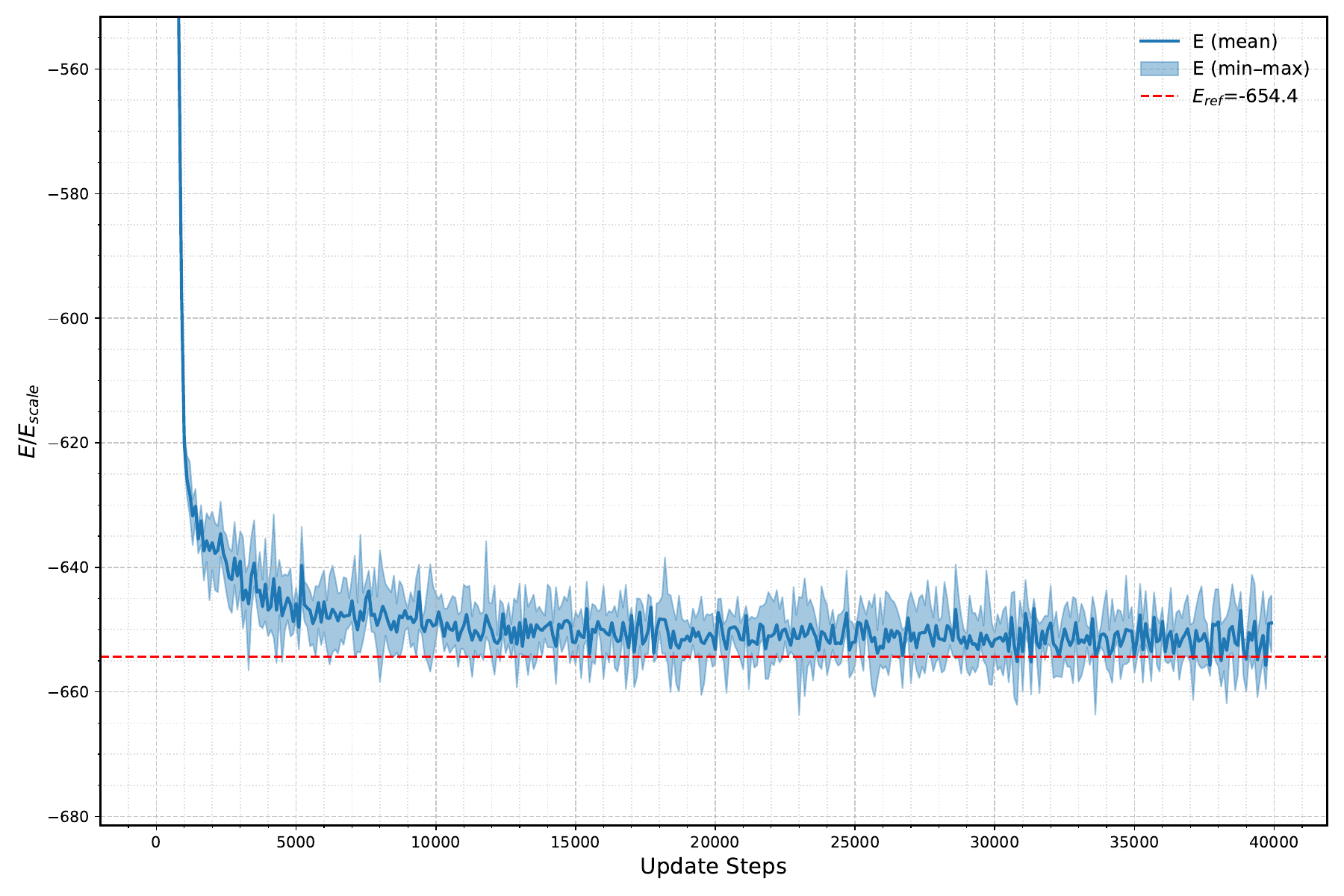}
        \caption{\centering
            \texttt{GELU-RW}.\\
            $E/E_{scale} = -652.84 \pm 4.24$.\\
            r-range (99.9\% CI): $[-0.27,\,0.27]$.
        }
        \label{fig:7p-gelu-rw}
    \end{subfigure}
    \hfill
    \begin{subfigure}[b]{0.4\textwidth}
        \centering
        \includegraphics[width=\textwidth, height=6cm, keepaspectratio]{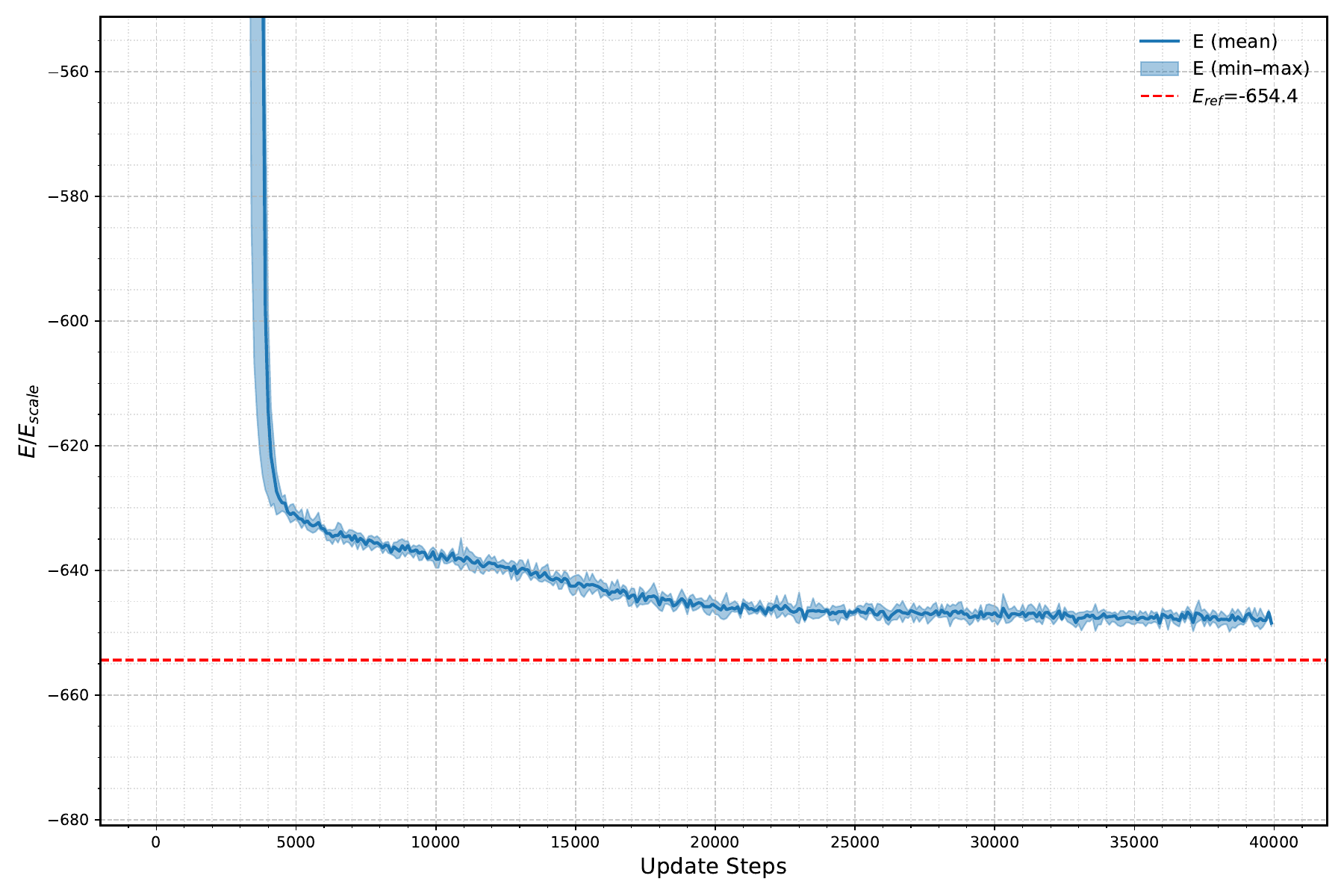}
        \caption{\centering
            \texttt{tanh-ARW}.\\
            $E/E_{scale} = -647.86 \pm 2.10$.\\
            r-range (99.9\% CI): $[-0.27,\,0.27]$.
        }
        \label{fig:7p-tanh-arw}
    \end{subfigure}

    \caption{7-particle harmonic-potential system with two-body interactions.}
    \label{fig:7particles-all-methods}
\end{figure}

\clearpage
\begin{figure}[htbp]
    \centering
    \begin{subfigure}[b]{0.32\textwidth}
        \centering
        \includegraphics[width=\textwidth, height=6cm, keepaspectratio]{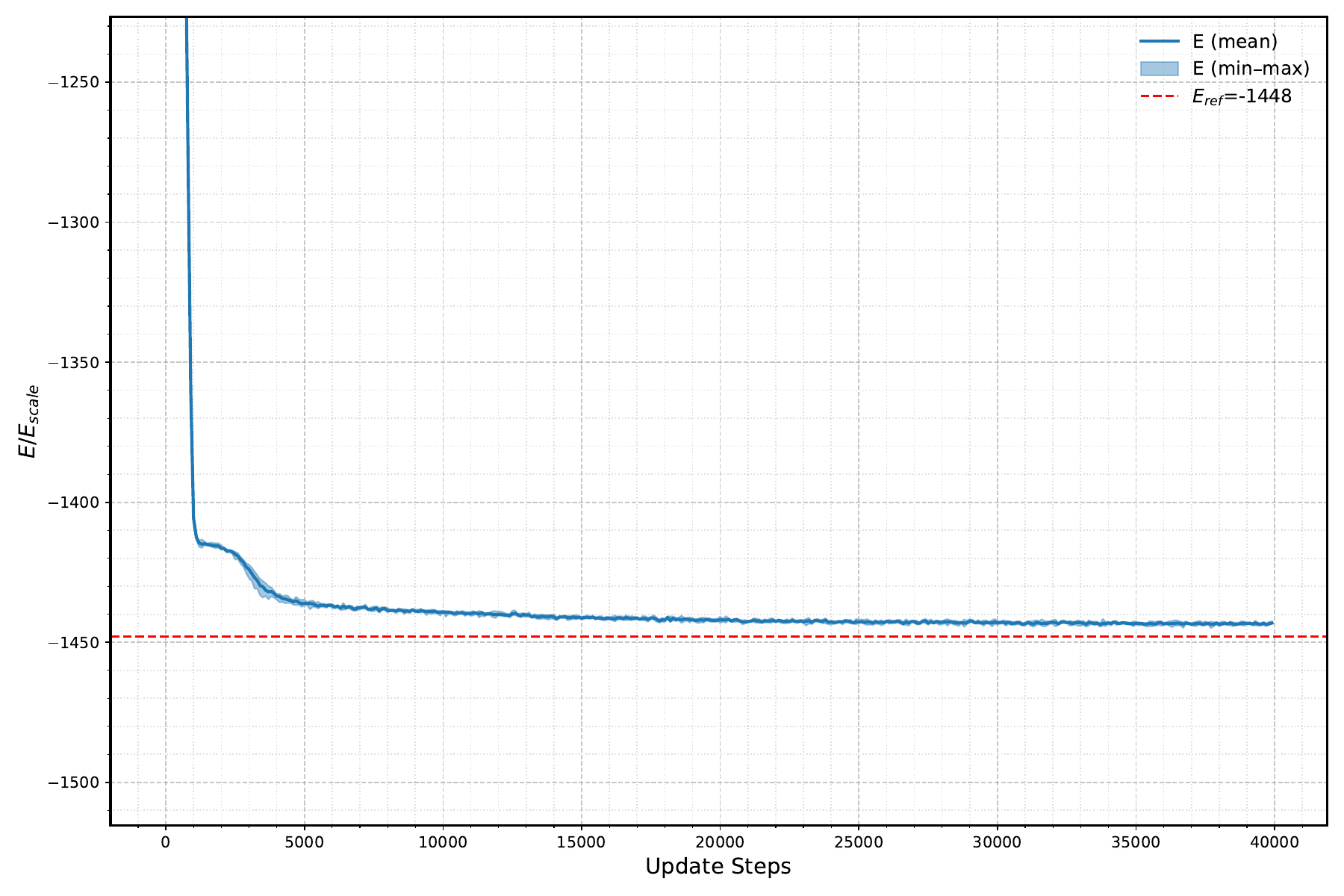}
        \caption{\centering
            \texttt{GELU-MALA}.\\
            $E/E_{scale} = -1443.54 \pm 2.04$.\\
            r-range (99.9\% CI): $[-0.24,\,0.24]$.
        }
        \label{fig:9p-gelu-mala}
    \end{subfigure}
    \hfill
    \begin{subfigure}[b]{0.32\textwidth}
        \centering
        \includegraphics[width=\textwidth, height=6cm, keepaspectratio]{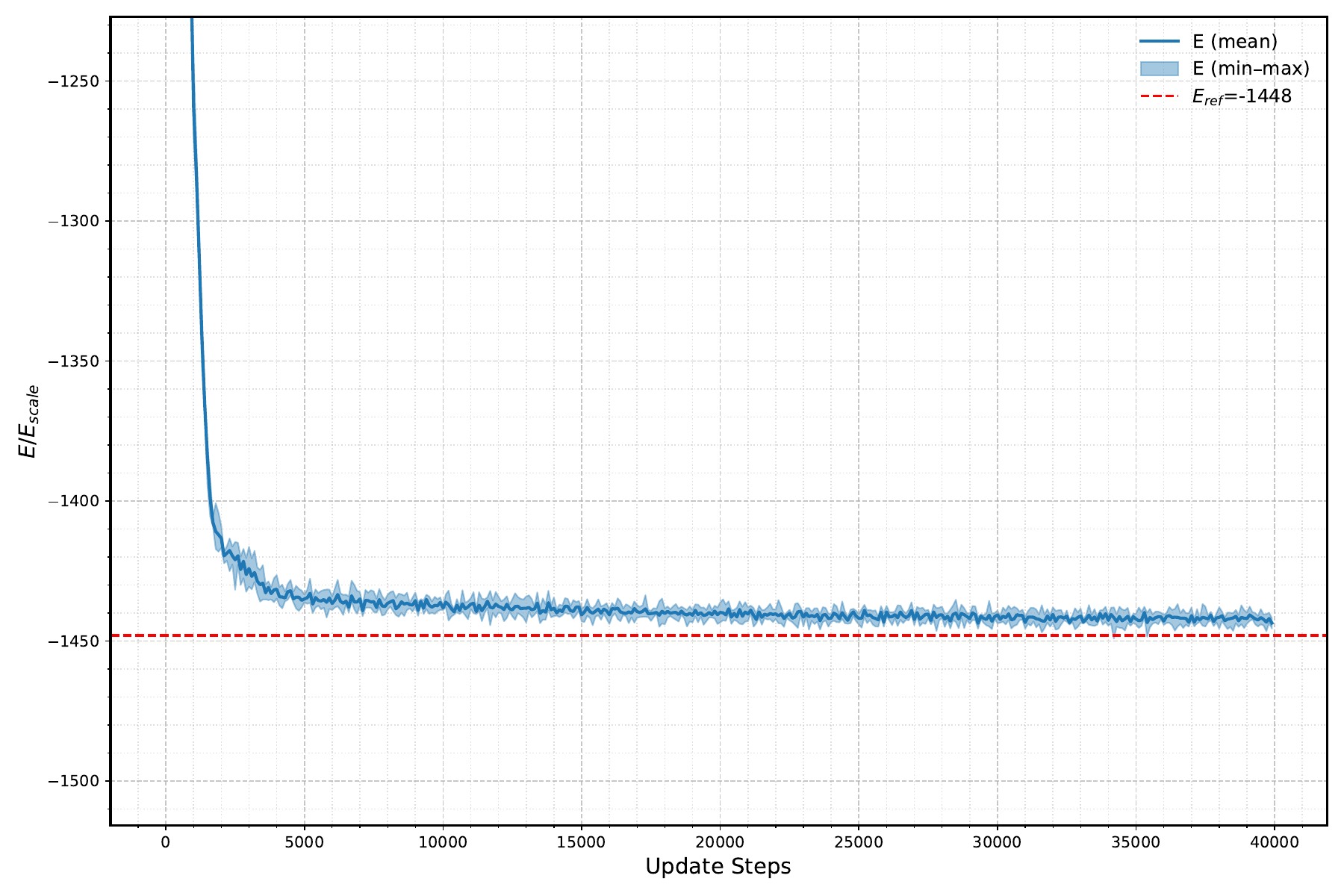}
        \caption{\centering
            \texttt{GELU-ARW}.\\
            $E/E_{scale} = -1447.32 \pm 16.04$.\\
            r-range (99.9\% CI): $[-0.24,\,0.24]$.
        }
        \label{fig:9p-gelu-arw}
    \end{subfigure}
    \hfill
    \begin{subfigure}[b]{0.32\textwidth}
        \centering
        \includegraphics[width=\textwidth, height=6cm, keepaspectratio]{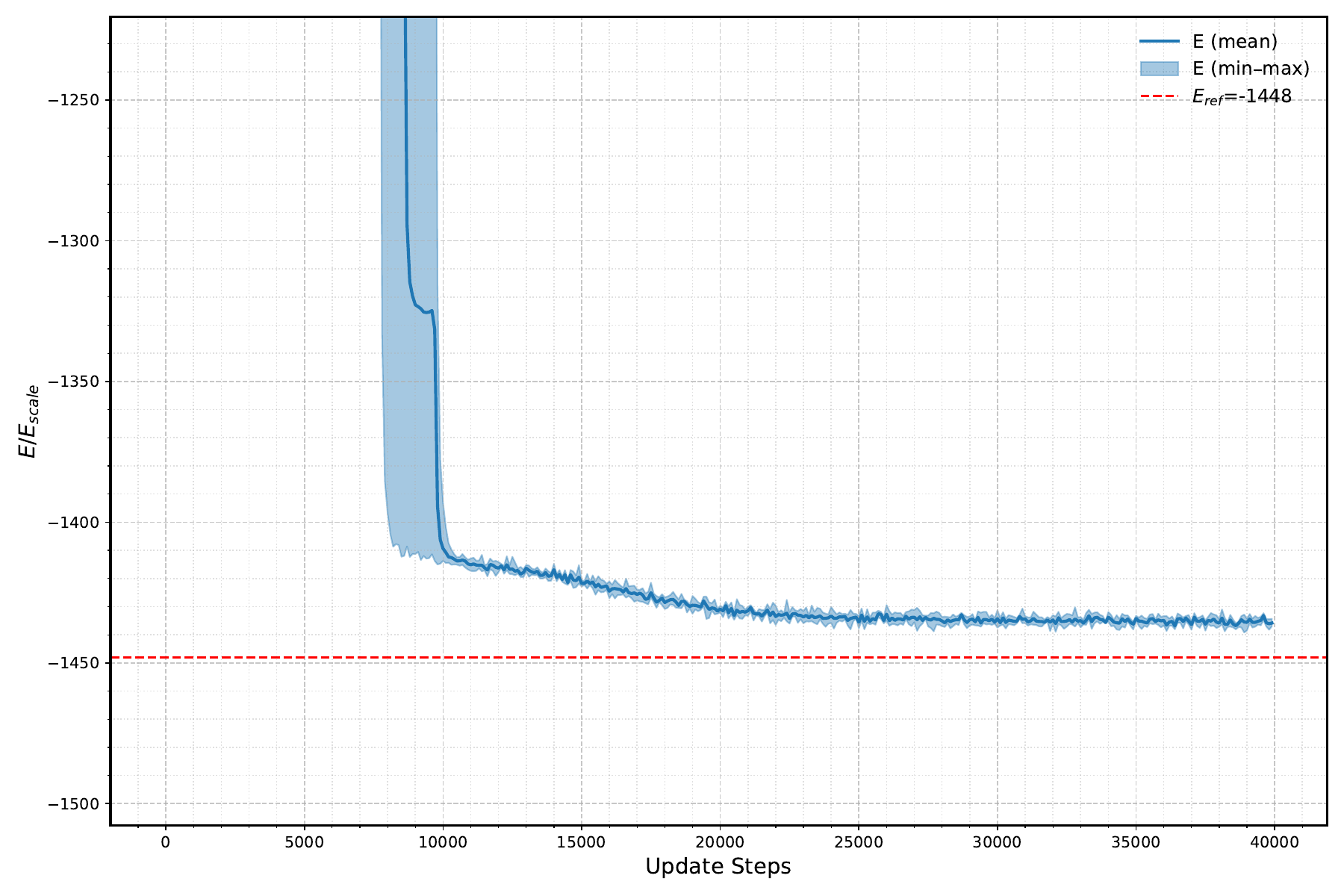}
        \caption{\centering
            \texttt{tanh-ARW}.\\
            $E/E_{scale} = -1438.94 \pm 4.46$.\\
            r-range (99.9\% CI): $[-0.24,\,0.24]$.
        }
        \label{fig:9p-tanh-arw}
    \end{subfigure}

    \caption{9-particle harmonic-potential system with two-body interactions.}
    \label{fig:9particles-all-methods}
\end{figure}

\begin{figure}[htbp]
    \centering
    \begin{subfigure}[b]{0.32\textwidth}
        \centering
        \includegraphics[width=\textwidth, height=6cm, keepaspectratio]{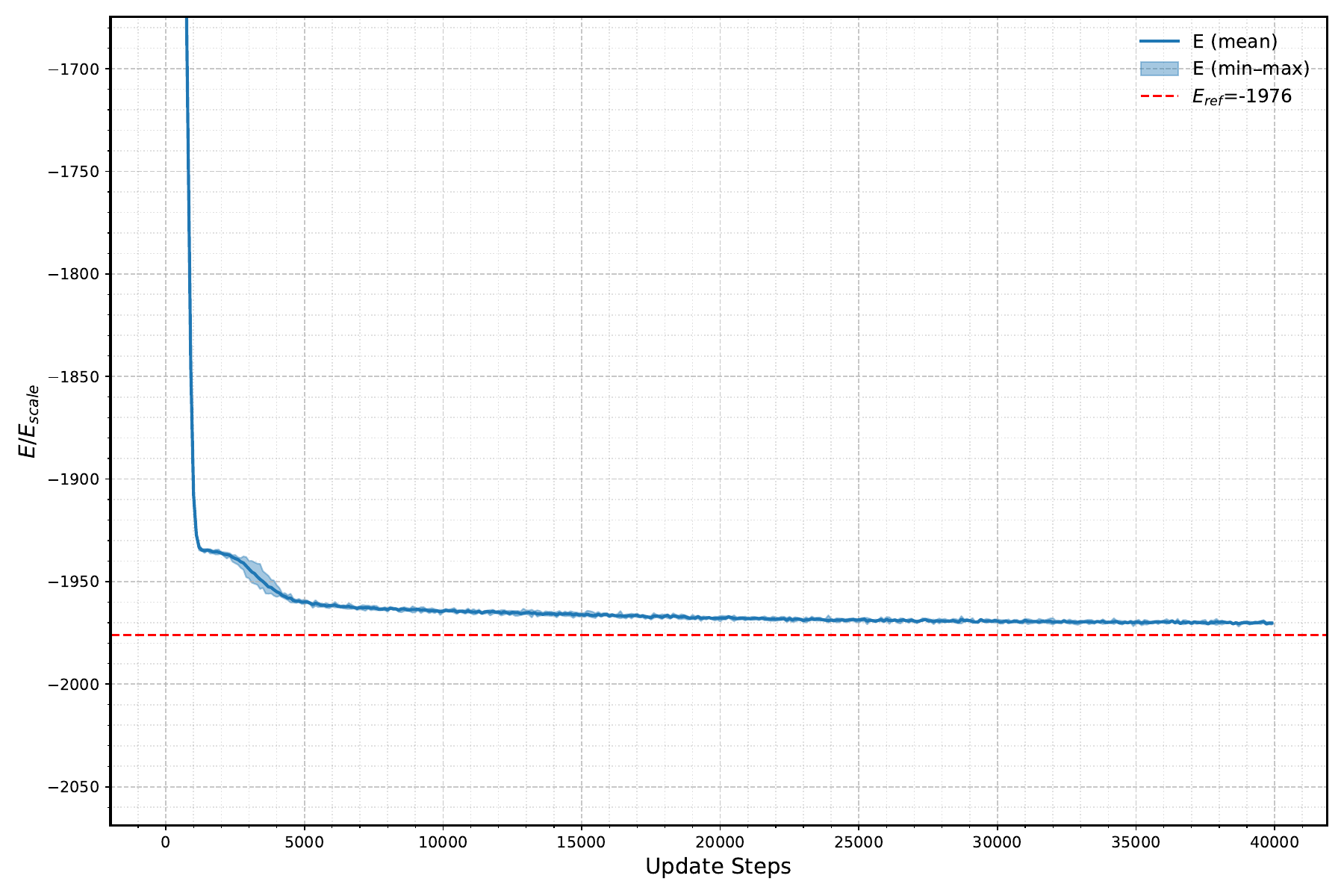}
        \caption{\centering
            \texttt{GELU-MALA}.\\
            $E/E_{scale} = -1972.02 \pm 3.65$.\\
            r-range (99.9\% CI): $[-0.23,\,0.23]$.
        }
        \label{fig:10p-gelu-mala}
    \end{subfigure}
    \hfill
    \begin{subfigure}[b]{0.32\textwidth}
        \centering
        \includegraphics[width=\textwidth, height=6cm, keepaspectratio]{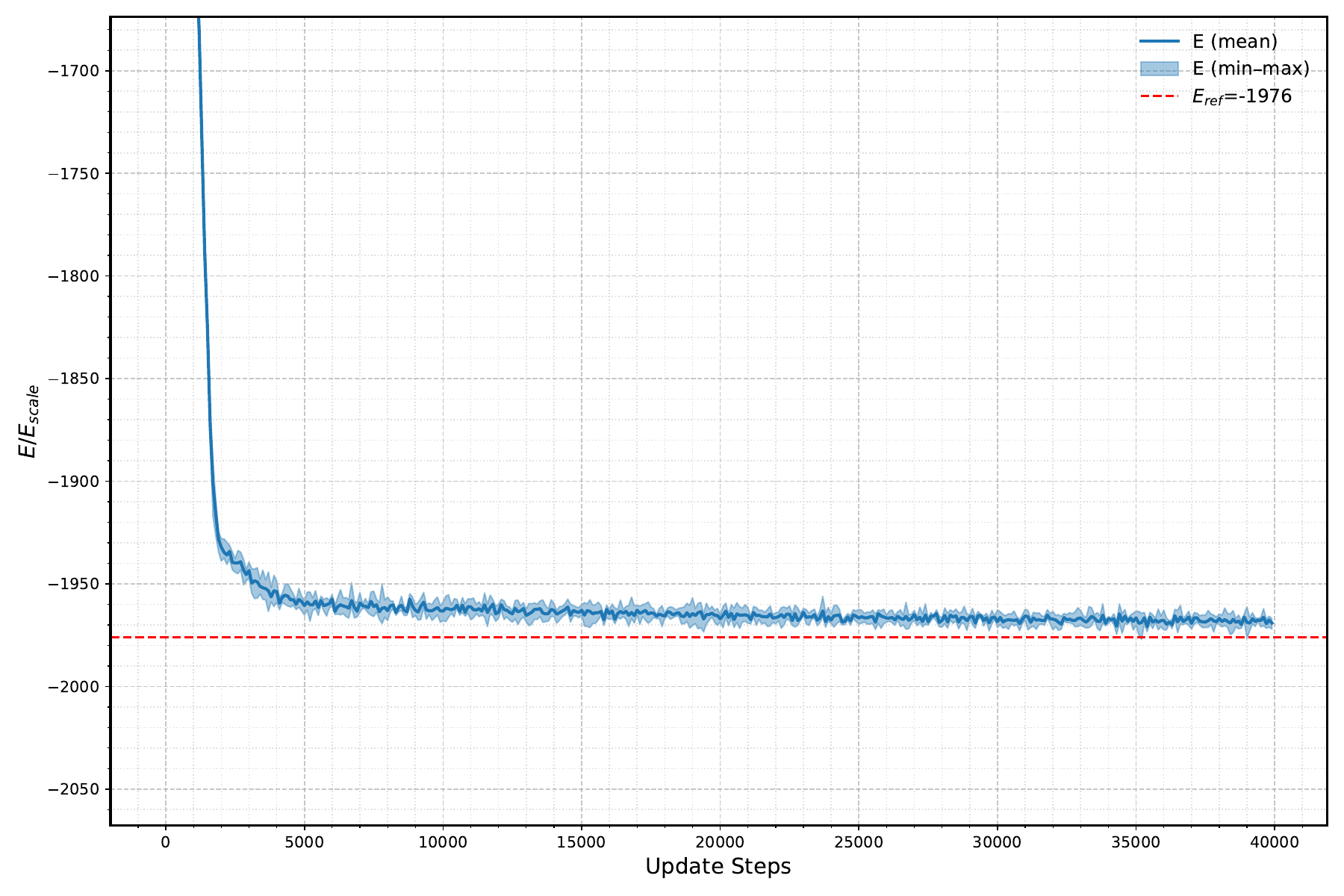}
        \caption{\centering
            \texttt{GELU-ARW}.\\
            $E/E_{scale} = -1979.01 \pm 13.41$.\\
            r-range (99.9\% CI): $[-0.23,\,0.23]$.
        }
        \label{fig:10p-gelu-arw}
    \end{subfigure}
    \hfill
    \begin{subfigure}[b]{0.32\textwidth}
        \centering
        \includegraphics[width=\textwidth, height=6cm, keepaspectratio]{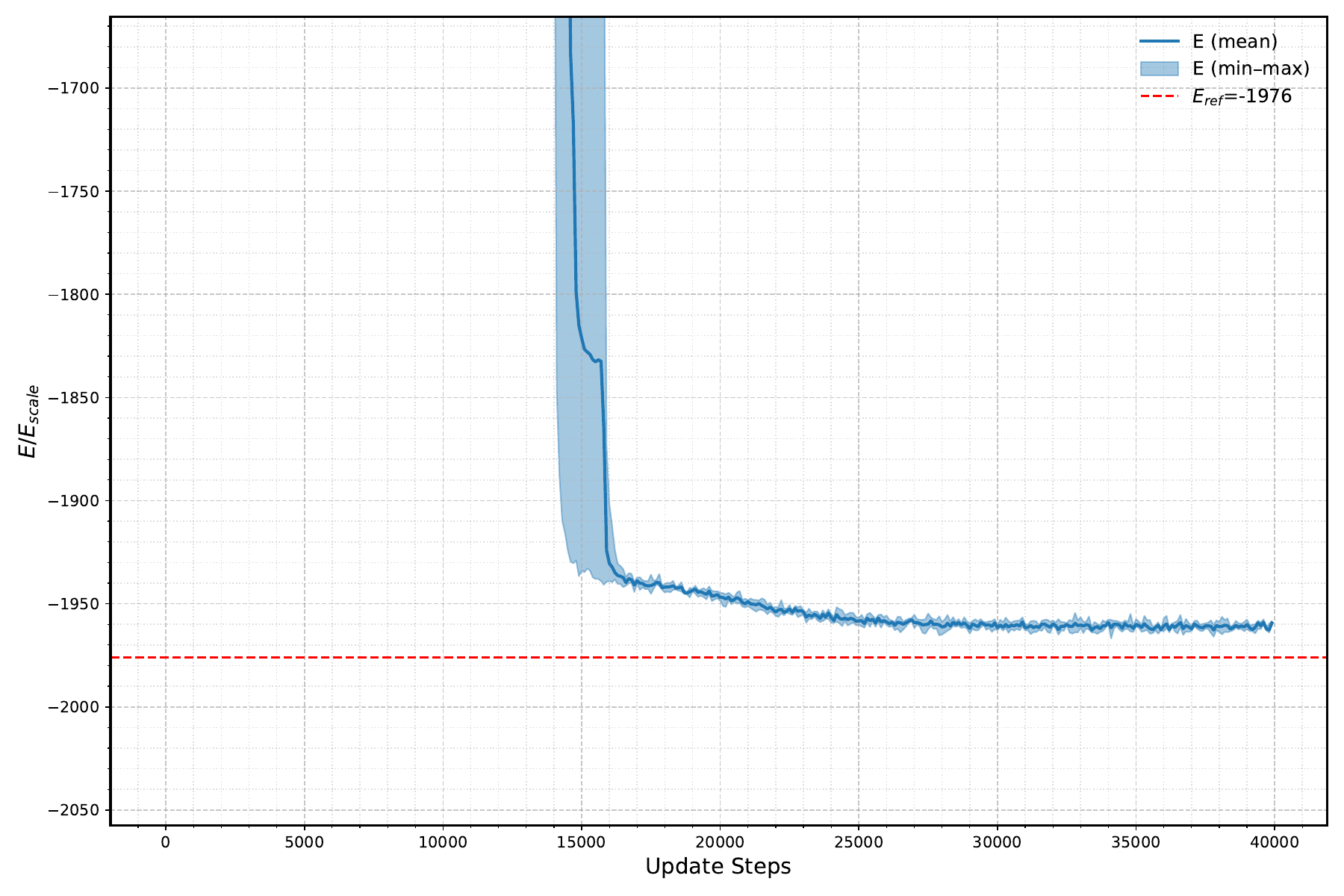}
        \caption{\centering
            \texttt{tanh-ARW}.\\
            $E/E_{scale} = -1956.61 \pm 10.46$.\\
            r-range (99.9\% CI): $[-0.23,\,0.23]$.
        }
        \label{fig:10p-tanh-arw}
    \end{subfigure}

    \caption{10-particle harmonic-potential system with two-body interactions.}
    \label{fig:10particles-all-methods}
\end{figure}

%
%
%
\clearpage
\subsection{System B: 3-20 particles}
\label{appendix:systemB}

\begin{figure}[htbp]
    \centering
    \begin{subfigure}[b]{0.32\textwidth}
        \centering
        \includegraphics[width=\textwidth, height=6cm, keepaspectratio]{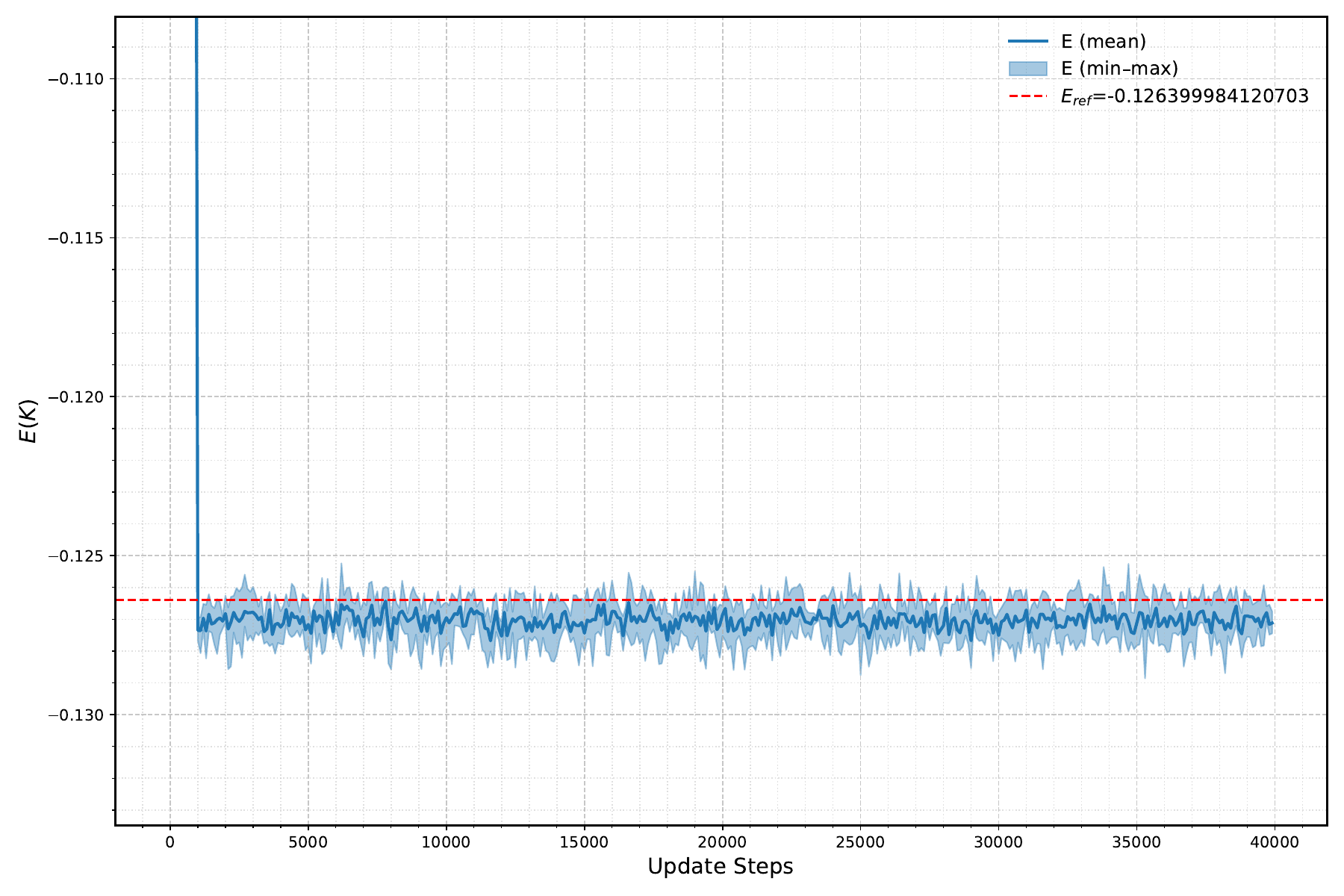}
        \caption{\centering
            \texttt{GELU-MALA}.\\
            E(K) = -0.126 ± 0.003.\\
            r-range (99.9\% CI): $[-60.00,\,59.59]$.
        }
        \label{fig:3p-tb-gelu-mala}
    \end{subfigure}
    \hfill
    \begin{subfigure}[b]{0.32\textwidth}
        \centering
        \includegraphics[width=\textwidth, height=6cm, keepaspectratio]{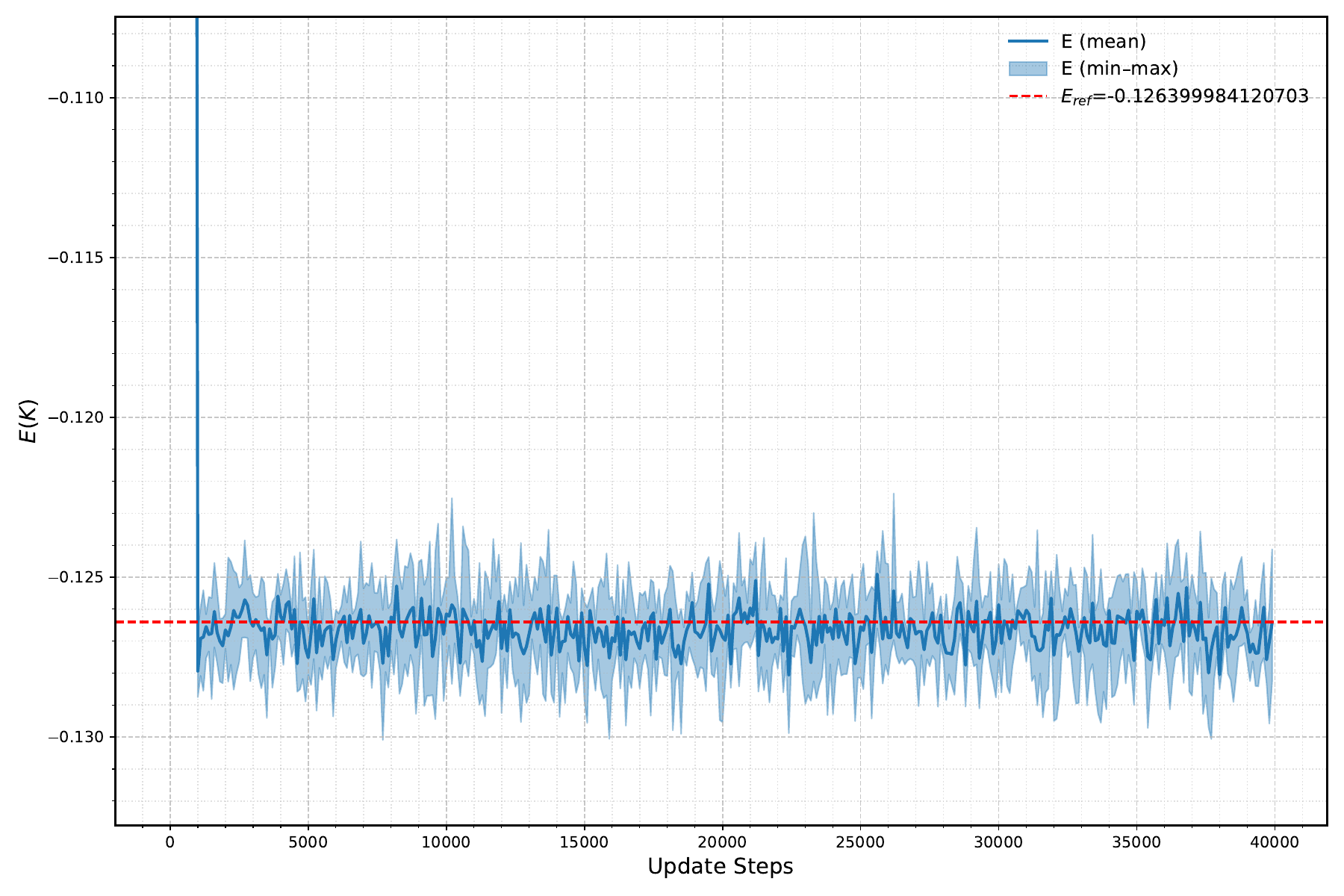}
        \caption{\centering
            \texttt{GELU-ARW}.\\
            E(K) = -0.124 $\pm$ 0.006.\\
            r-range (99.9\% CI): $[-59.37,\,59.76]$.
        }
        \label{fig:3p-tb-gelu-arw}
    \end{subfigure}
    \hfill
    \begin{subfigure}[b]{0.32\textwidth}
        \centering
        \includegraphics[width=\textwidth, height=6cm, keepaspectratio]{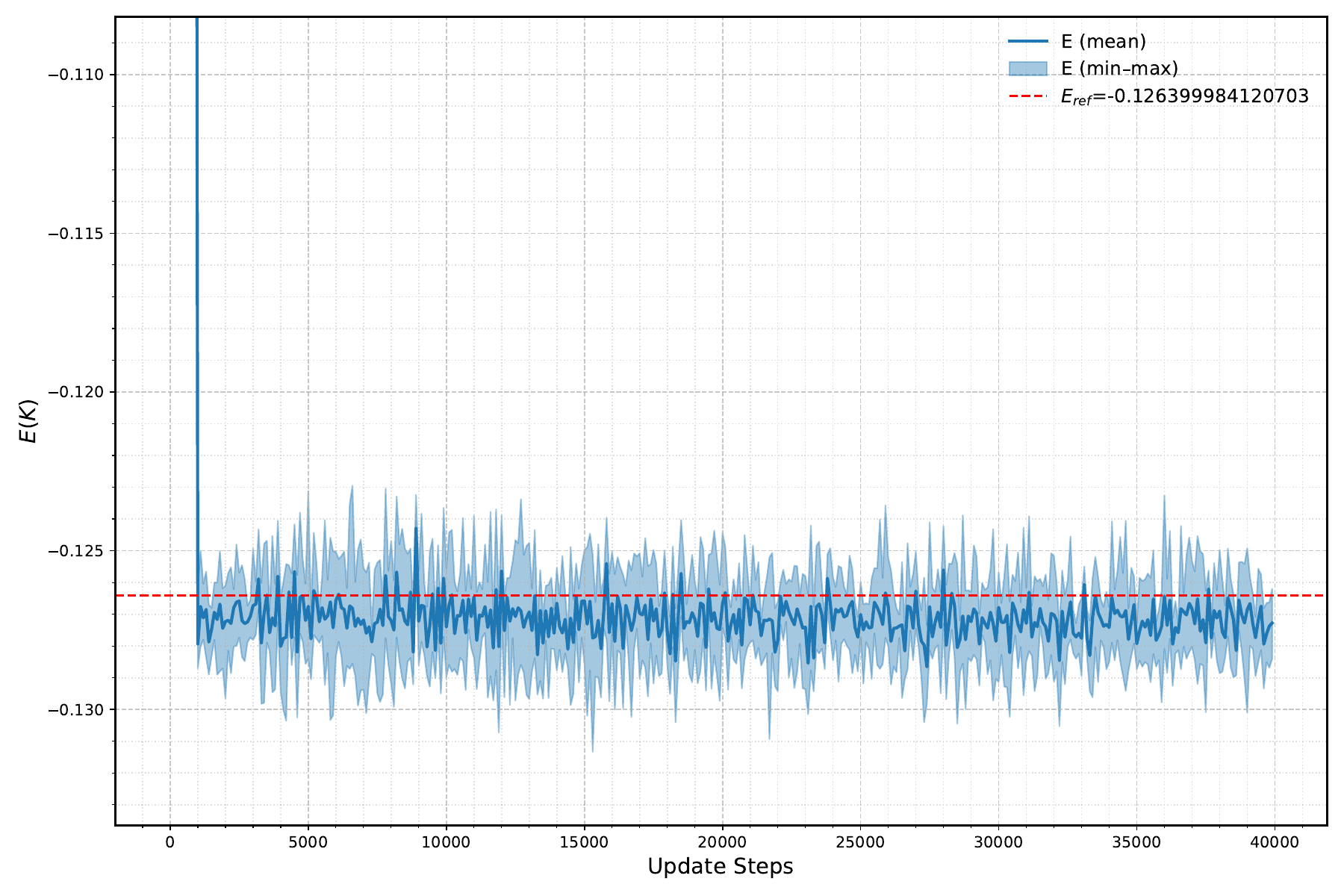}
        \caption{\centering
            \texttt{GELU-RW}.\\
            E(K) = -0.13 $\pm$ 0.01.\\
            r-range (99.9\% CI): $[-58.71,\,59.05]$.
        }
        \label{fig:3p-tb-gelu-rw}
    \end{subfigure}

    \caption{3-particle system with two-body and three-body interactions.}
    \label{fig:3particles-tb-inte-methods}
\end{figure}

\begin{figure}[htbp]
    \centering
    \begin{subfigure}[b]{0.32\textwidth}
        \centering
        \includegraphics[width=\textwidth, height=6cm, keepaspectratio]{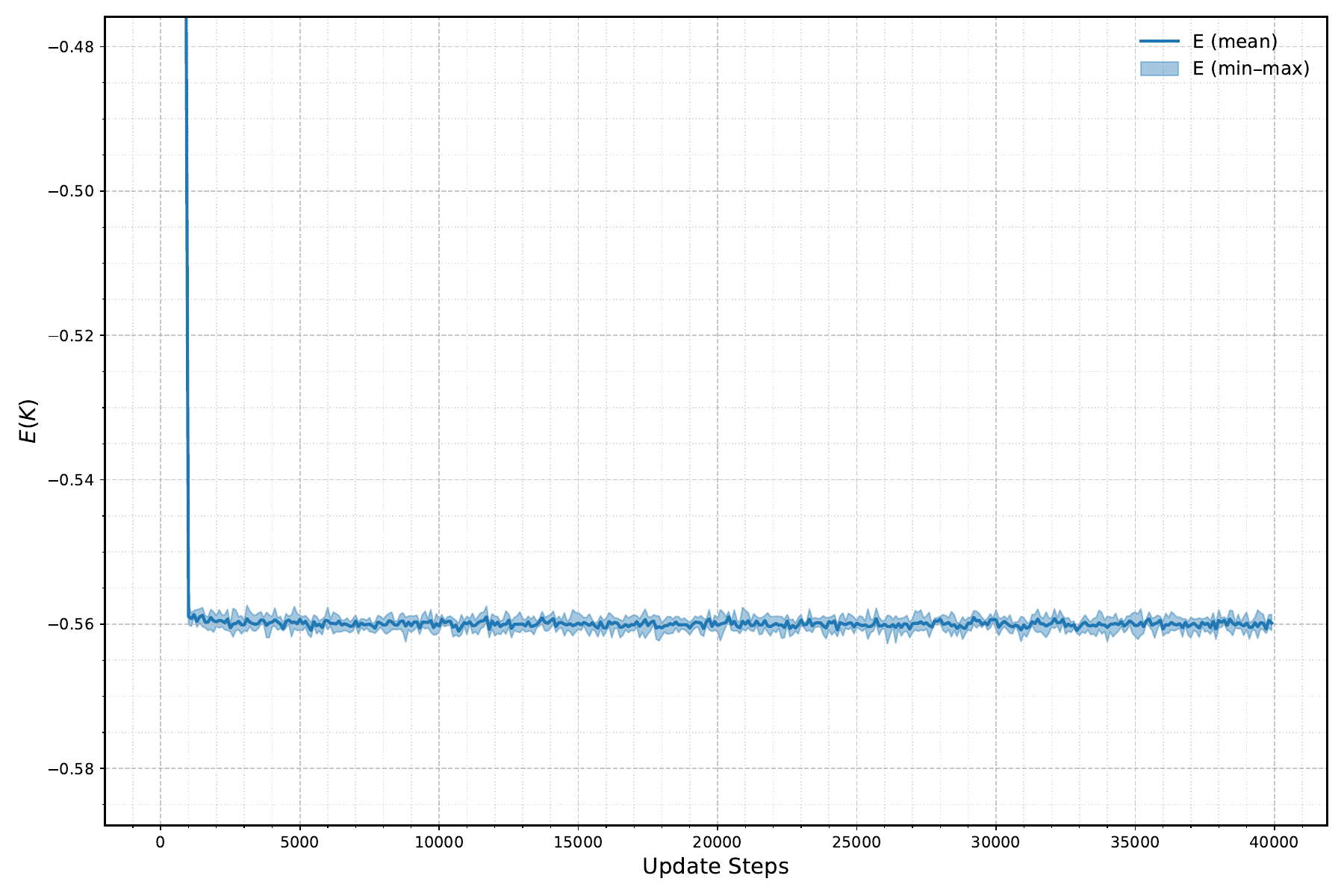}
        \caption{\centering
            \texttt{GELU-MALA}.\\
            E(K) = -0.561 ± 0.003.\\
            r-range (99.9\% CI): $[-34.16,\,34.22]$.
        }
        \label{fig:4p-tb-gelu-mala}
    \end{subfigure}
    \hfill
    \begin{subfigure}[b]{0.32\textwidth}
        \centering
        \includegraphics[width=\textwidth, height=6cm, keepaspectratio]{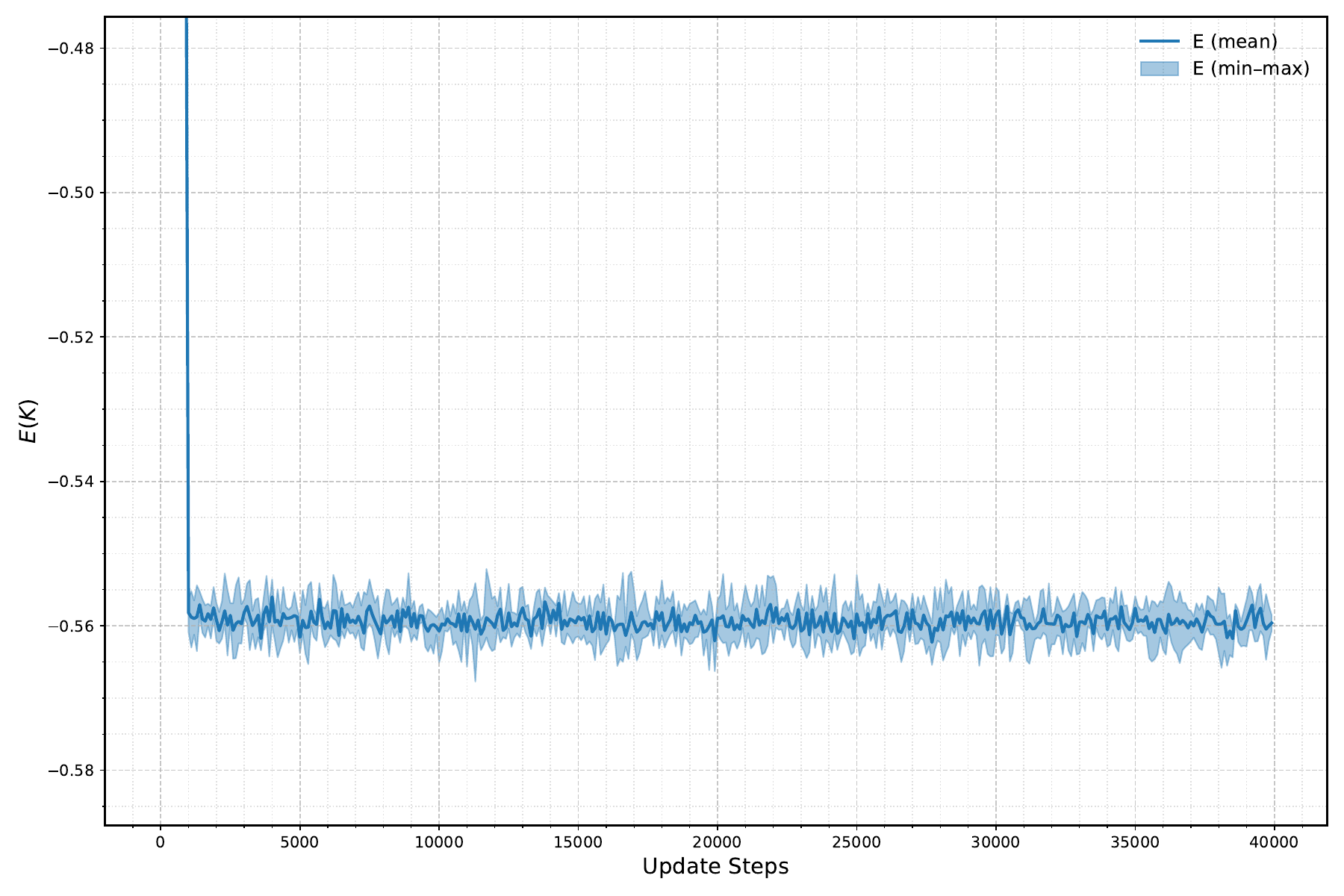}
        \caption{\centering
            \texttt{GELU-ARW}.\\
            E(K) = -0.56 $\pm$ 0.01.\\
            r-range (99.9\% CI): $[-34.60,\,34.31]$.
        }
        \label{fig:4p-tb-gelu-arw}
    \end{subfigure}
    \hfill
    \begin{subfigure}[b]{0.32\textwidth}
        \centering
        \includegraphics[width=\textwidth, height=6cm, keepaspectratio]{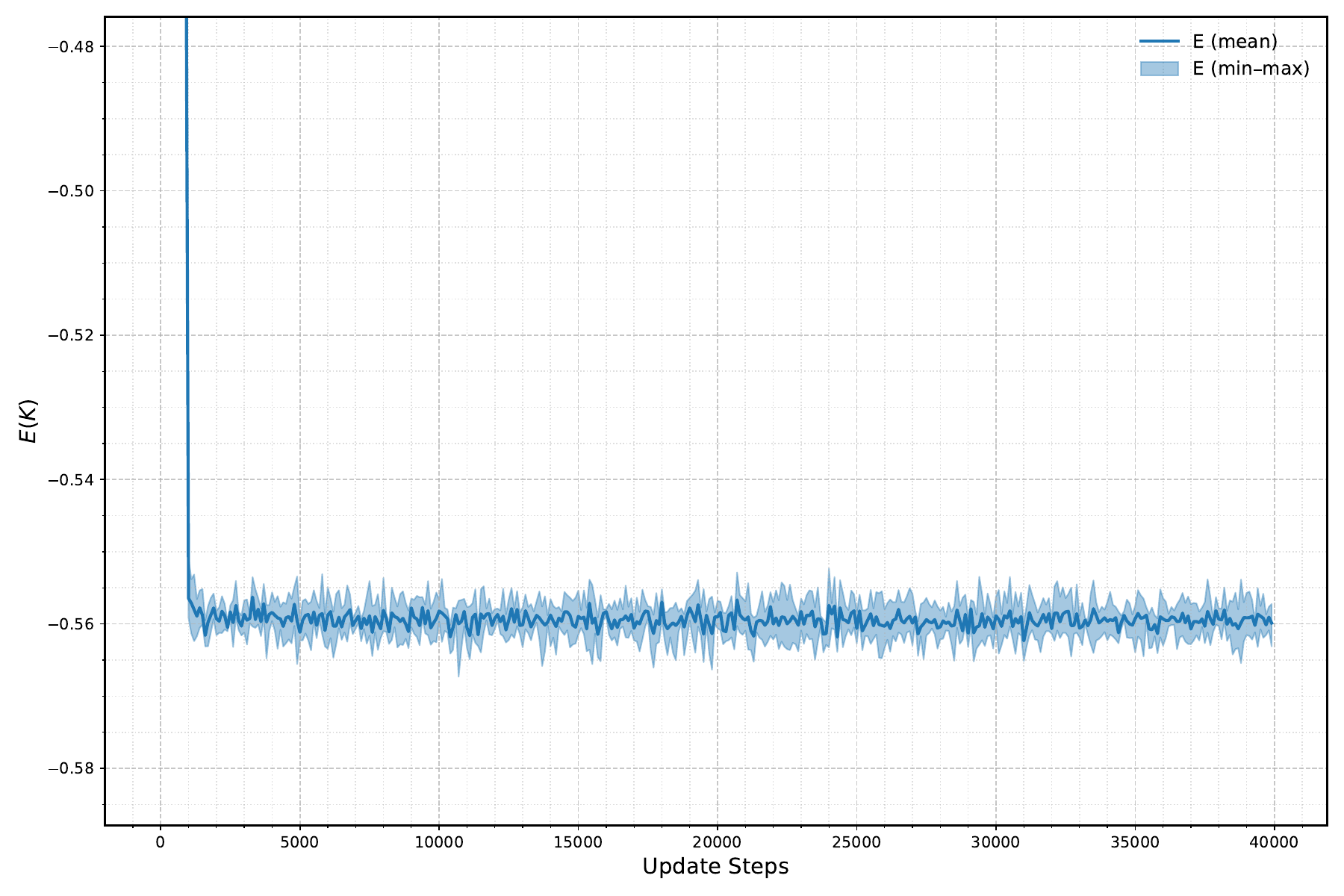}
        \caption{\centering
            \texttt{GELU-RW}.\\
            E(K) = -0.55 $\pm$ 0.01.\\
            r-range (99.9\% CI): $[-34.44,\,33.90]$.
        }
        \label{fig:4p-tb-gelu-rw}
    \end{subfigure}

    \caption{4-particle system with two-body and three-body interactions.}
    \label{fig:4particles-tb-inte-methods}
\end{figure}

\begin{figure}[htbp]
    \centering
    \begin{subfigure}[b]{0.32\textwidth}
        \centering
        \includegraphics[width=\textwidth, height=6cm, keepaspectratio]{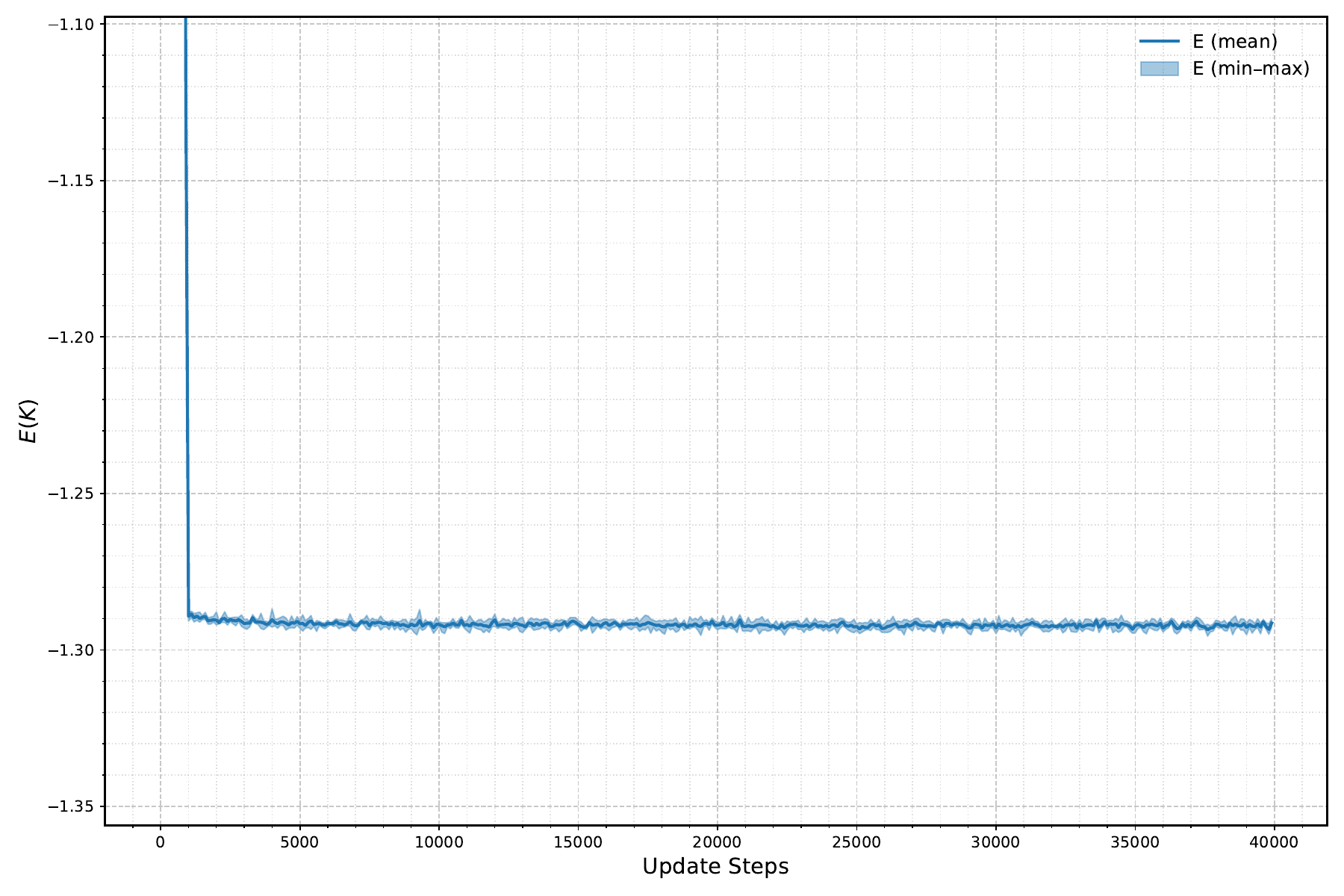}
        \caption{\centering
            \texttt{GELU-MALA}.\\
            E(K) = -1.29 ± 0.02.\\
            r-range (99.9\% CI): $[-29.33,\,29.27]$.
        }
        \label{fig:5p-tb-gelu-mala}
    \end{subfigure}
    \hfill
    \begin{subfigure}[b]{0.32\textwidth}
        \centering
        \includegraphics[width=\textwidth, height=6cm, keepaspectratio]{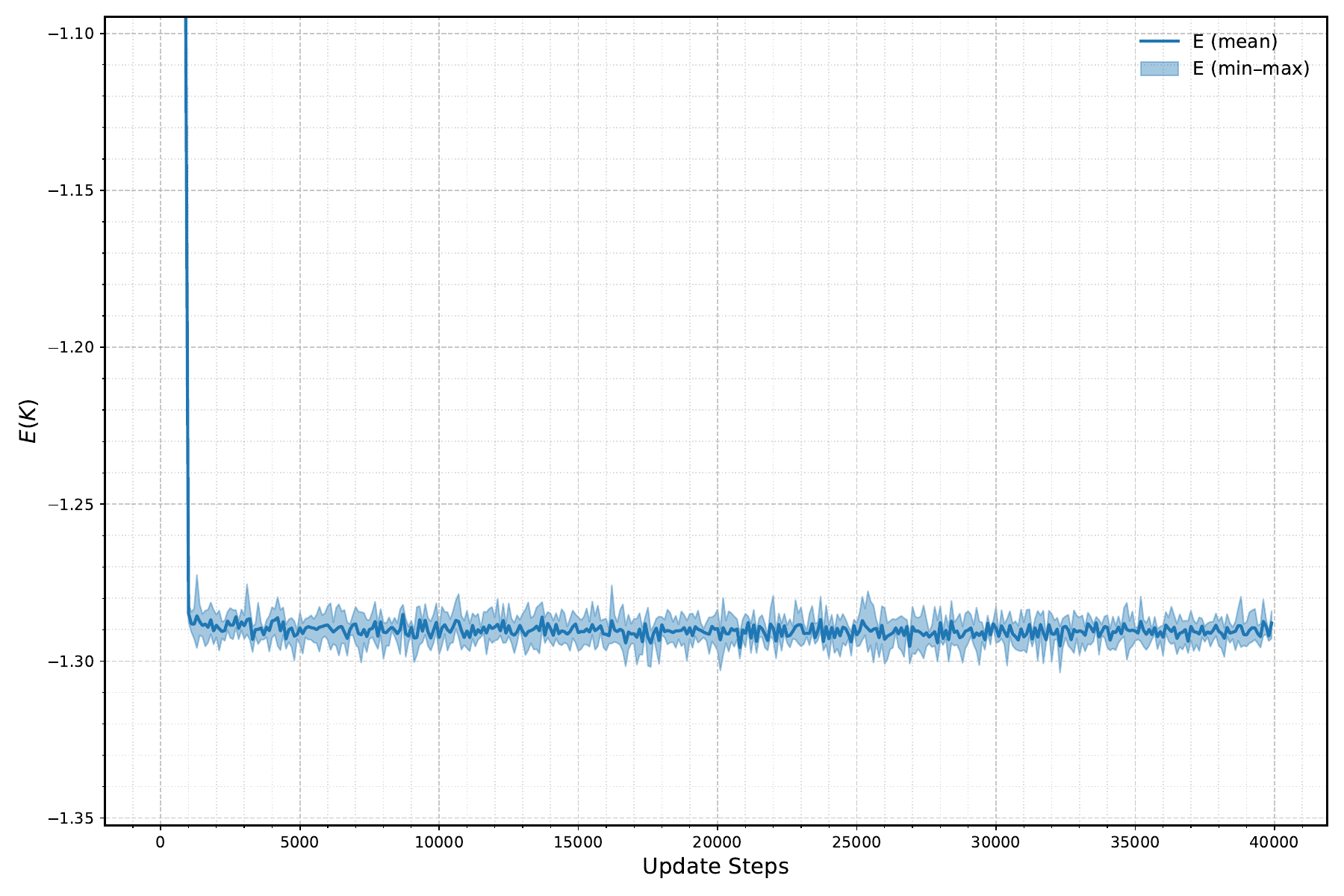}
        \caption{\centering
            \texttt{GELU-ARW}.\\
            E(K) = -1.3 $\pm$ 0.04.\\
            r-range (99.9\% CI): $[-29.28,\,29.15]$.
        }
        \label{fig:5p-tb-gelu-arw}
    \end{subfigure}
    \hfill
    \begin{subfigure}[b]{0.32\textwidth}
        \centering
        \includegraphics[width=\textwidth, height=6cm, keepaspectratio]{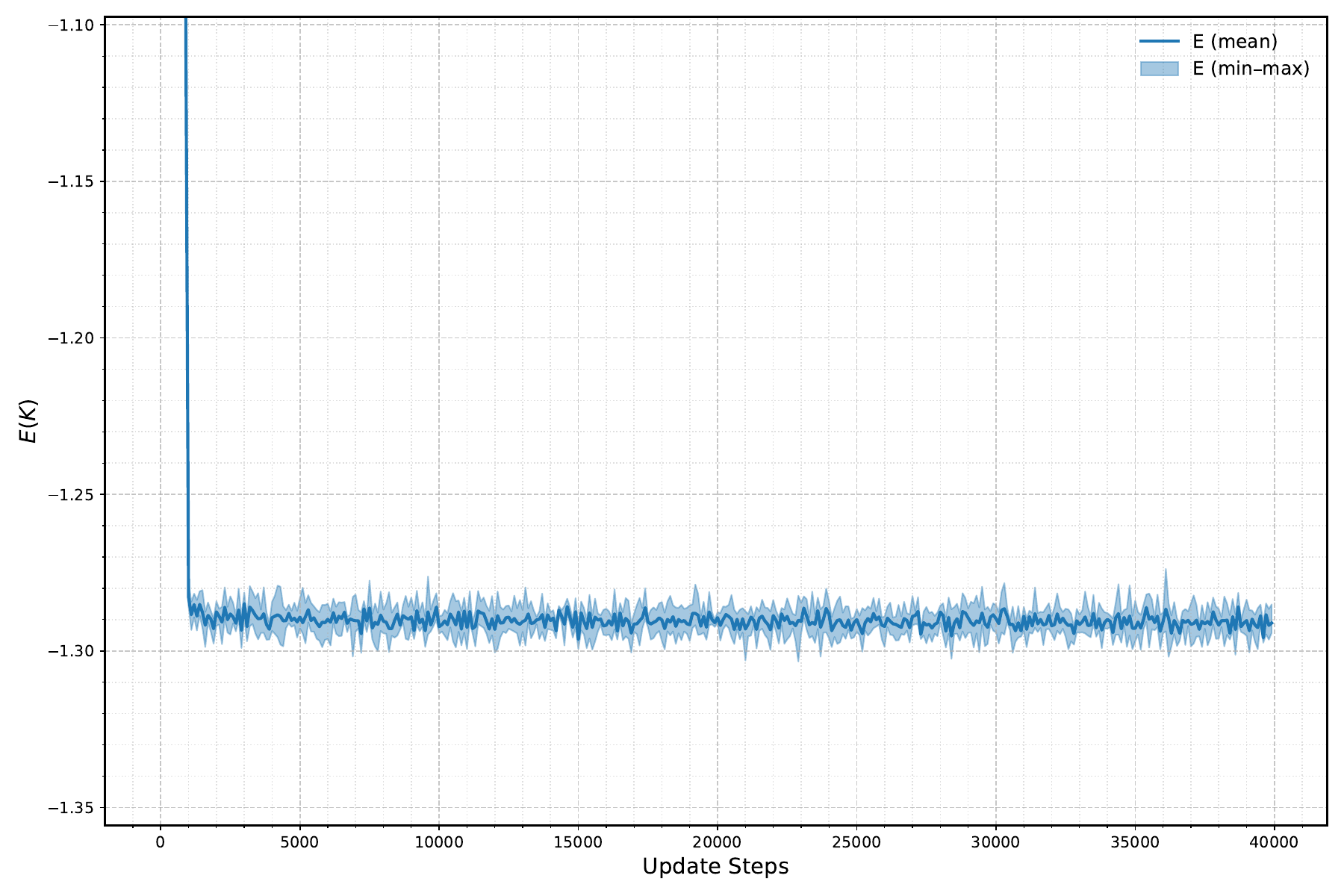}
        \caption{\centering
            \texttt{GELU-RW}.\\
            E(K) = -1.28 $\pm$ 0.03.\\
            r-range (99.9\% CI): $[-29.15,\,29.10]$.
        }
        \label{fig:5p-tb-gelu-rw}
    \end{subfigure}

    \caption{5-particle system with two-body and three-body interactions.}
    \label{fig:5particles-tb-inte-methods}
\end{figure}

\begin{figure}[htbp]
    \centering
    \begin{subfigure}[b]{0.32\textwidth}
        \centering
        \includegraphics[width=\textwidth, height=6cm, keepaspectratio]{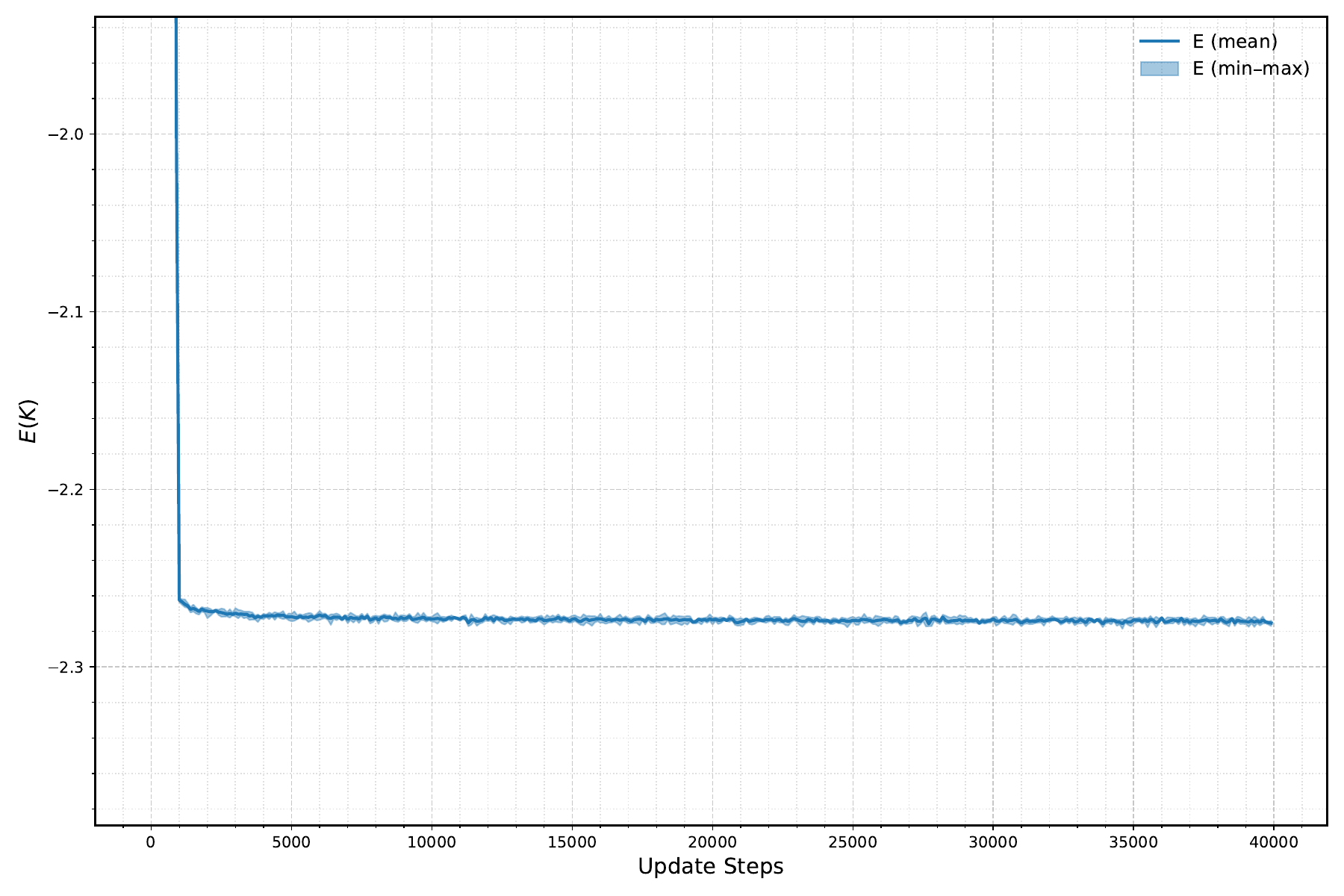}
        \caption{\centering
            \texttt{GELU-MALA}.\\
            E(K) = -2.28 $\pm$ 0.003.\\
            r-range (99.9\% CI): $[-27.20,\,27.18]$.
        }
        \label{fig:6p-tb-gelu-mala}
    \end{subfigure}
    \hfill
    \begin{subfigure}[b]{0.32\textwidth}
        \centering
        \includegraphics[width=\textwidth, height=6cm, keepaspectratio]{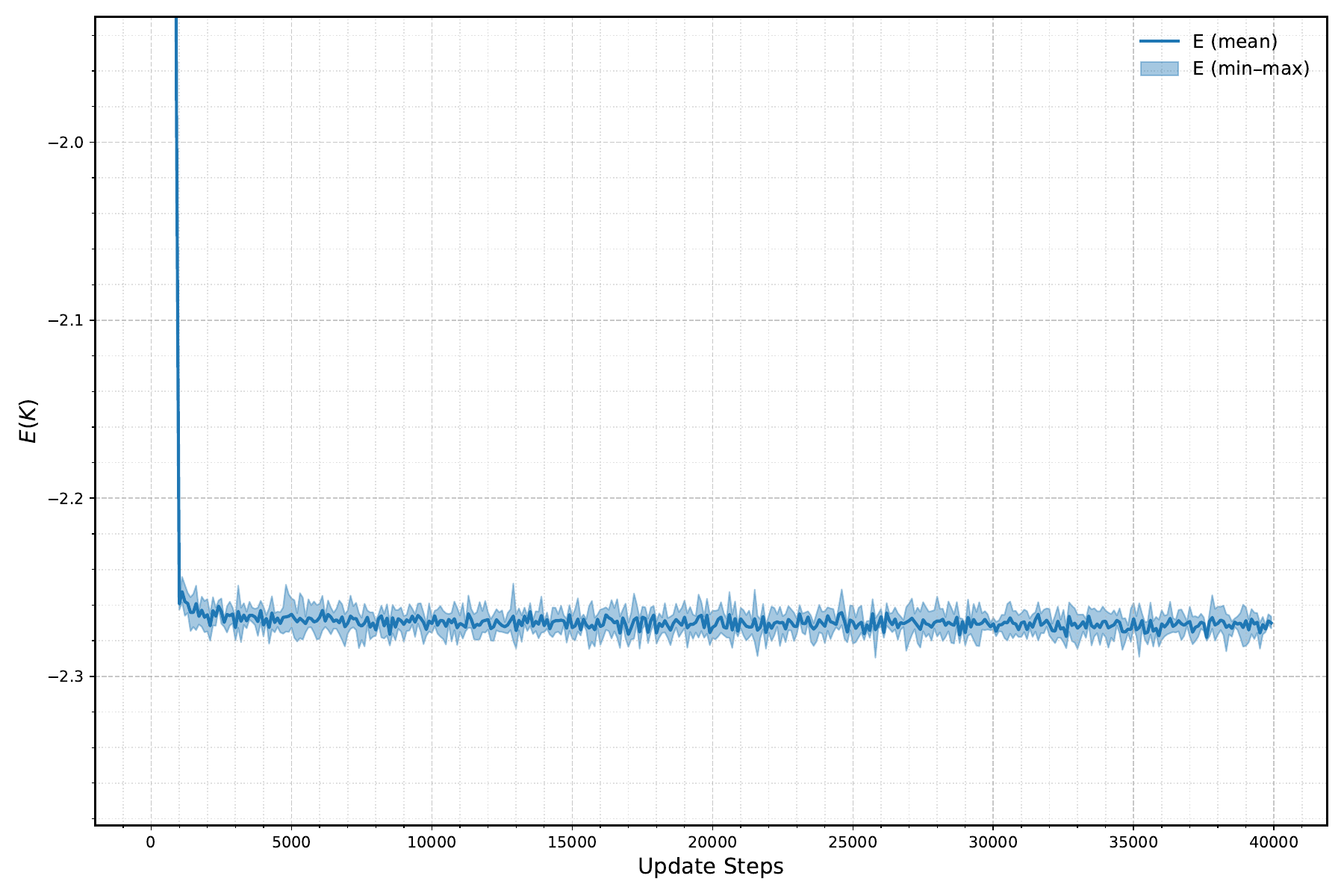}
        \caption{\centering
            \texttt{GELU-ARW}.\\
            E(K) = -2.23 $\pm$ 0.02.\\
            r-range (99.9\% CI): $[-27.26,\,27.20]$.
        }
        \label{fig:6p-tb-gelu-arw}
    \end{subfigure}
    \hfill
    \begin{subfigure}[b]{0.32\textwidth}
        \centering
        \includegraphics[width=\textwidth, height=6cm, keepaspectratio]{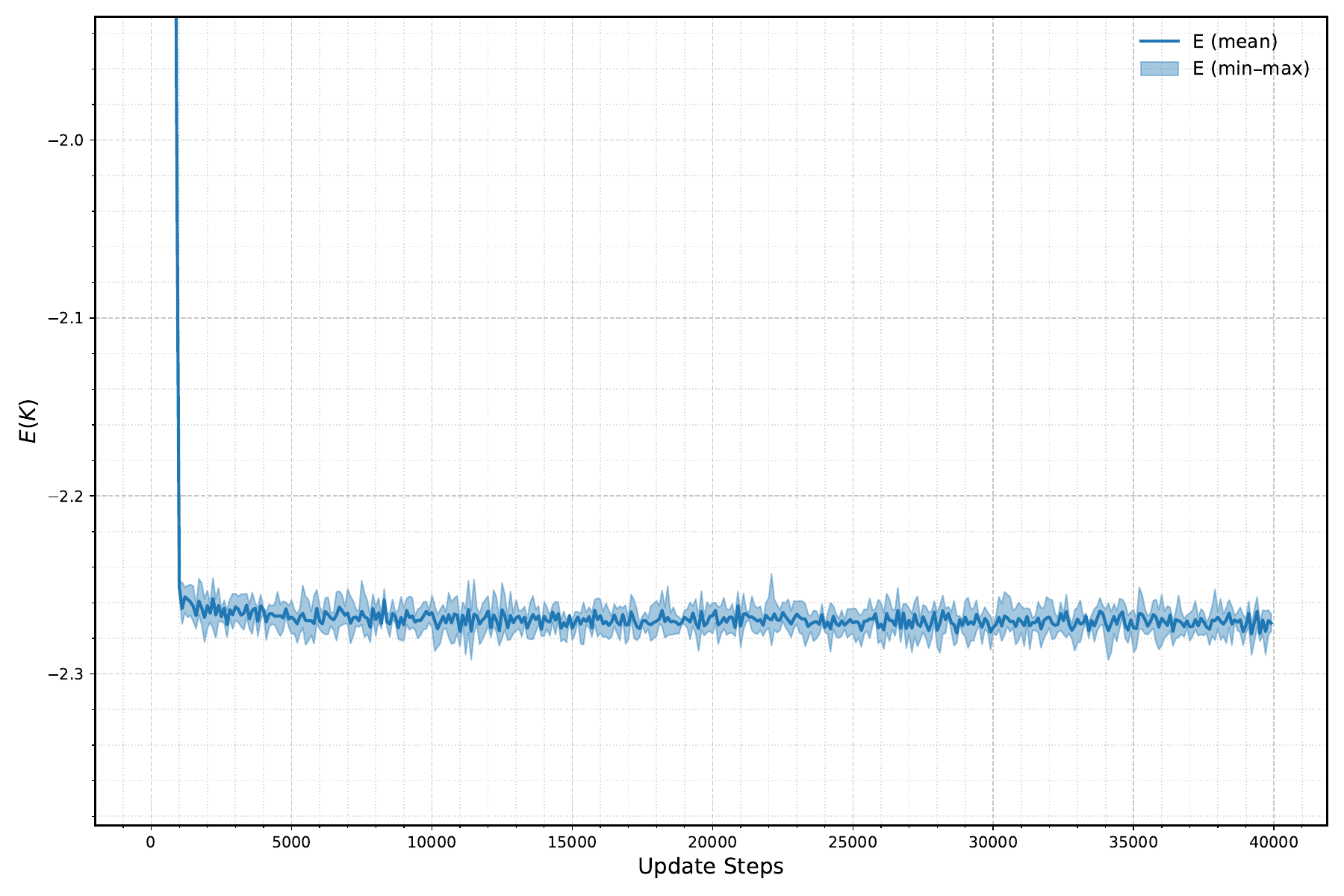}
        \caption{\centering
            \texttt{GELU-RW}.\\
            E(K) = -2.25 $\pm$ 0.05.\\
            r-range (99.9\% CI): $[-27.36,\,27.08]$.
        }
        \label{fig:6p-tb-gelu-rw}
    \end{subfigure}

    \caption{6-particle system with two-body and three-body interactions.}
    \label{fig:6particles-tb-inte-methods}
\end{figure}

\begin{figure}[htbp]
    \centering
    \begin{subfigure}[b]{0.32\textwidth}
        \centering
        \includegraphics[width=\textwidth, height=6cm, keepaspectratio]{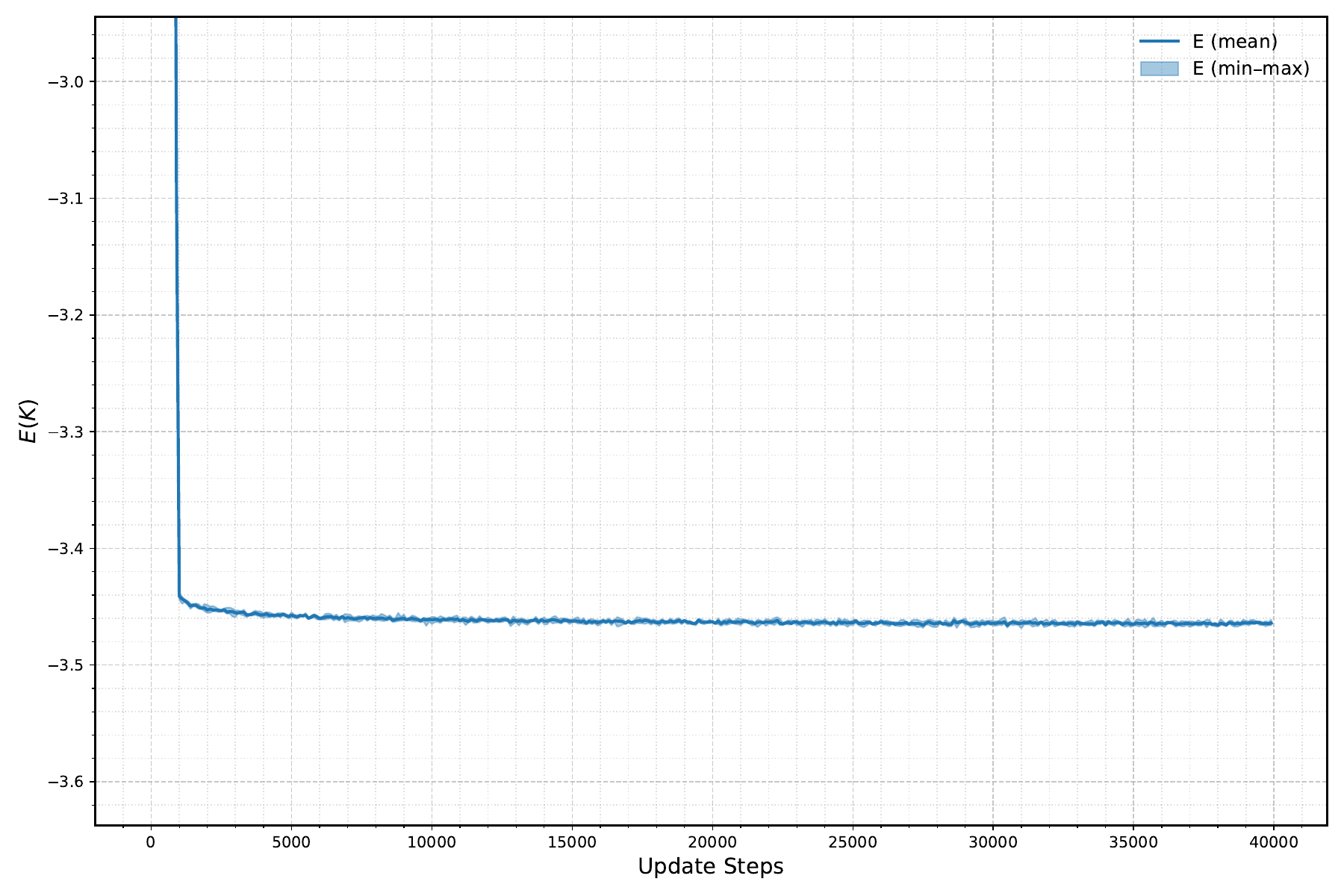}
        \caption{\centering
            \texttt{GELU-MALA}.\\
            E(K) = -3.47 $\pm$ 0.01.\\
            r-range (99.9\% CI): $[-26.11,\,26.05]$.
        }
        \label{fig:7p-tb-gelu-mala}
    \end{subfigure}
    \hfill
    \begin{subfigure}[b]{0.32\textwidth}
        \centering
        \includegraphics[width=\textwidth, height=6cm, keepaspectratio]{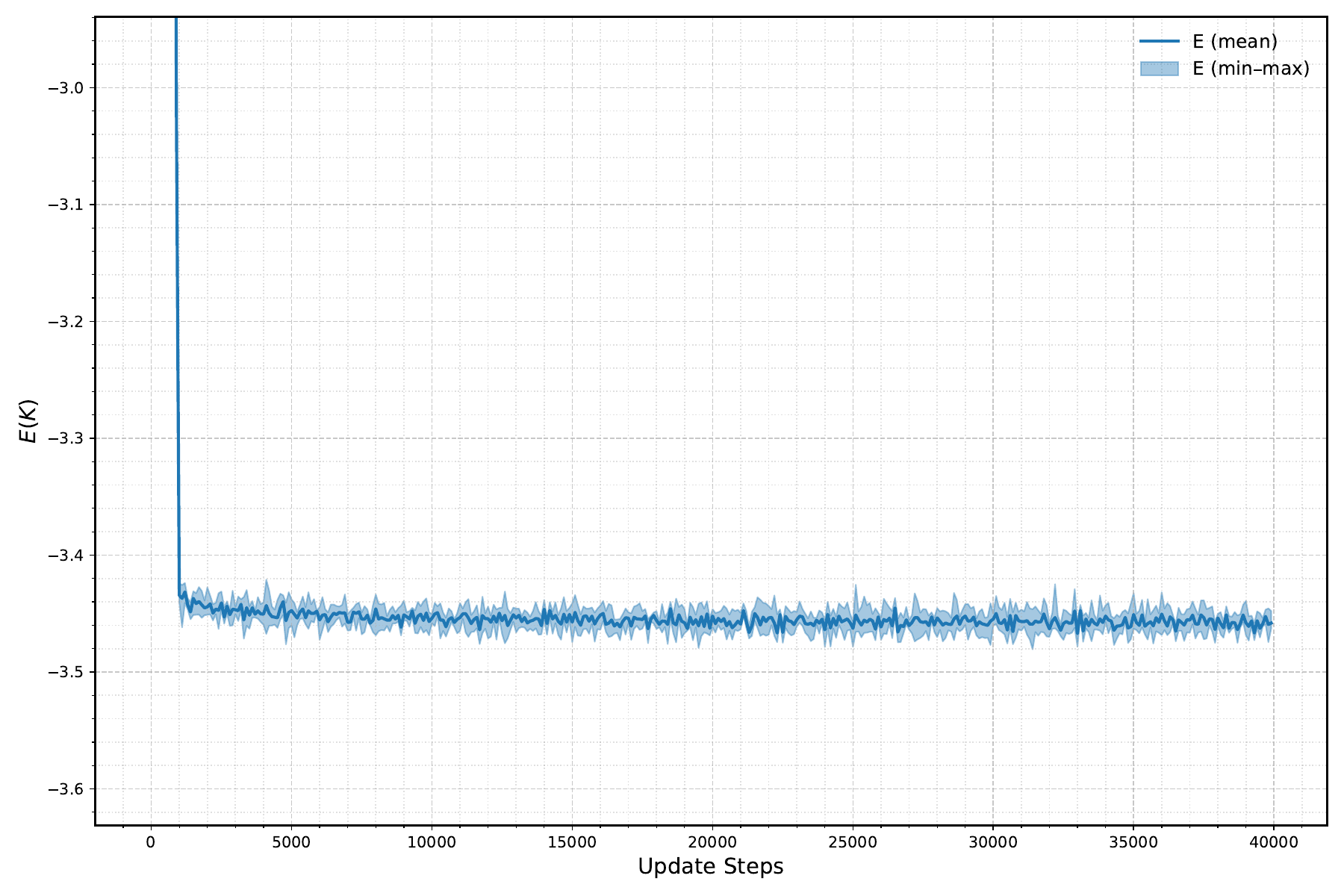}
        \caption{\centering
            \texttt{GELU-ARW}.\\
            E(K) = -3.5 $\pm$ 0.08.\\
            r-range (99.9\% CI): $[-26.19,\,26.25]$.
        }
        \label{fig:7p-tb-gelu-arw}
    \end{subfigure}
    \hfill
    \begin{subfigure}[b]{0.32\textwidth}
        \centering
        \includegraphics[width=\textwidth, height=6cm, keepaspectratio]{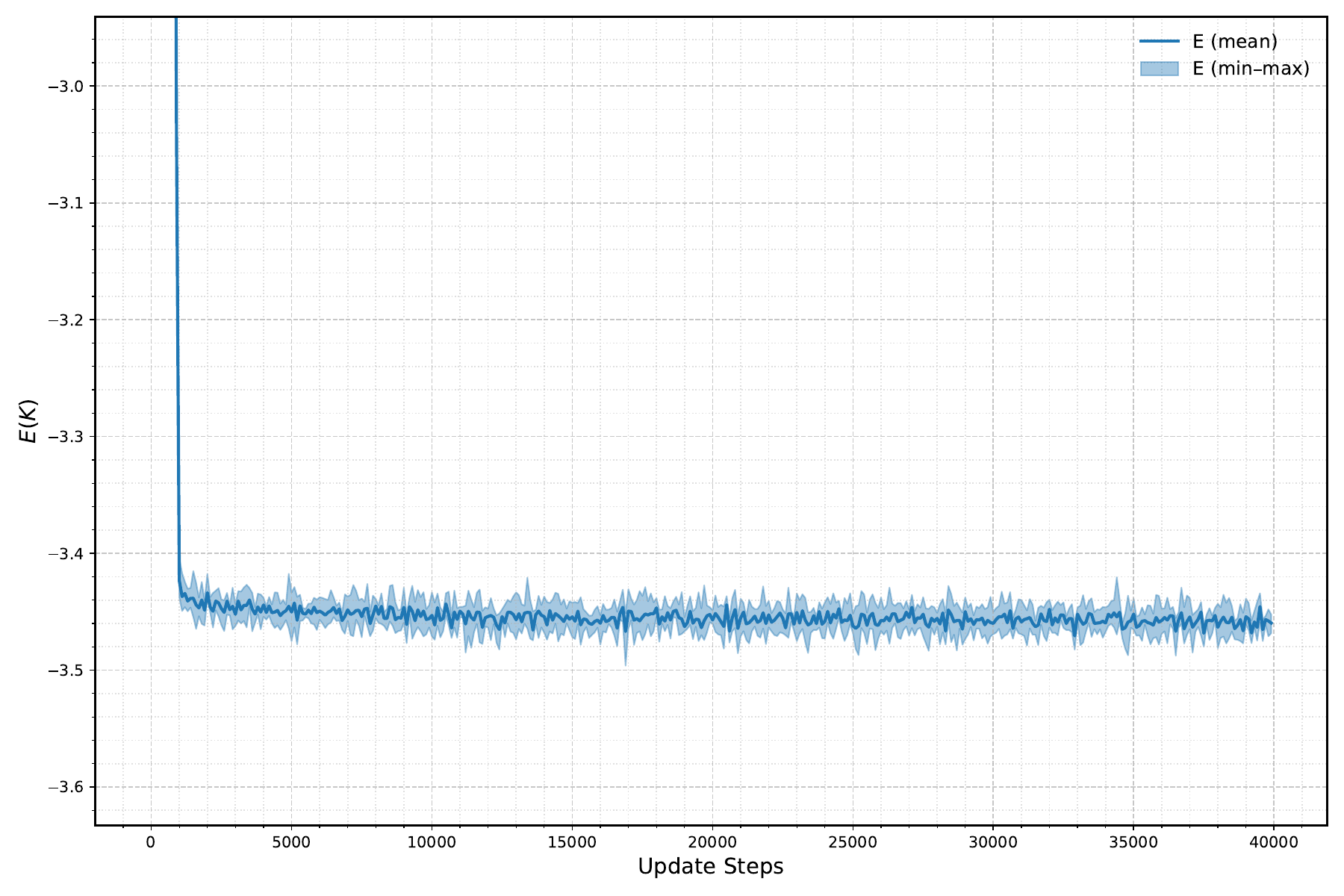}
        \caption{\centering
            \texttt{GELU-RW}.\\
            E(K) = -3.55 $\pm$ 0.06.\\
            r-range (99.9\% CI): $[-26.19,\,26.17]$.
        }
        \label{fig:7p-tb-gelu-rw}
    \end{subfigure}

    \caption{7-particle system with two-body and three-body interactions.}
    \label{fig:7particles-tb-inte-methods}
\end{figure}

\begin{figure}[htbp]
    \centering
    \begin{subfigure}[b]{0.32\textwidth}
        \centering
        \includegraphics[width=\textwidth, height=6cm, keepaspectratio]{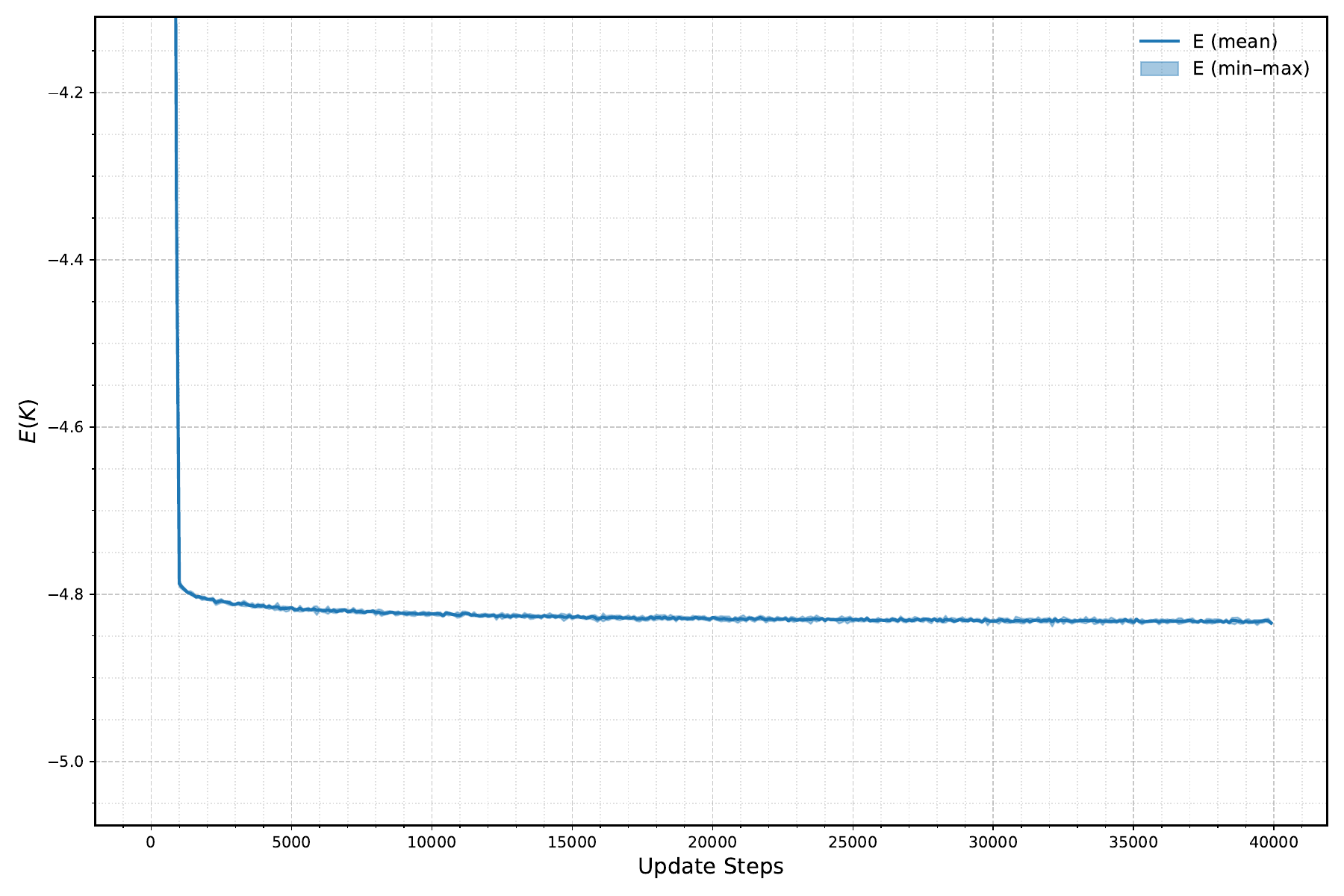}
        \caption{\centering
            \texttt{GELU-MALA}.\\
            E(K) = -4.82 $\pm$ 0.01.\\
            r-range (99.9\% CI): $[-25.70,\,25.68]$.
        }
        \label{fig:8p-tb-gelu-mala}
    \end{subfigure}
    \hfill
    \begin{subfigure}[b]{0.32\textwidth}
        \centering
        \includegraphics[width=\textwidth, height=6cm, keepaspectratio]{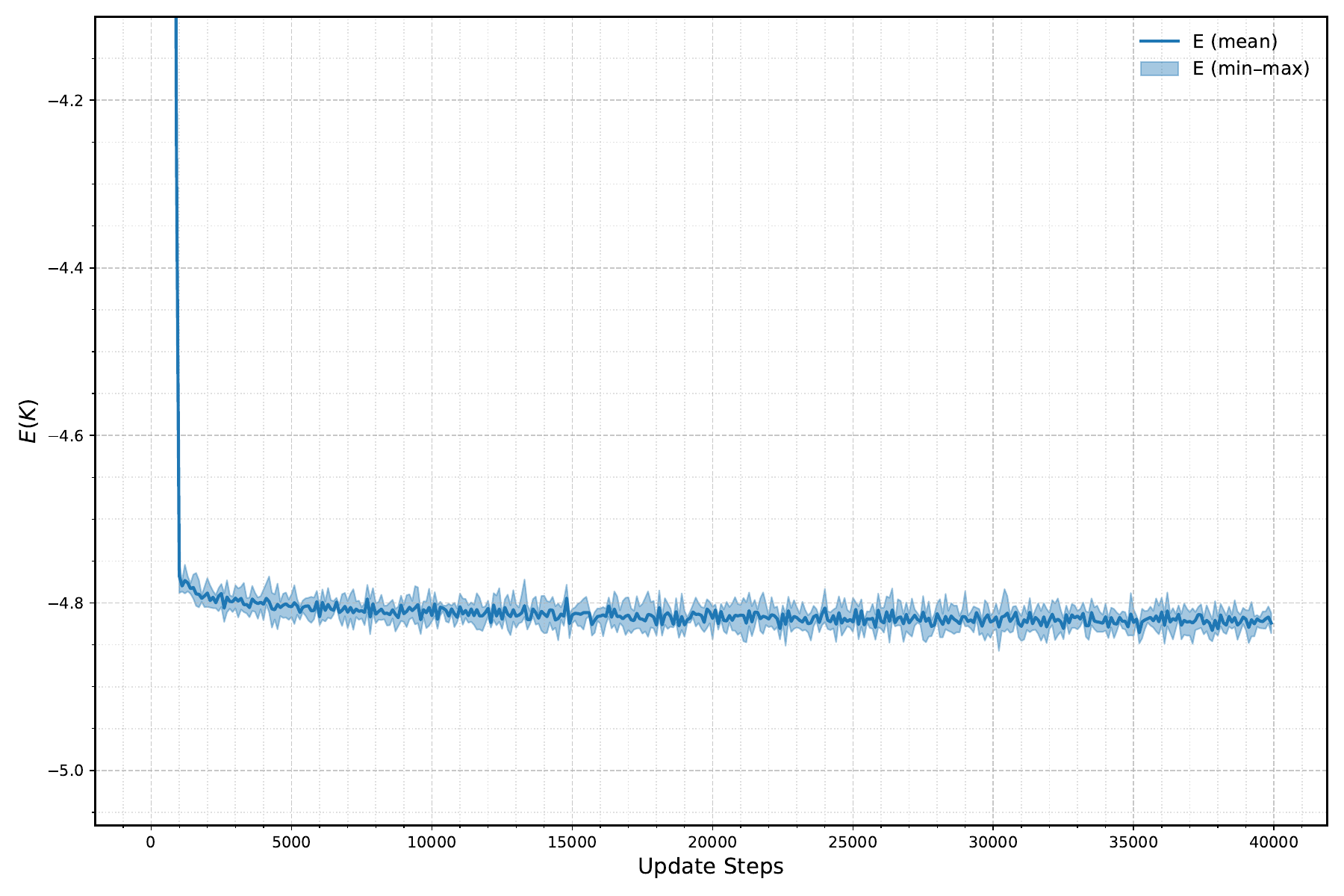}
        \caption{\centering
            \texttt{GELU-ARW}.\\
            E(K) = -4.9 $\pm$ 0.2.\\
            r-range (99.9\% CI): $[-25.46,\,25.48]$.
        }
        \label{fig:8p-tb-gelu-arw}
    \end{subfigure}
    \hfill
    \begin{subfigure}[b]{0.32\textwidth}
        \centering
        \includegraphics[width=\textwidth, height=6cm, keepaspectratio]{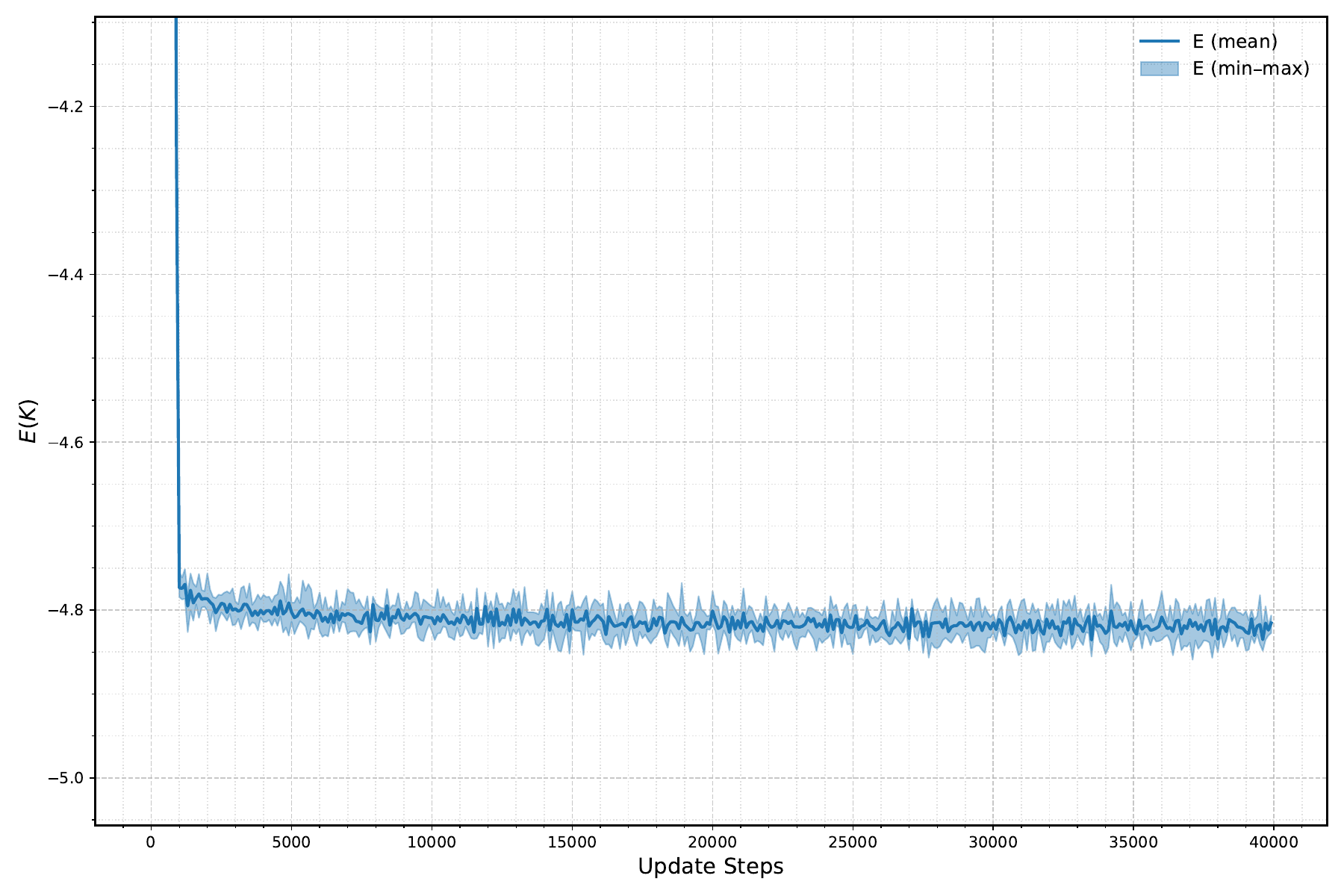}
        \caption{\centering
            \texttt{GELU-RW}.\\
            E(K) = -4.83 $\pm$ 0.07.\\
            r-range (99.9\% CI): $[-25.71,\,25.63]$.
        }
        \label{fig:8p-tb-gelu-rw}
    \end{subfigure}

    \caption{8-particle system with two-body and three-body interactions.}
    \label{fig:8particles-tb-inte-methods}
\end{figure}

\begin{figure}[htbp]
    \centering
    \begin{subfigure}[b]{0.32\textwidth}
        \centering
        \includegraphics[width=\textwidth, height=6cm, keepaspectratio]{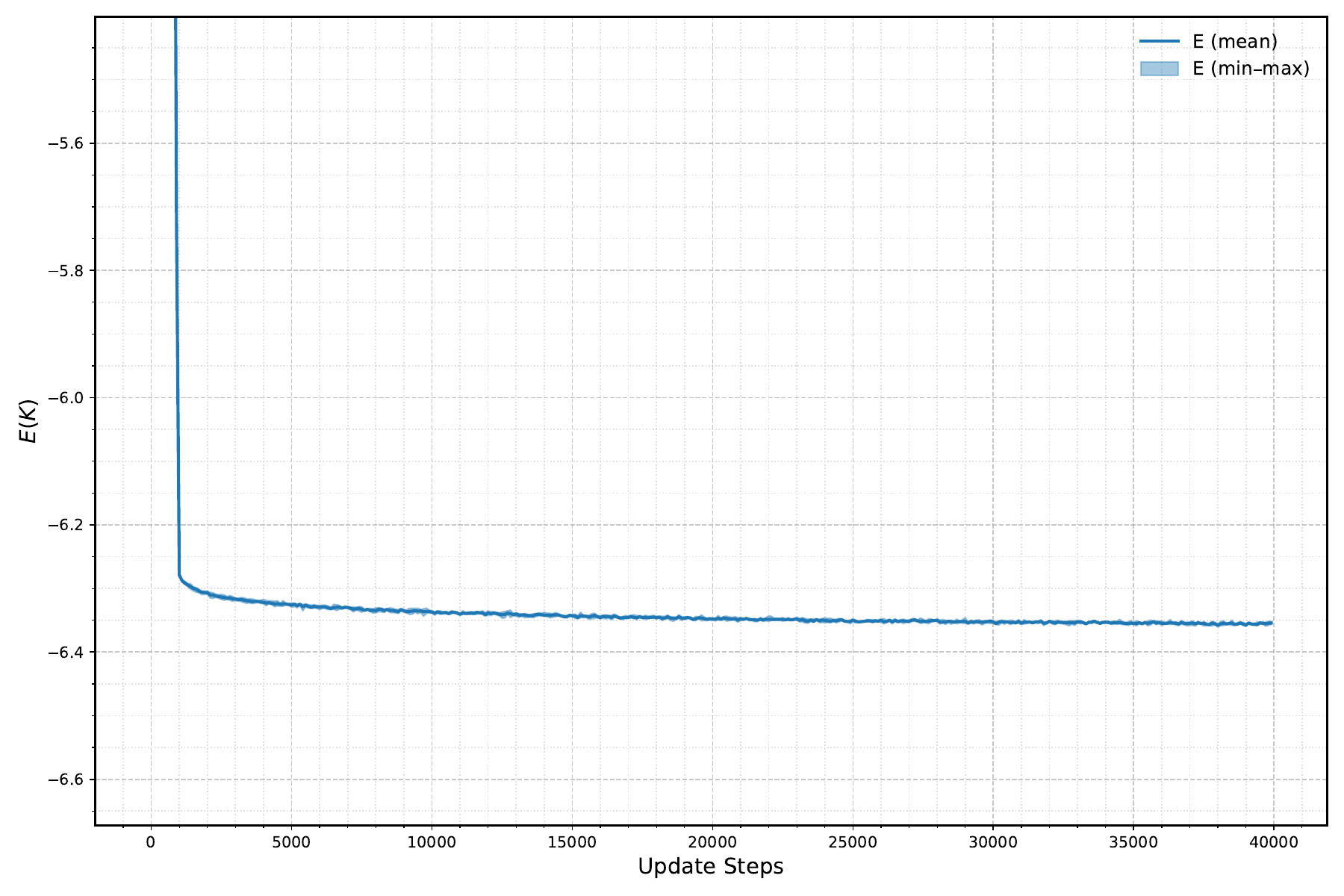}
        \caption{\centering
            \texttt{GELU-MALA}.\\
            E(K) = -6.35 $\pm$ 0.02.\\
            r-range (99.9\% CI): $[-25.41,\,25.44]$.
        }
        \label{fig:9p-tb-gelu-mala}
    \end{subfigure}
    \hfill
    \begin{subfigure}[b]{0.32\textwidth}
        \centering
        \includegraphics[width=\textwidth, height=6cm, keepaspectratio]{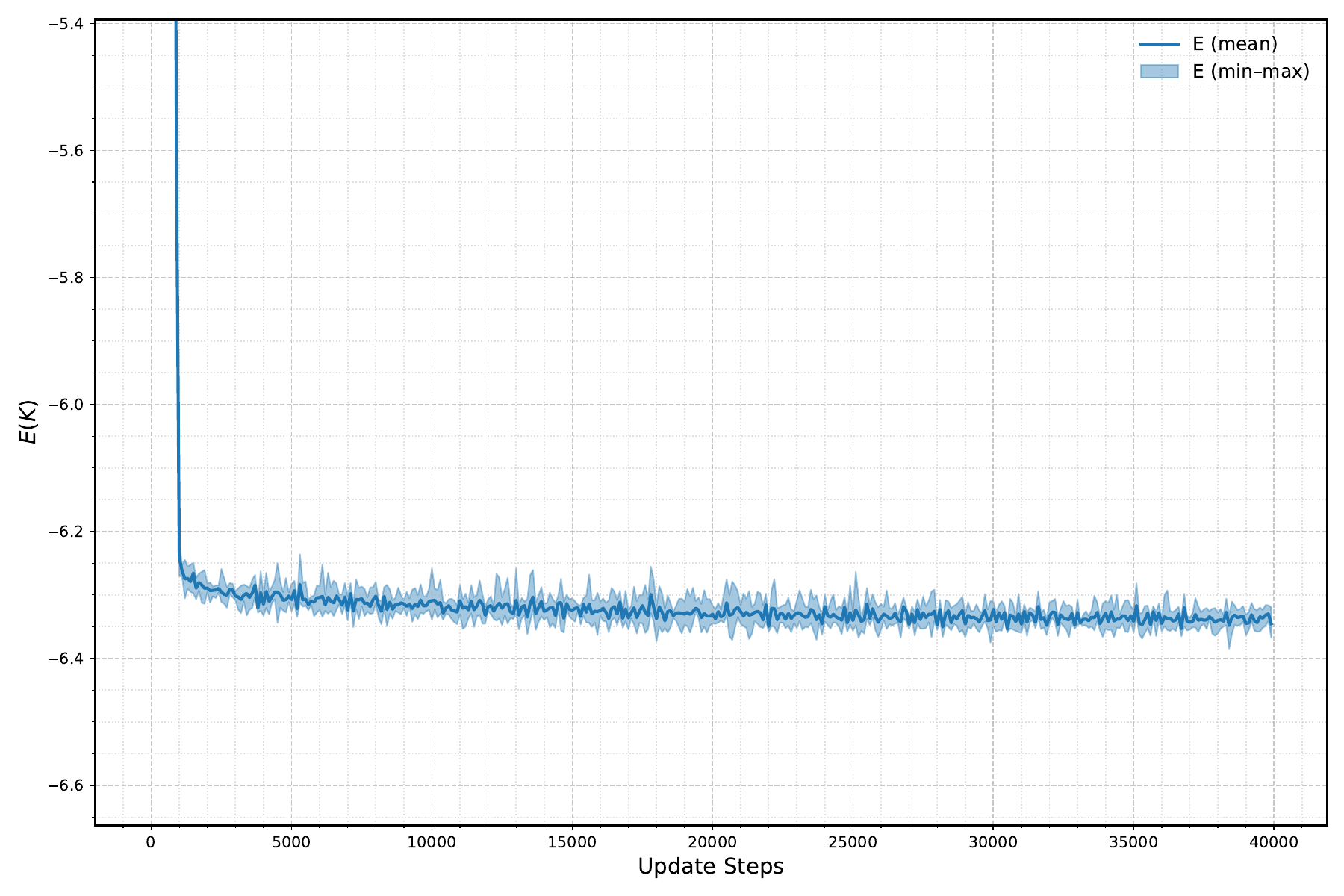}
        \caption{\centering
            \texttt{GELU-ARW}.\\
            E(K) = -6.2 $\pm$ 0.1.\\
            r-range (99.9\% CI): $[-25.32,\,25.34]$.
        }
        \label{fig:9p-tb-gelu-arw}
    \end{subfigure}
    \hfill
    \begin{subfigure}[b]{0.32\textwidth}
        \centering
        \includegraphics[width=\textwidth, height=6cm, keepaspectratio]{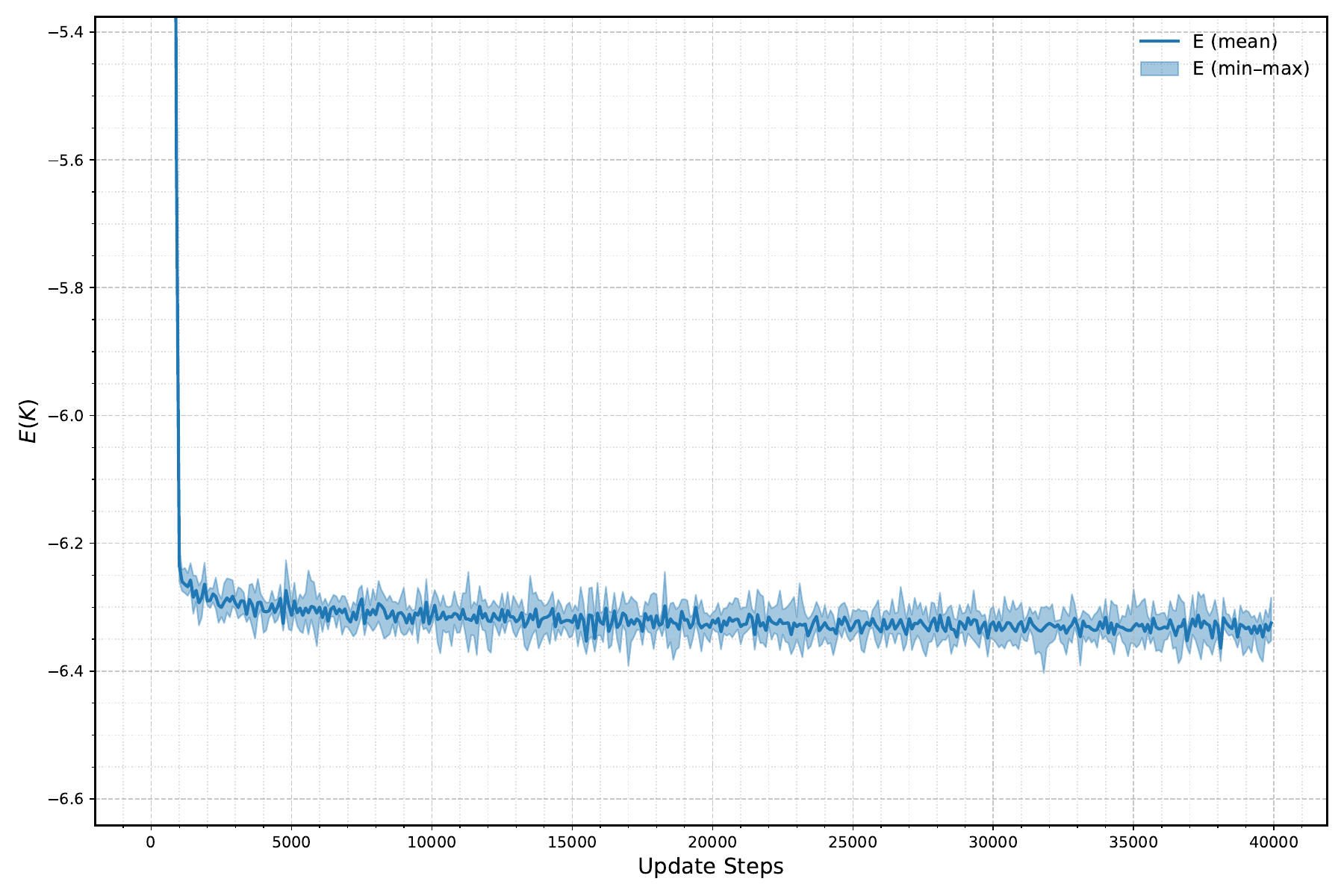}
        \caption{\centering
            \texttt{GELU-RW}.\\
            E(K) = -6.3 $\pm$ 0.1.\\
            r-range (99.9\% CI): $[-25.42,\,25.25]$.
        }
        \label{fig:9p-tb-gelu-rw}
    \end{subfigure}

    \caption{9-particle system with two-body and three-body interactions.}
    \label{fig:9particles-tb-inte-methods}
\end{figure}

\begin{figure}[htbp]
    \centering

    \begin{subfigure}[b]{0.4\textwidth}
        \centering
        \includegraphics[width=\textwidth, height=4cm, keepaspectratio]{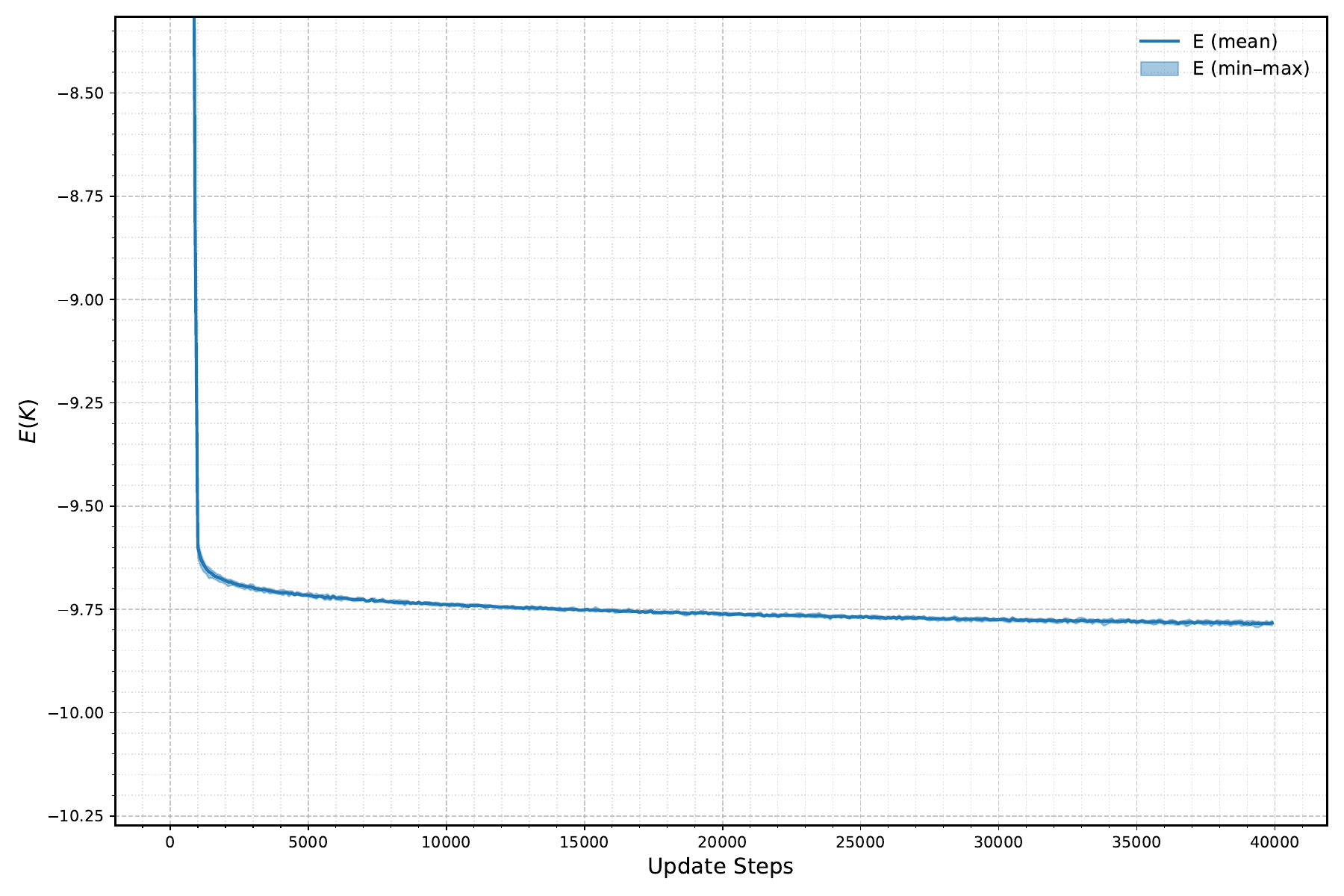}
        \caption{\centering
            11 particles.\\
            E(K) = -9.80 $\pm$ 0.02.\\
            r-range (99.9\% CI): $[-25.34,\,25.30]$.
        }
        \label{fig:11p-tb-gelu-mala}
    \end{subfigure}
    \hfill
    \begin{subfigure}[b]{0.4\textwidth}
        \centering
        \includegraphics[width=\textwidth, height=4cm, keepaspectratio]{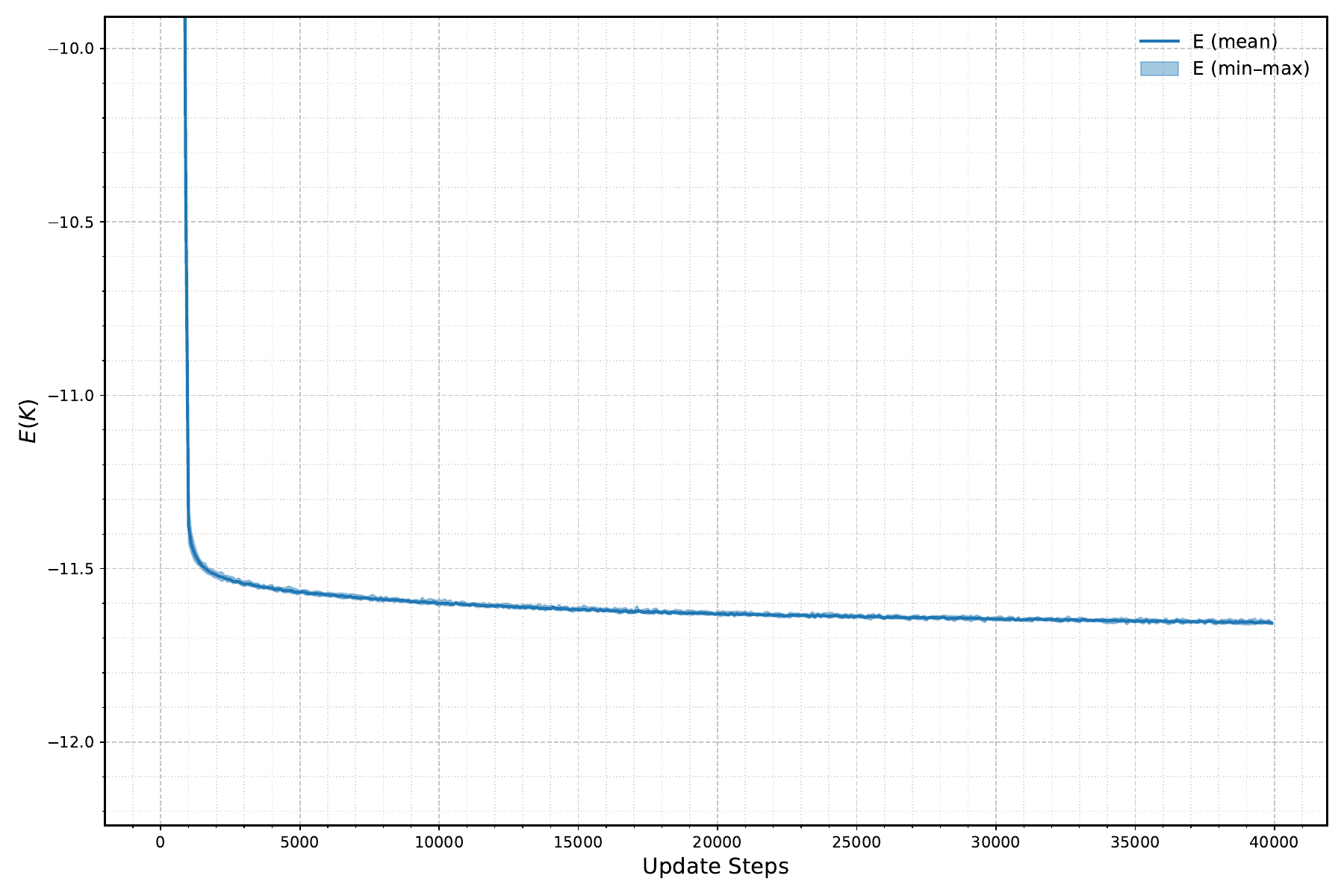}
        \caption{\centering
            12 particles.\\
            E(K) = -11.65 $\pm$ 0.01.\\
            r-range (99.9\% CI): $[-25.19,\,25.26]$.
        }
        \label{fig:12p-tb-gelu-mala}
    \end{subfigure}

    \vspace{0.5em}

    \begin{subfigure}[b]{0.4\textwidth}
        \centering
        \includegraphics[width=\textwidth, height=4cm, keepaspectratio]{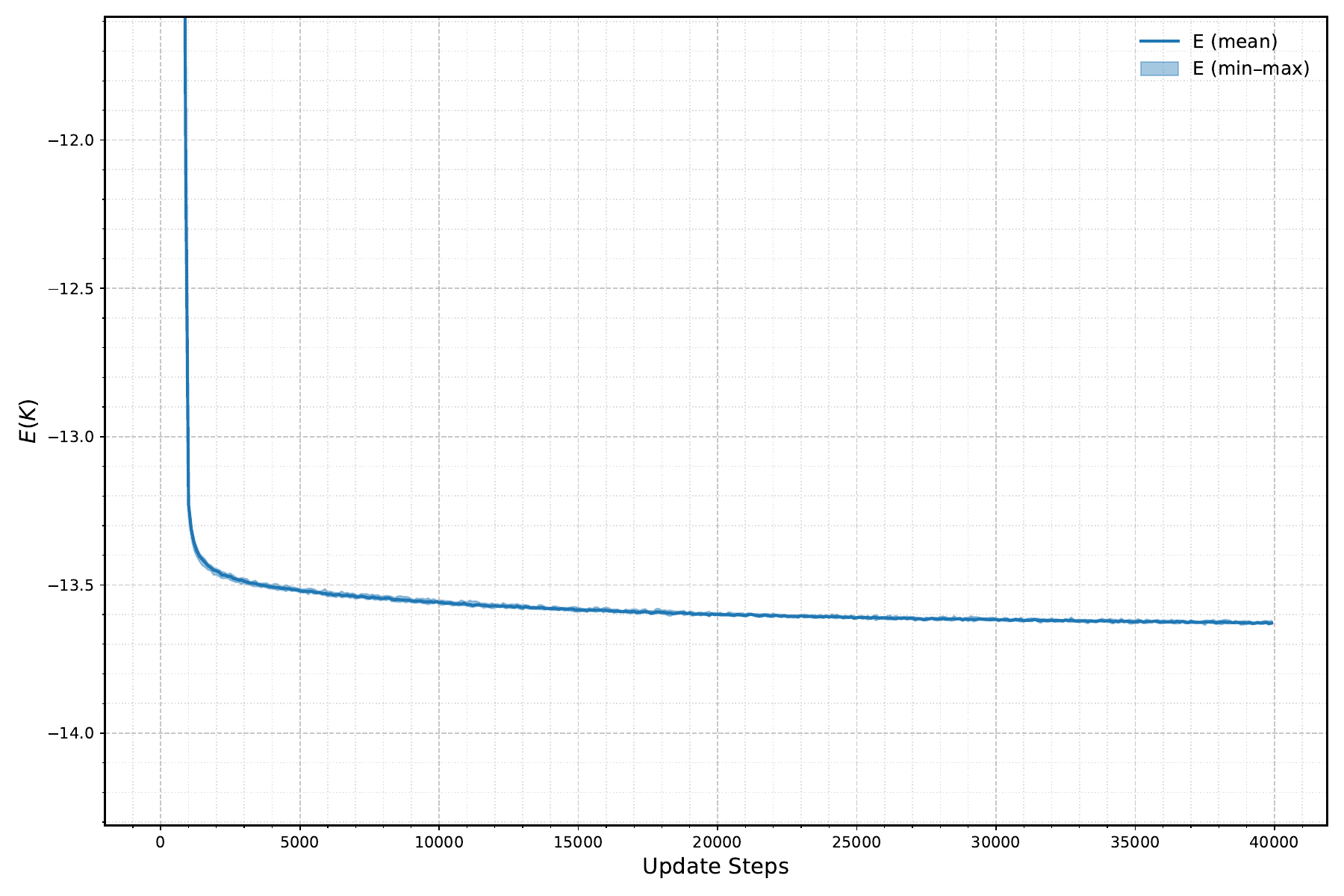}
        \caption{\centering
            13 particles.\\
            E(K) = -13.63 $\pm$ 0.03.\\
            r-range (99.9\% CI): $[-25.20,\,25.19]$.
        }
        \label{fig:13p-tb-gelu-mala}
    \end{subfigure}
    \hfill
    \begin{subfigure}[b]{0.4\textwidth}
        \centering
        \includegraphics[width=\textwidth, height=4cm, keepaspectratio]{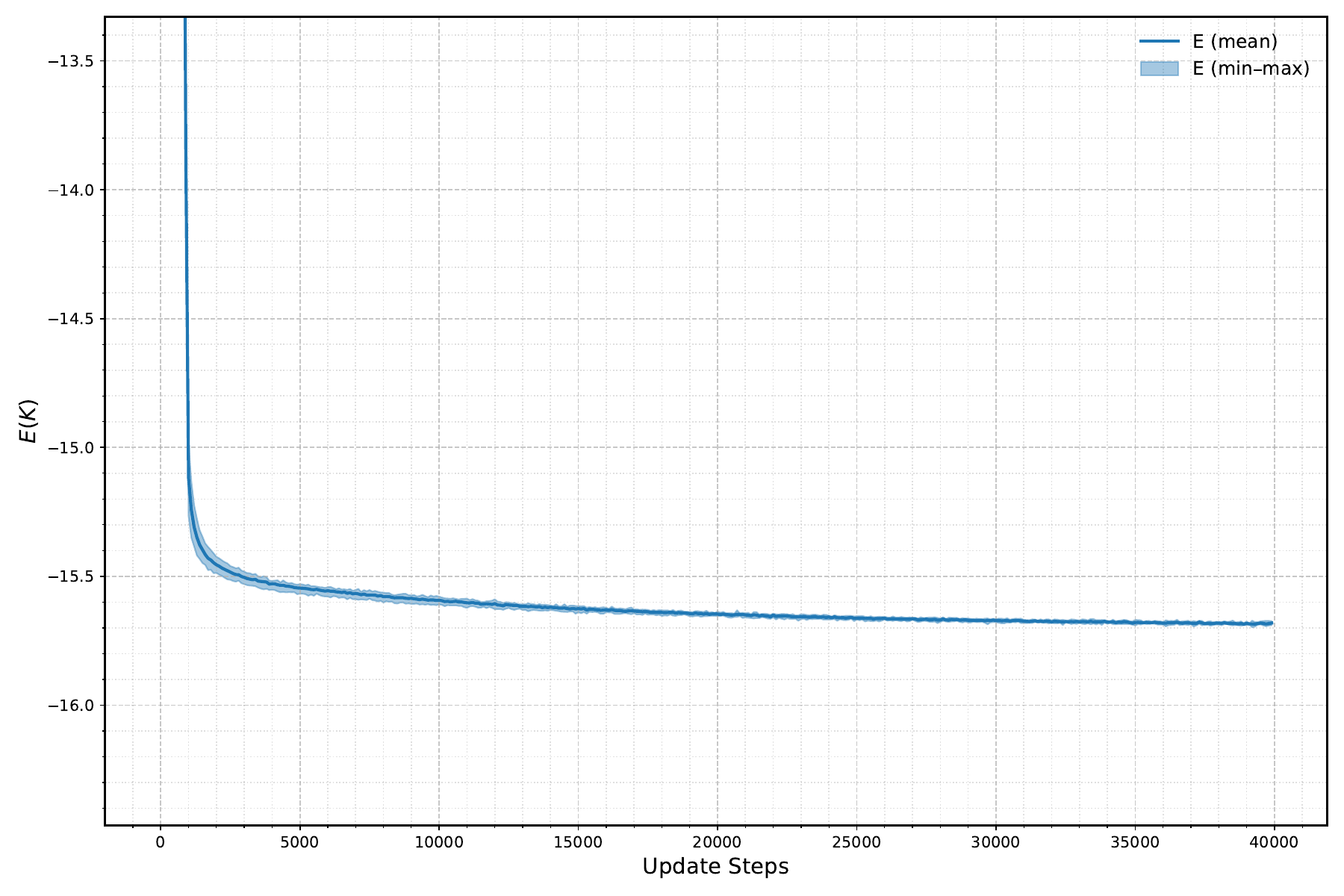}
        \caption{\centering
            14 particles.\\
            E(K) = -15.67 $\pm$ 0.01.\\
            r-range (99.9\% CI): $[-25.26,\,25.31]$.
        }
        \label{fig:14p-tb-gelu-mala}
    \end{subfigure}

    \vspace{0.5em}

    \begin{subfigure}[b]{0.4\textwidth}
        \centering
        \includegraphics[width=\textwidth, height=4cm, keepaspectratio]{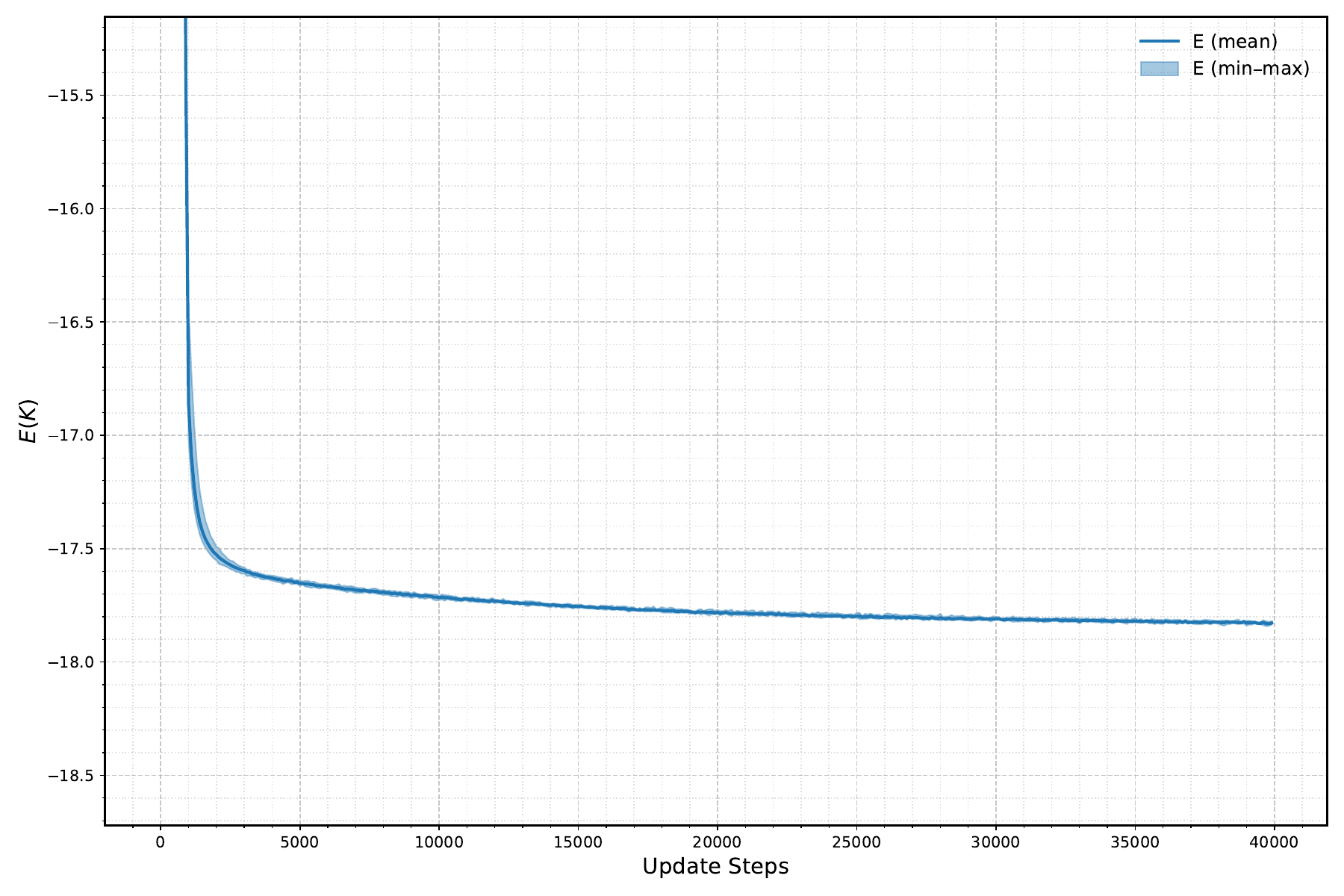}
        \caption{\centering
            15 particles.\\
            E(K) = -17.83 $\pm$ 0.02.\\
            r-range (99.9\% CI): $[-25.35,\,25.48]$.
        }
        \label{fig:15p-tb-gelu-mala}
    \end{subfigure}
    \hfill
    \begin{subfigure}[b]{0.4\textwidth}
        \centering
        \includegraphics[width=\textwidth, height=4cm, keepaspectratio]{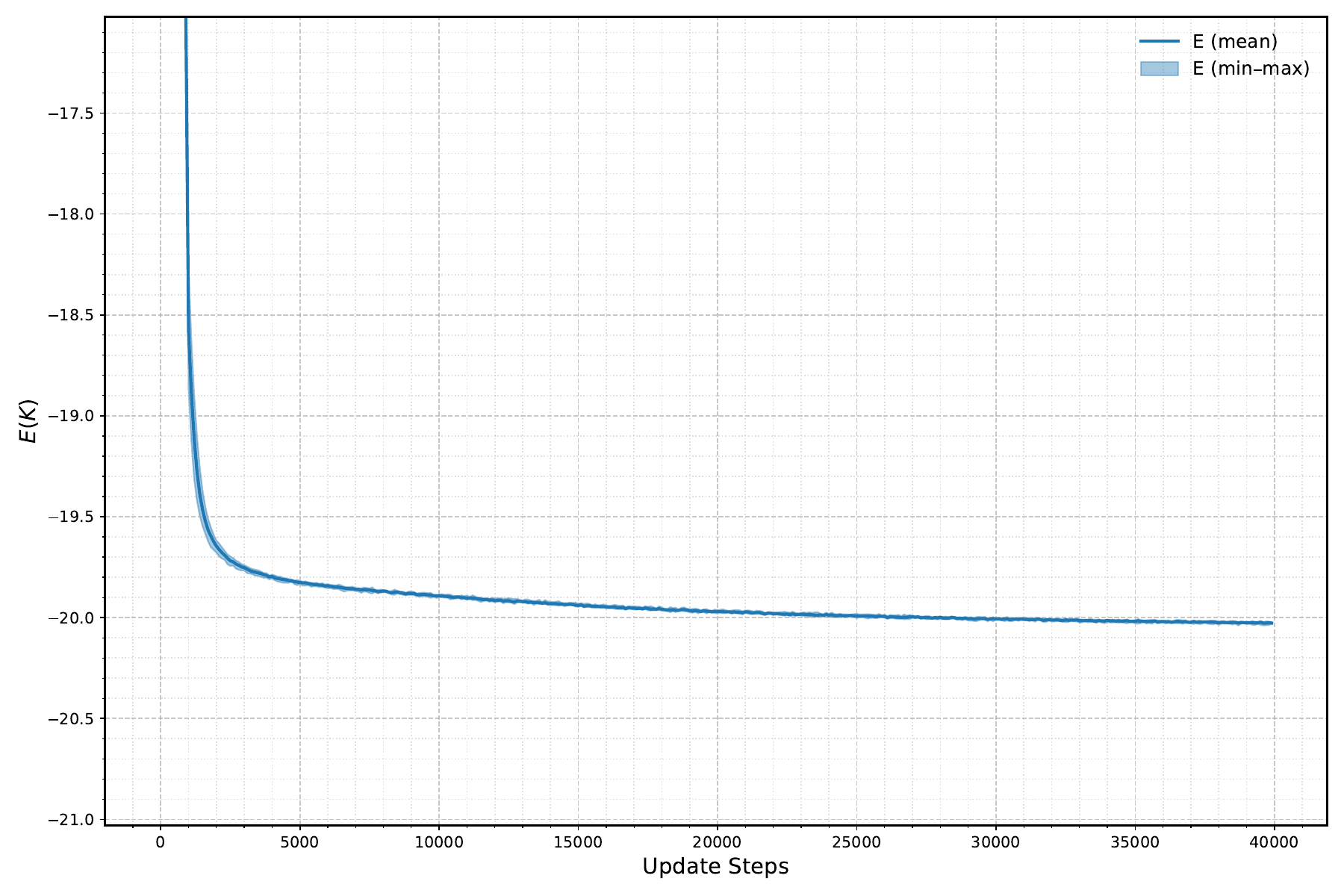}
        \caption{\centering
            16 particles.\\
            E(K) = -20.04 $\pm$ 0.03.\\
            r-range (99.9\% CI): $[-25.61,\,25.63]$.
        }
        \label{fig:16p-tb-gelu-mala}
    \end{subfigure}

    \vspace{0.5em}

    \begin{subfigure}[b]{0.4\textwidth}
        \centering
        \includegraphics[width=\textwidth, height=4cm, keepaspectratio]{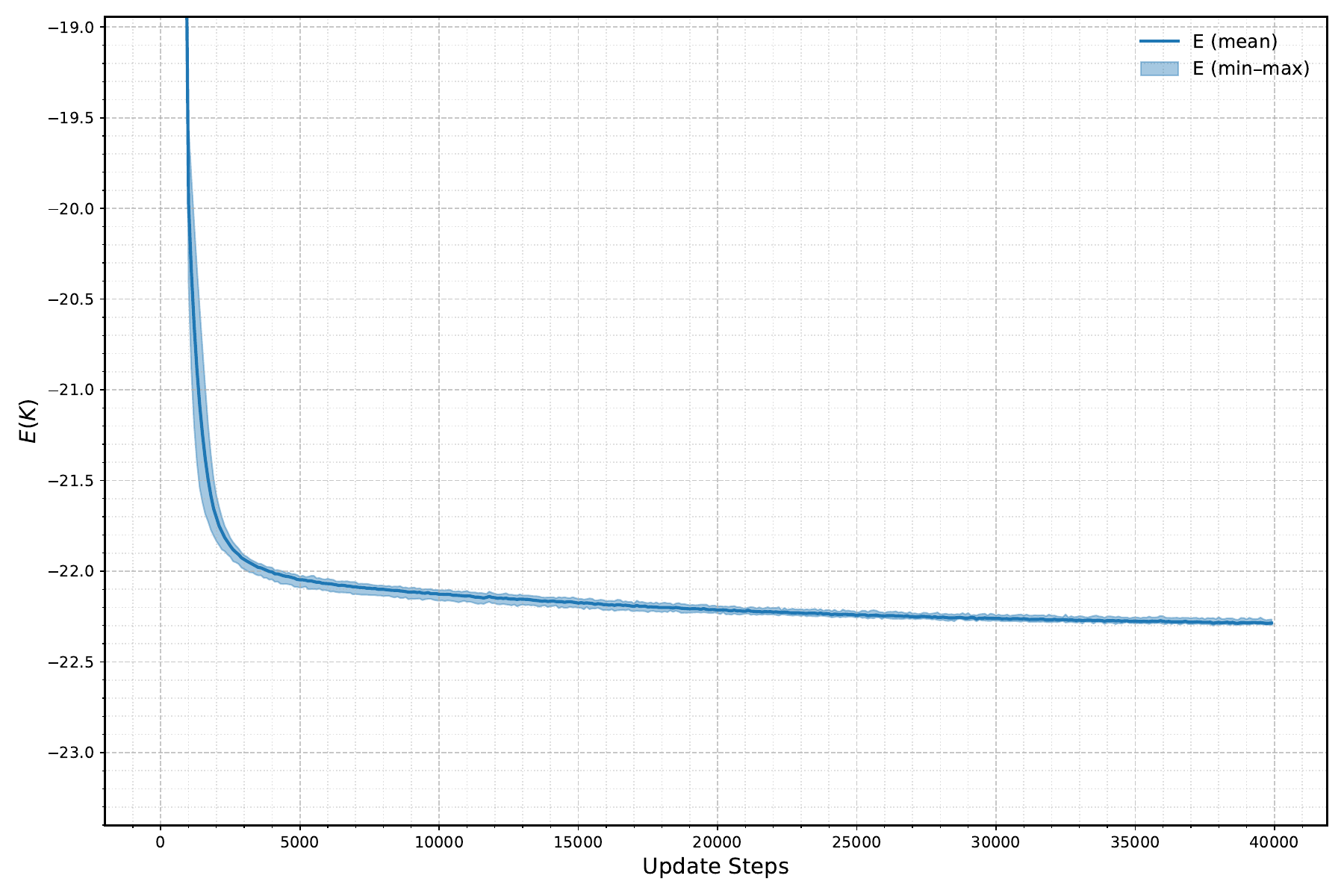}
        \caption{\centering
            17 particles.\\
            E(K) = -22.3 $\pm$ 0.03.\\
            r-range (99.9\% CI): $[-25.64,\,25.64]$.
        }
        \label{fig:17p-tb-gelu-mala}
    \end{subfigure}
    \hfill
    \begin{subfigure}[b]{0.4\textwidth}
        \centering
        \includegraphics[width=\textwidth, height=4cm, keepaspectratio]{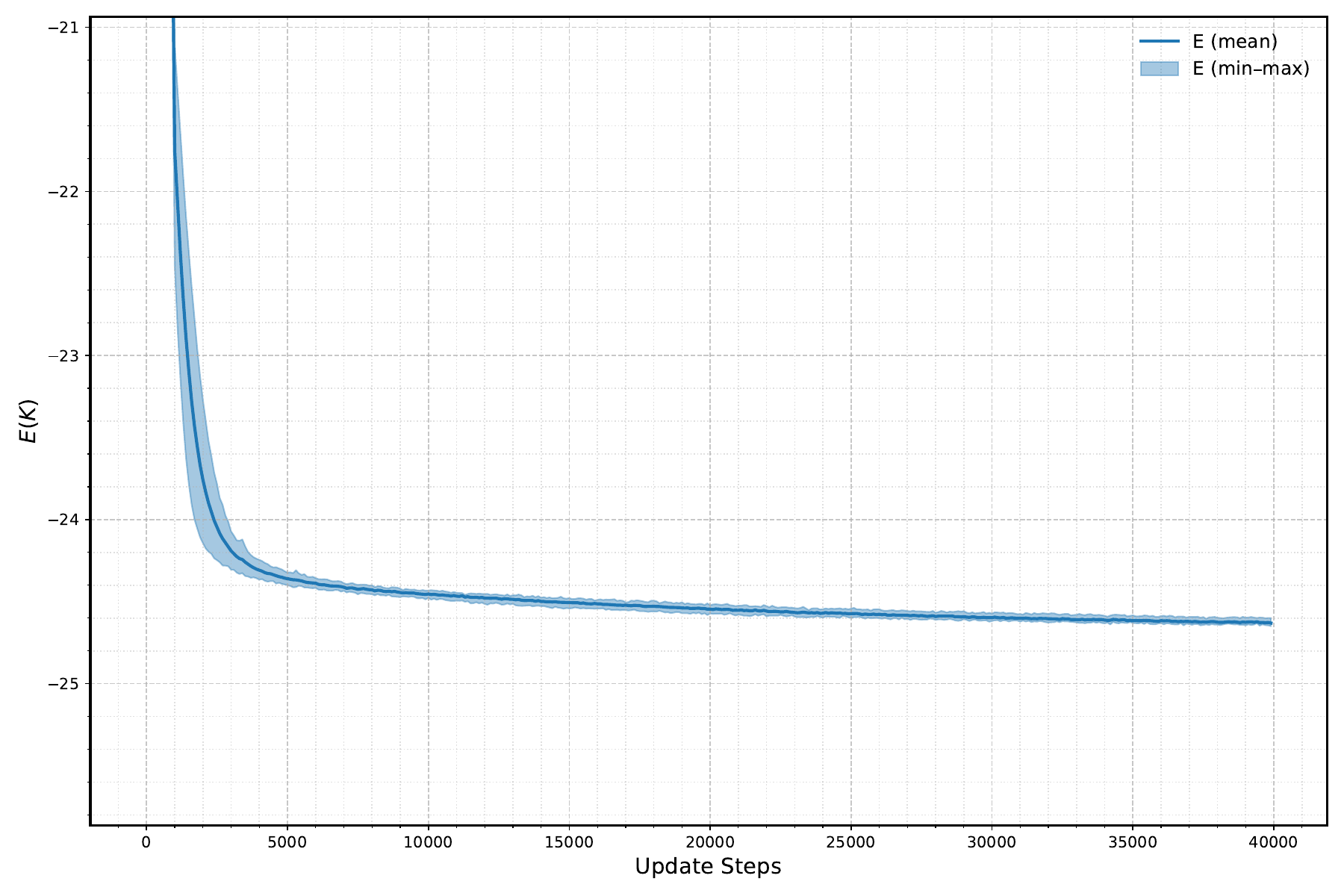}
        \caption{\centering
            18 particles.\\
            E(K) = -24.62 $\pm$ 0.03.\\
            r-range (99.9\% CI): $[-25.95,\,25.94]$.
        }
        \label{fig:18p-tb-gelu-mala}
    \end{subfigure}

    \vspace{0.5em}

    \begin{subfigure}[b]{0.4\textwidth}
        \centering
        \includegraphics[width=\textwidth, height=4cm, keepaspectratio]{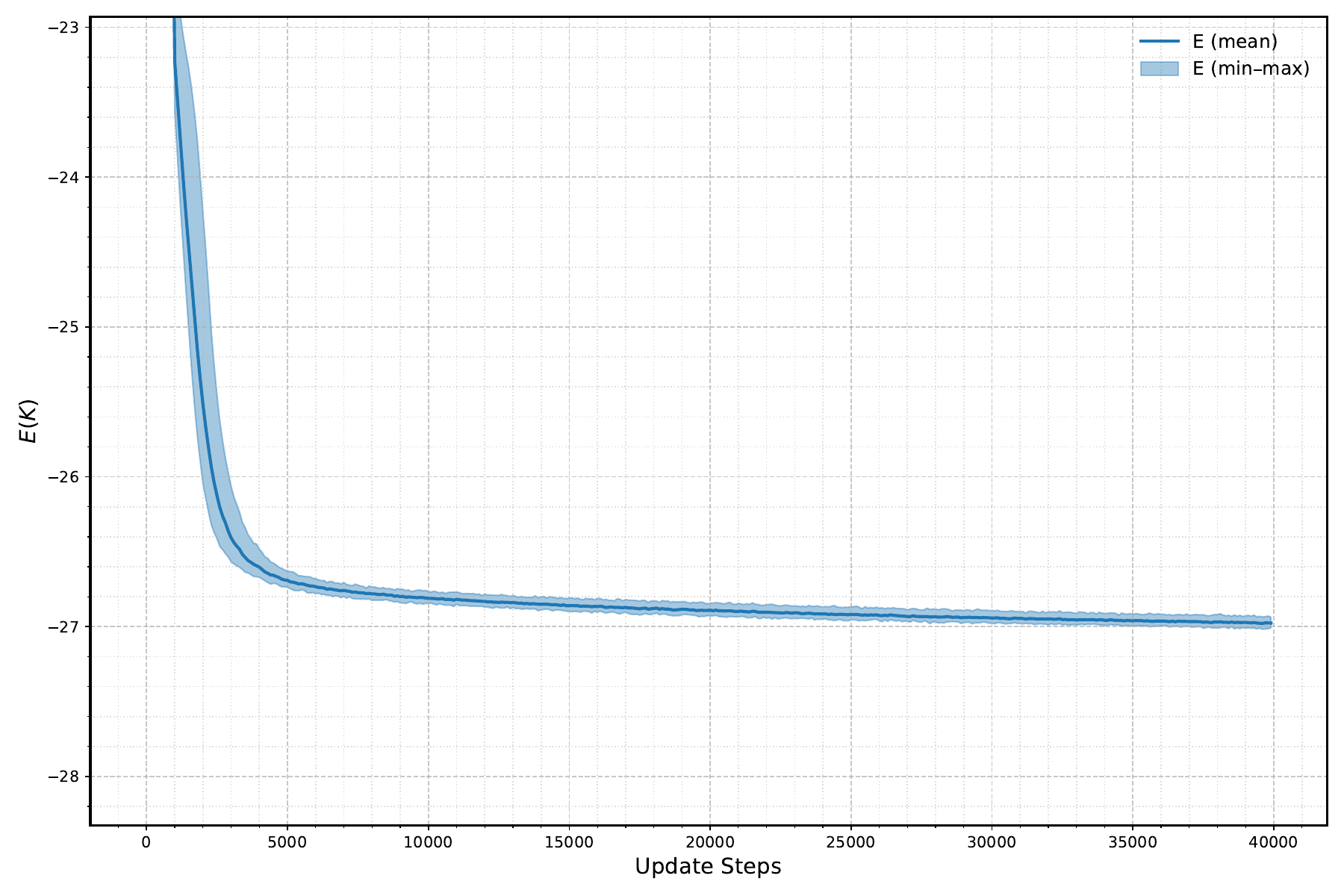}
        \caption{\centering
            19 particles.\\
            E(K) = -26.98 $\pm$ 0.06.\\
            r-range (99.9\% CI): $[-26.08,\,26.11]$.
        }
        \label{fig:19p-tb-gelu-mala}
    \end{subfigure}
    \hfill
    \begin{subfigure}[b]{0.4\textwidth}
        \centering
        \includegraphics[width=\textwidth, height=4cm, keepaspectratio]{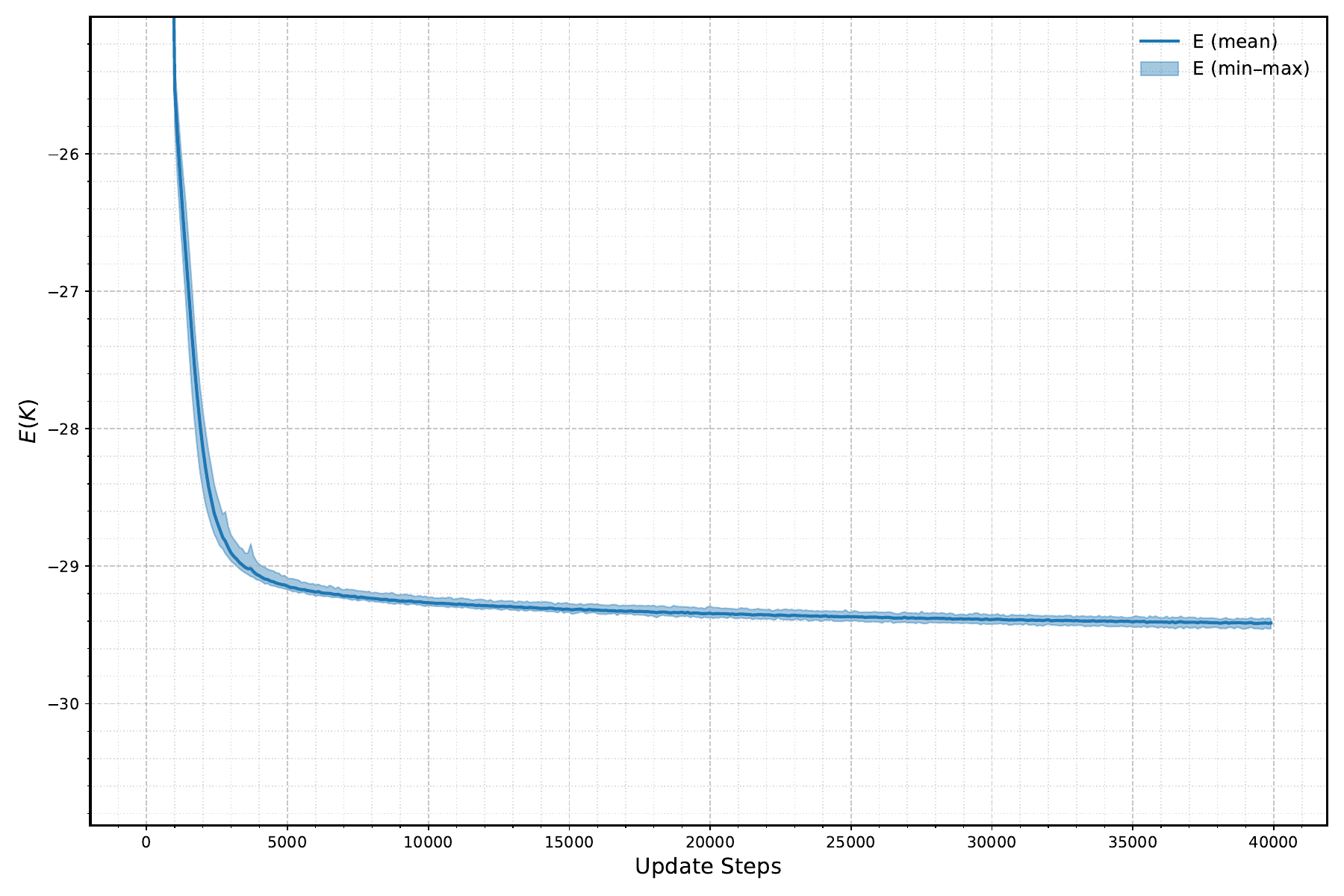}
        \caption{\centering
            20 particles.\\
            E(K) = -29.41 $\pm$ 0.03.\\
            r-range (99.9\% CI): $[-26.24,\,26.22]$.
        }
        \label{fig:20p-tb-gelu-mala}
    \end{subfigure}

    \caption{Training convergence for 11--20 particle systems using \texttt{GELU-MALA}.}
    \label{fig:11-20particles-tb-gelu-mala}
\end{figure}

\clearpage
\subsection{System C}
\label{appendix:systemC}

\begin{figure}[htbp]
    \centering
    \begin{subfigure}[b]{0.32\textwidth}
        \centering
        \includegraphics[width=\textwidth, height=6cm, keepaspectratio]{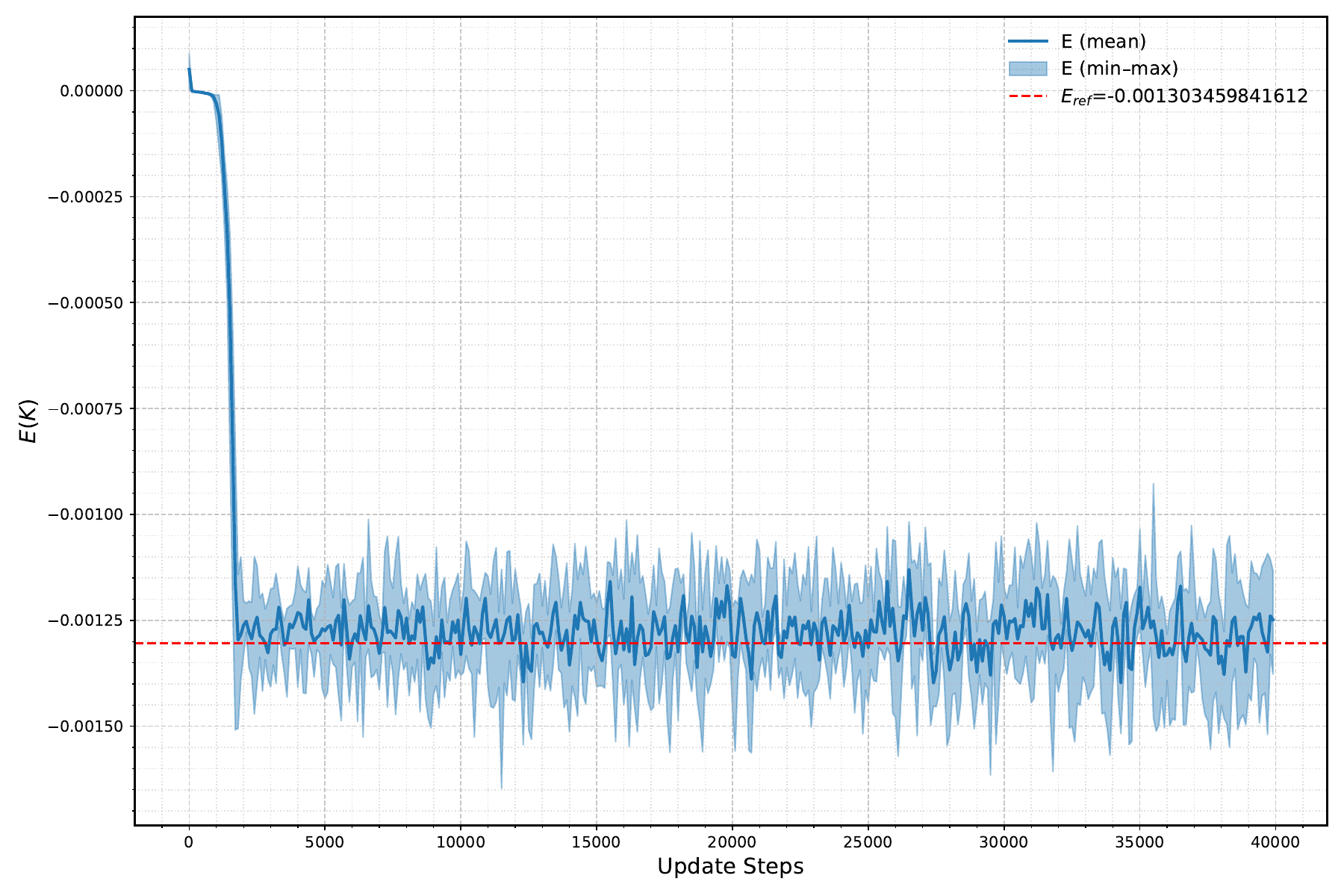}
        \caption{\centering
            \texttt{GELU-MALA}.\\
            E(K) = $-1.1\times10^{-3} \ \pm\ 0.5\times10^{-3}$.\\
            r-range (99.9\% CI): $[-468.83,\,468.25]$.
        }
        \label{fig:2p-tb-gelu-mala}
    \end{subfigure}
    \hfill
    \begin{subfigure}[b]{0.32\textwidth}
        \centering
        \includegraphics[width=\textwidth, height=6cm, keepaspectratio]{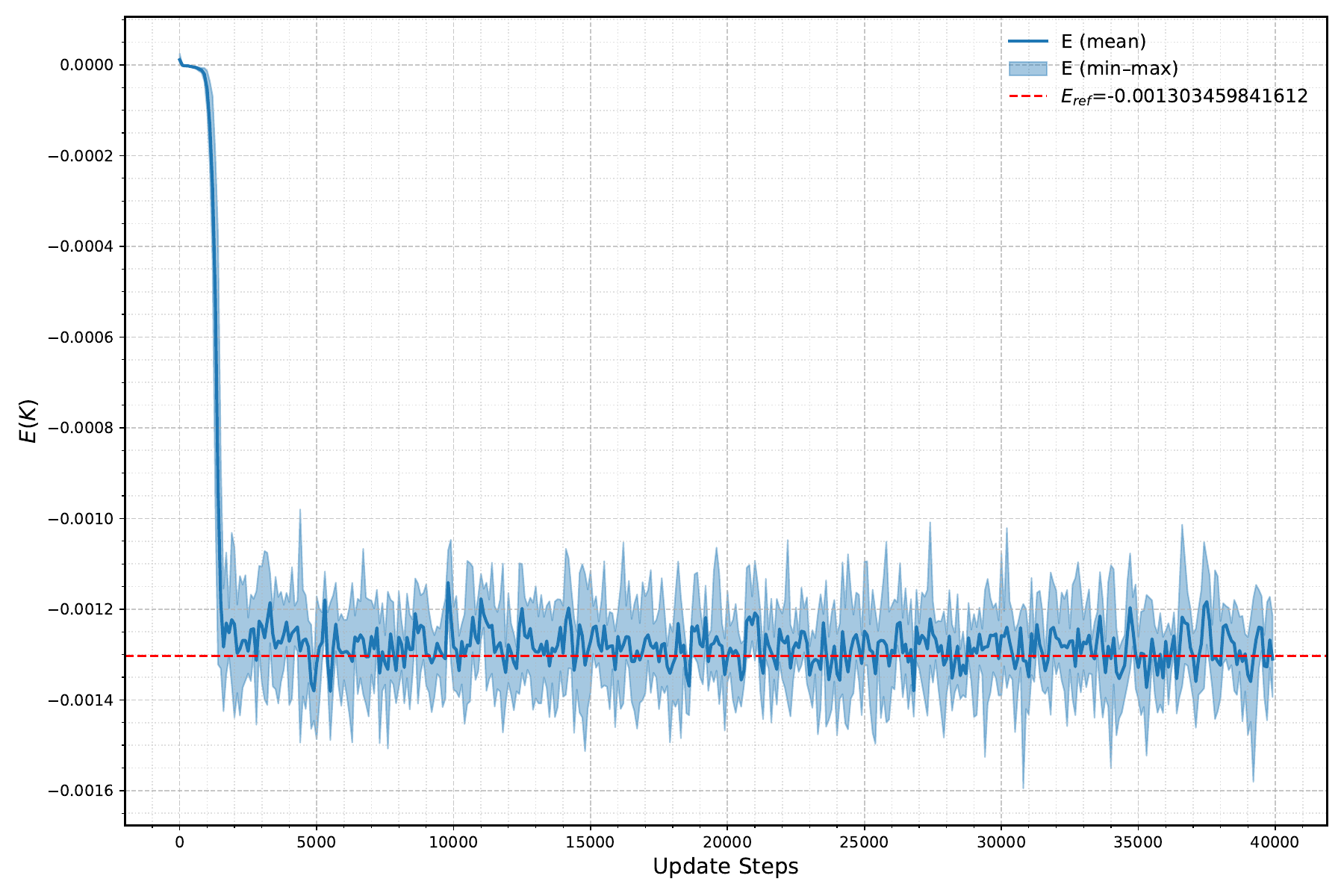}
        \caption{\centering
            \texttt{GELU-ARW}.\\
            E = $-2.1\times10^{-3} \ \pm\ 1.5\times10^{-3}$.\\
            r-range (99.9\% CI): $[-480.05,\,477.62]$.
        }
        \label{fig:2p-tb-gelu-arw}
    \end{subfigure}
    \hfill
    \begin{subfigure}[b]{0.32\textwidth}
        \centering
        \includegraphics[width=\textwidth, height=6cm, keepaspectratio]{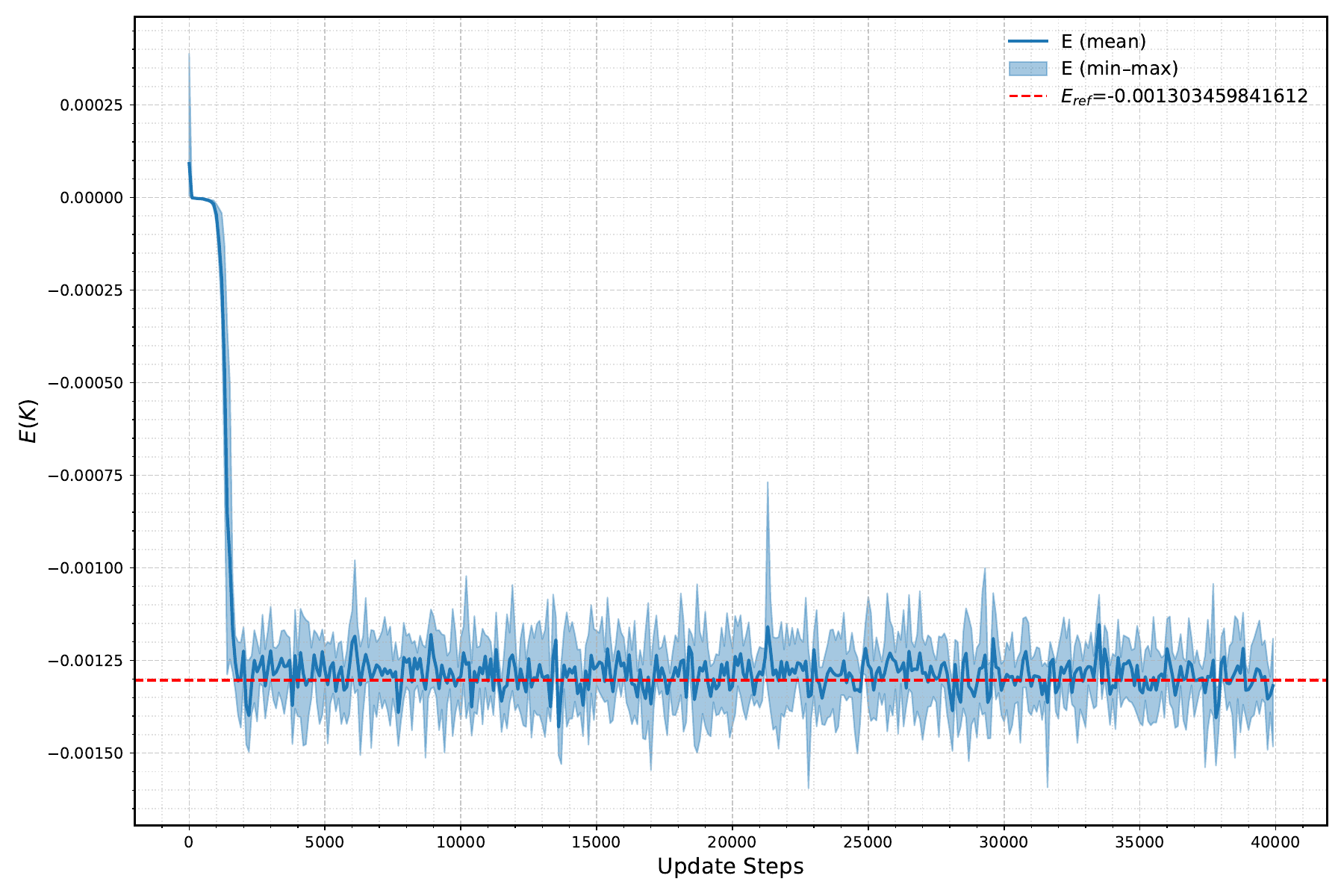}
        \caption{\centering
            \texttt{GELU-RW}.\\
            E = $-0.9\times10^{-3} \ \pm\ 0.6\times10^{-3}$.\\
            r-range (99.9\% CI): $[-477.98,\,474.03]$.
        }
        \label{fig:2p-tb-gelu-rw}
    \end{subfigure}

    \caption{2-particle system (System C) with two-body interactions only.}
    \label{fig:2particles-tb-inte-methods}
\end{figure}

\begin{figure}[htbp]
    \centering
    \begin{subfigure}[b]{0.32\textwidth}
        \centering
        \includegraphics[width=\textwidth, height=6cm, keepaspectratio]{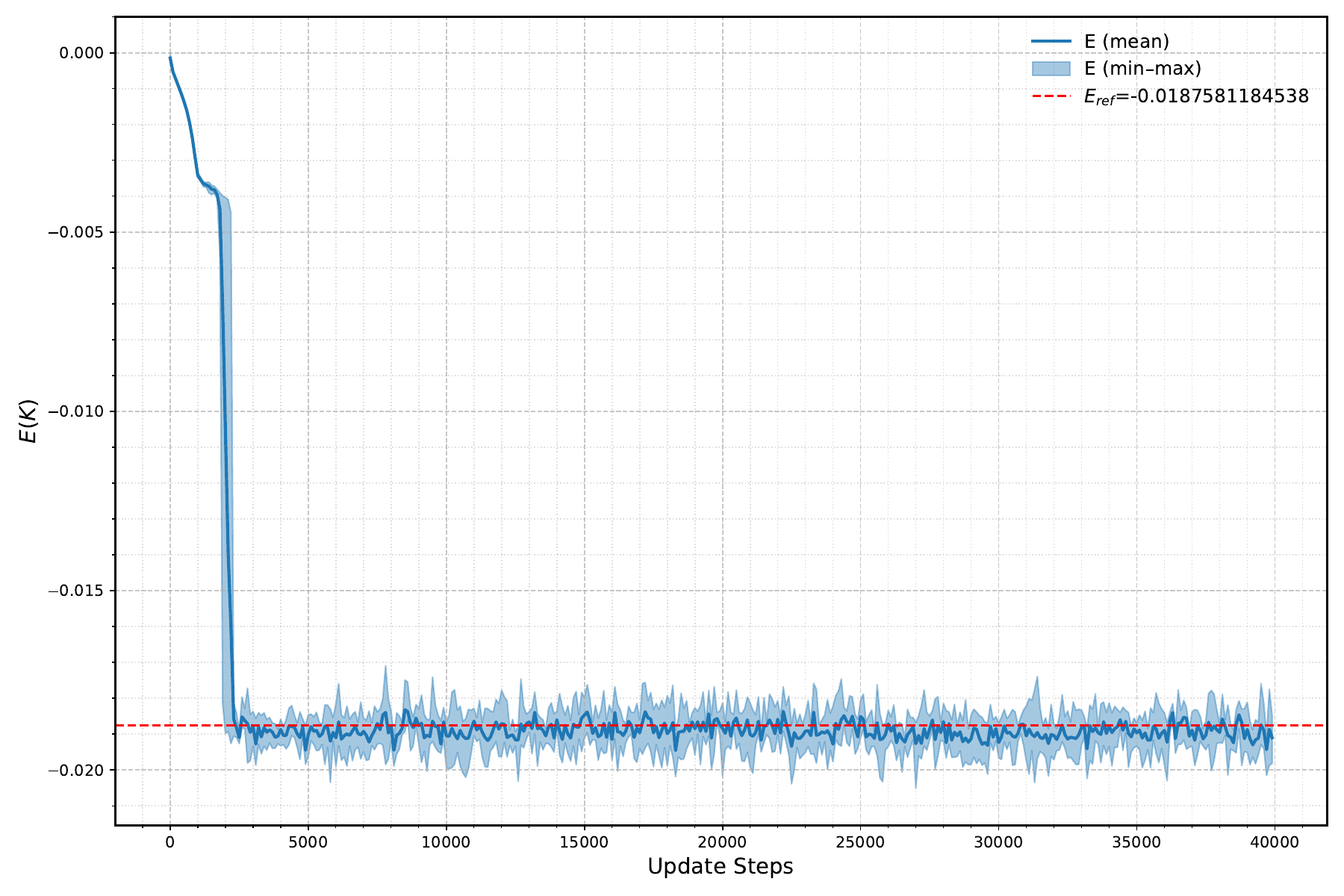}
        \caption{\centering
            \texttt{GELU-MALA}.\\
            E = $-18.3\times10^{-3} \ \pm\ 0.9\times10^{-3}$.\\
            r-range (99.9\% CI): $[-122.71,\,124.66]$.
        }
        \label{fig:3p-mass-gelu-mala}
    \end{subfigure}
    \hfill
    \begin{subfigure}[b]{0.32\textwidth}
        \centering
        \includegraphics[width=\textwidth, height=6cm, keepaspectratio]{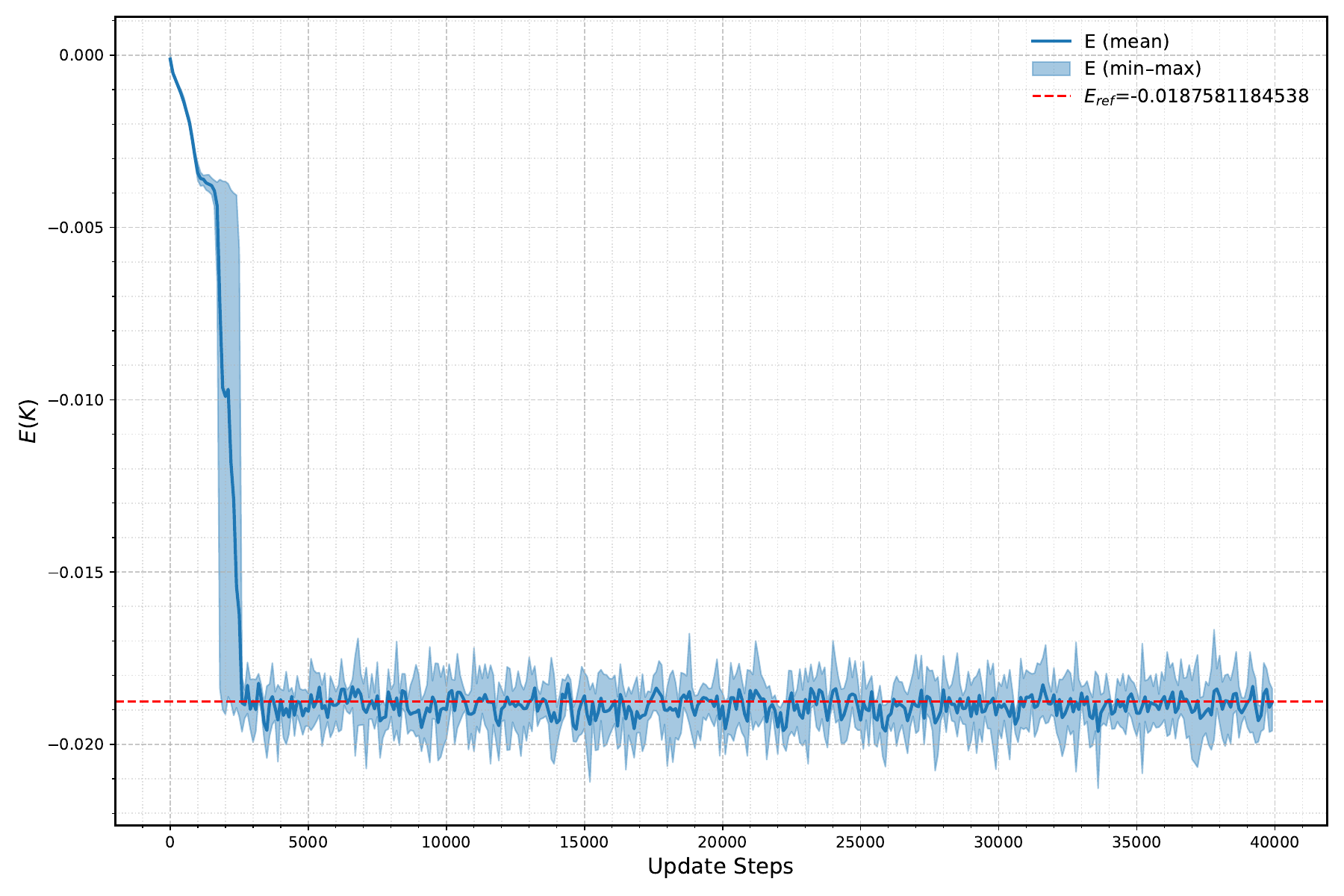}
        \caption{\centering
            \texttt{GELU-ARW}.\\
            E = $-17.9\times10^{-3} \ \pm\ 3.5\times10^{-3}$.\\
            r-range (99.9\% CI): $[-118.73,\,118.35]$.
        }
        \label{fig:3p-mass-gelu-arw}
    \end{subfigure}
    \hfill
    \begin{subfigure}[b]{0.32\textwidth}
        \centering
        \includegraphics[width=\textwidth, height=6cm, keepaspectratio]{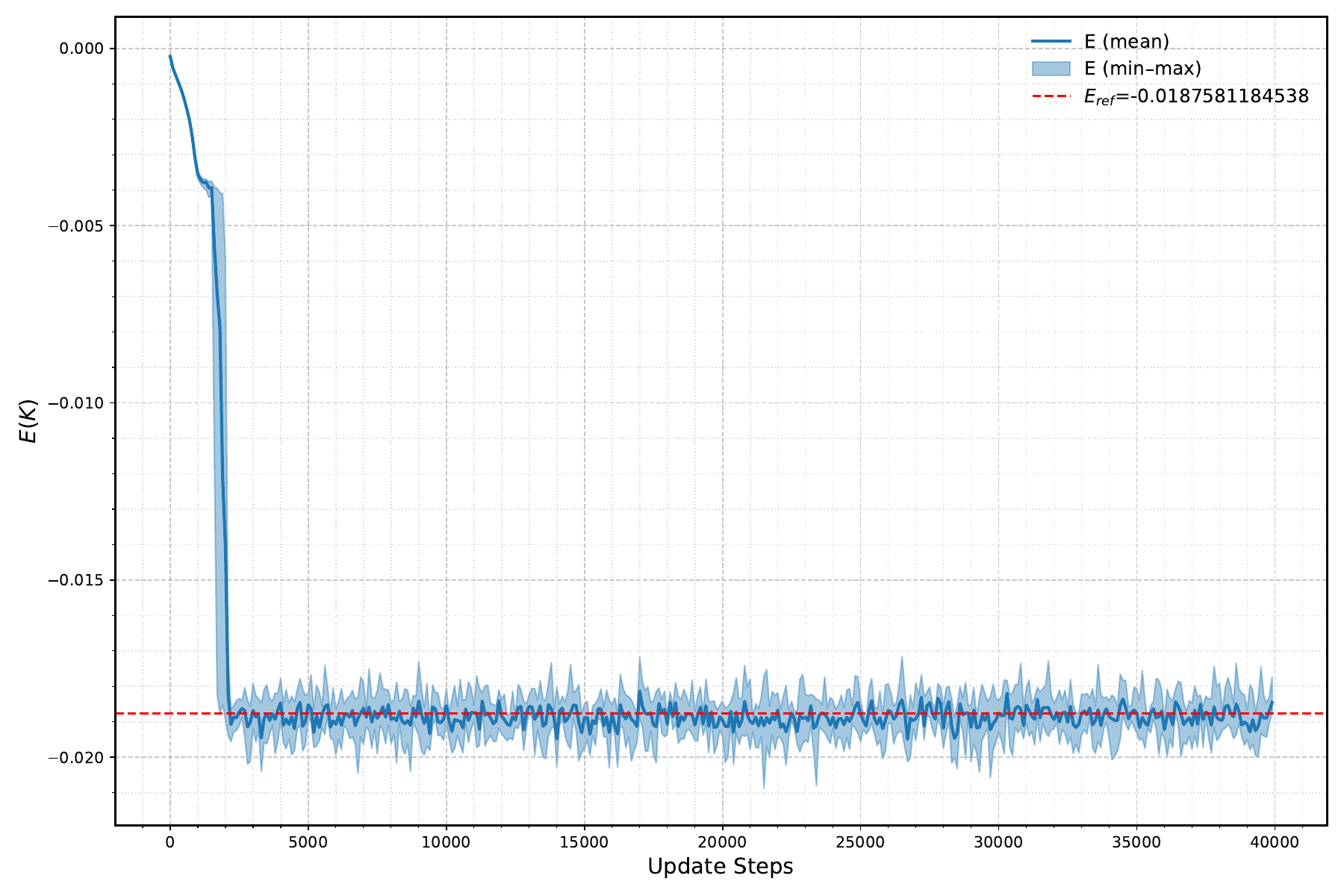}
        \caption{\centering
            \texttt{GELU-RW}.\\
            E = $-18.0\times10^{-3} \ \pm\ 4.2\times10^{-3}$.\\
            r-range (99.9\% CI): $[-120.72,\,121.26]$.
        }
        \label{fig:3p-mass-gelu-rw}
    \end{subfigure}

    \caption{3-particle system (System C with unequal masses) under two-body interactions.}
    \label{fig:3particles-mass-methods}
\end{figure}

\clearpage
\section{Parameter Sensitivity Analysis of Sampling Methods}
\label{appendix:parameter-sensitivity}

\begin{figure}[htbp]
    \centering
    \includegraphics[width=1.0\textwidth]{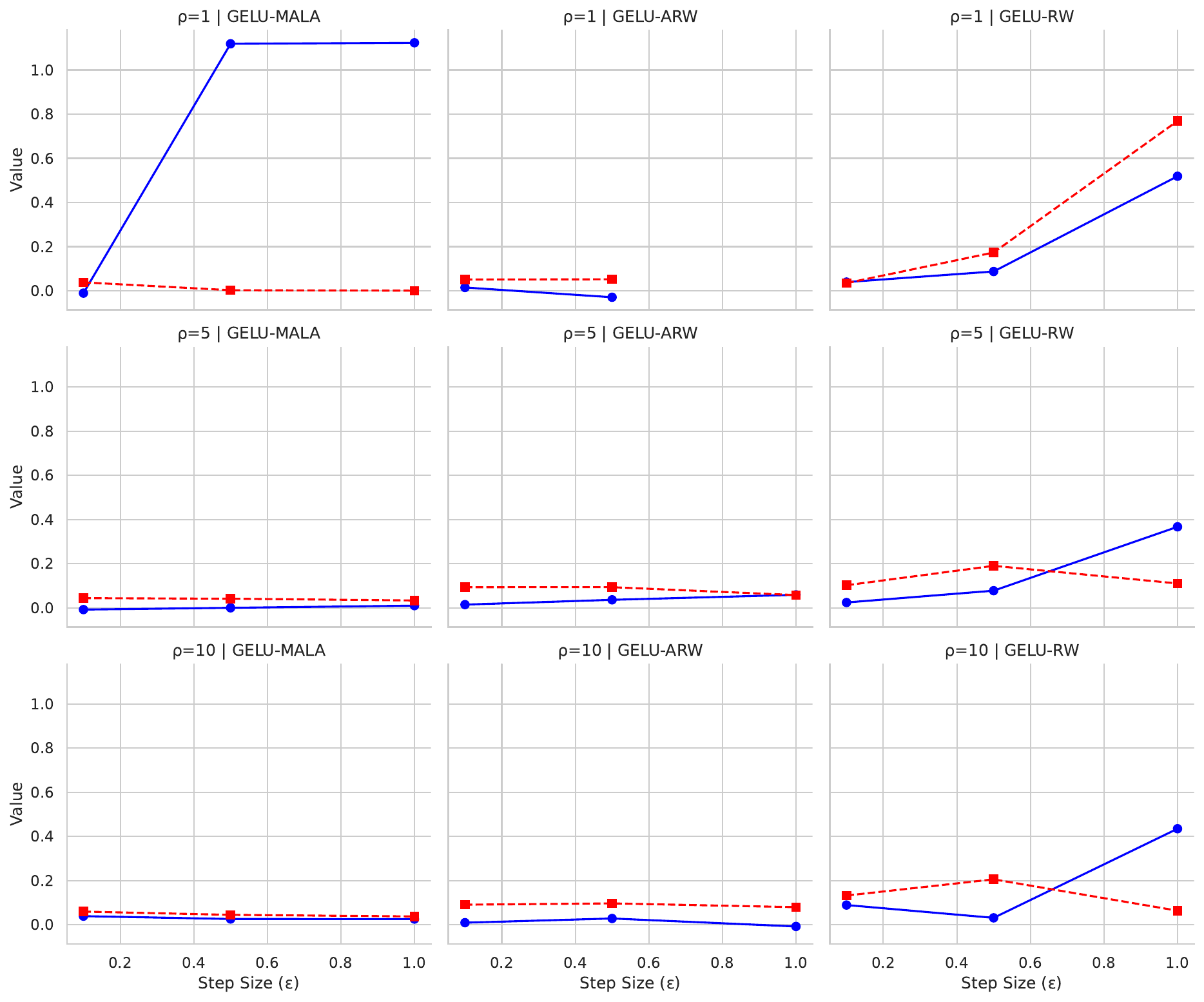}
    \caption{Mean error (blue) and error standard deviation (red) relative to $E_{\text{ref}}$ from~\cite{Yan_2014} for the three-particle harmonic system with two-body interactions. 
    Results are averaged over five independent trials for each combination of range factor $\rho \in \{1, 5, 10\}$ and step size $\epsilon \in \{0.1, 0.5, 1.0\}$. 
    The blue lines correspond to the mean relative error $\frac{|E_{\mathrm{final}} - E_{\mathrm{ref}}|}{|E_{\mathrm{ref}}|}$, while the red dashed lines represent the standard deviation of this error across runs. 
    (\texttt{GELU-ARW} failed for $\rho = 1$ and $\epsilon = 1$ due to numerical overflow.)}
    \label{fig:error_std_facet}
\end{figure}

To evaluate the robustness of our proposed methods, we conducted extensive sensitivity testing on a harmonic system with two-body interactions with $N = 3$. Our experiments systematically varied two key parameters: the range factor $\rho \in \{1, 5, 10\}$ and the step size $\epsilon \in \{0.1, 0.5, 1.0\}$. The results, illustrated in Figure~\ref{fig:error_std_facet}, reveal several important insights about the performance characteristics of each method. 

As evident from Figure~\ref{fig:error_std_facet}, adaptive methods yield stable results, with two notable exceptions. The \texttt{GELU-MALA} method fails when $\rho = 1$ combined with either $\epsilon = 0.5$ or $\epsilon = 1.0$. Similarly, \texttt{GELU-ARW} breaks down under the extreme condition of $\rho = 1$ and $\epsilon = 1.0$. These findings suggest that adaptive methods benefit from initialization with relatively smaller $\epsilon$ values within the sampling range $\rho$, as they can subsequently self-adjust to discover optimal sampling pattern.

The non-adaptive \texttt{GELU-RW} method demonstrates significantly higher error magnitudes and standard deviations across different parameters, confirming its inferior stability compared to adaptive approaches. This behavior is expected, as it lacks the self-correcting mechanisms present in the adaptive variants.

\begin{figure}[htbp]
    \centering
    \includegraphics[width=1.0\textwidth]{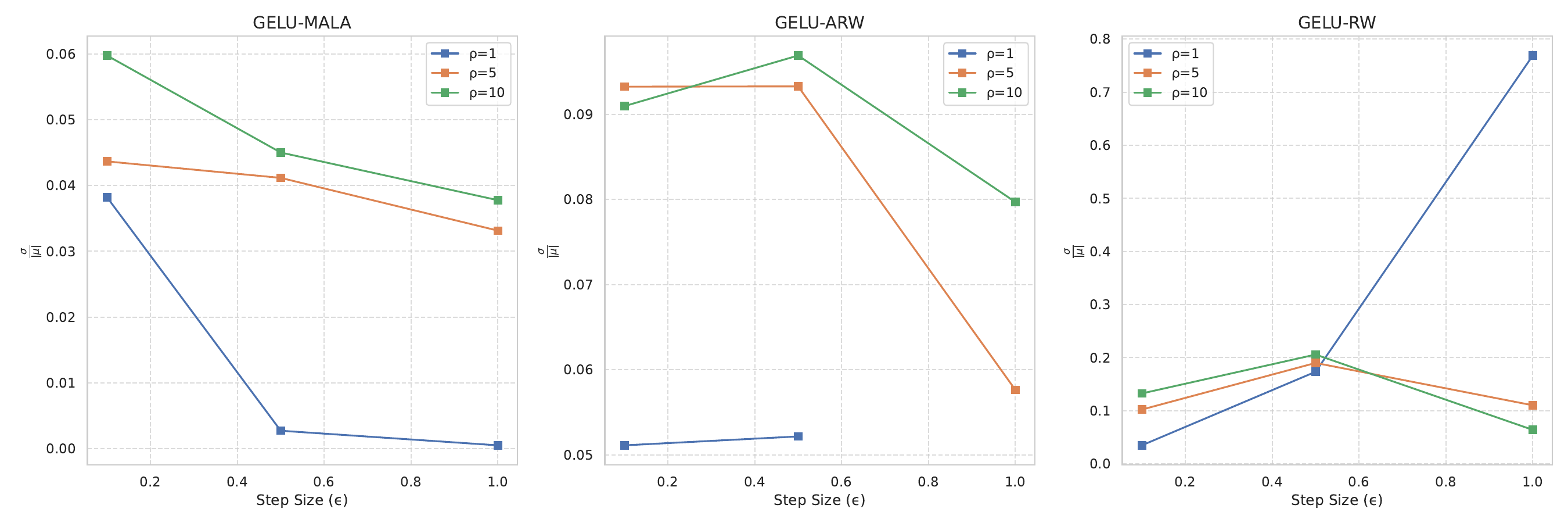}
    \caption{Coefficient of variation ($\sigma/\mu$) as a function of step size $\epsilon$, with separate panels for each sampling method. Different colors represent various range factors $\rho$. The \texttt{GELU-RW} method shows significantly higher variability across all configurations, whereas the adaptive methods exhibit lower variation.}
    \label{fig:std_by_sampling}
\end{figure}

Further analysis of method stability is presented in Figure~\ref{fig:std_by_sampling}, which illustrates the coefficient of variation for each sampling method plotted against step size. The coefficient of variation $\frac{\sigma}{|\mu|}$, provides a normalized measure of dispersion that enables direct comparison across different parameters.

Our results demonstrate why automatically adjusting parameters during the sampling process is so valuable, especially when working close to the edge of what produces good results. When optimal settings aren't known beforehand, the system's ability to fine-tune its own step sizes becomes particularly beneficial. This self-regulation helps maintain stable and reliable sampling even under challenging conditions where fixed parameters might fail.